\documentclass[onecolumn,sort&compress,numbers]{els-mrw} 

\usepackage{amsmath,amssymb,amsfonts,amsthm,makeidx,graphicx}
\usepackage{txfonts}
\usepackage{helvet}

\def\bea#1\eea{\begin{align}#1\end{align}}

\newcommand{\bef}{\begin{figure}[h!tb]\centering}
\newcommand{\eef}{\end{figure}}

\usepackage[table]{xcolor}
\usepackage{pifont}
\usepackage{amsmath}
\usepackage{array} 
\usepackage{slashed}
\usepackage{bm}

\definecolor{myblue}{RGB}{0,0,255}
\definecolor{myred}{RGB}{204,0,0}

\begin{document}


\chapter{Cold QCD in nuclear media: theory}\label{chap1}

\author[1,2]{Zhong-Bo Kang}%
\address[1]{\orgname{University of California, Los Angeles},
\orgdiv{Department of Physics and Astronomy}, 
\orgaddress{CA 90095, USA}}
\address[2]{\orgname{University of California, Los Angeles},
\orgdiv{Mani L. Bhaumik Institute for Theoretical Physics}, 
\orgaddress{CA 90095, USA}}

\author[3,4,5]{Farid Salazar}%
\address[3]{\orgname{Temple University}, 
\orgdiv{Department of Physics}, 
\orgaddress{Philadelphia, Pennsylvania 19122, USA}}
\address[4]{RIKEN-BNL Research Center, Brookhaven National Laboratory, Upton, New York 11973, USA}
\address[5]{Physics Department, Brookhaven National Laboratory, Upton, New York 11973, USA}

\author[6]{Xin-Nian Wang}%
\address[6]{\orgname{Central China Normal University}, 
\orgdiv{Institute of Particle Physics}, 
\orgaddress{Wuhan, China}}

\author[7]{Hongxi Xing}%
\address[7]{\orgname{South China Normal University}, 
\orgdiv{Institute of Quantum Matter}, \orgaddress{Guangzhou, China}}


\renewcommand{\articlecopyright}{}
\maketitle

\begin{abstract}[Abstract]
High-energy scattering off nuclei probes parton correlations that govern the propagation of energetic quarks and gluons through QCD matter. This chapter reviews multiple parton scattering in cold nuclear matter, focusing on medium-induced energy loss and transverse momentum broadening. The high-twist (HT) framework relates these effects to multiparton correlations and the jet transport coefficient $\hat q$ inside nuclei. We discuss Landau--Pomeranchuk--Migdal interference, next-to-leading-order factorization and universality tests, and phenomenological constraints on $\hat q$ from electron--nucleus and proton--nucleus reactions. We then examine the correspondence between the HT and Color Glass Condensate descriptions at small $x$. Using direct-photon production as an example, we demonstrate how retaining longitudinal momentum phases and photon emission between successive scatterings recovers the full HT result through twist 4 at leading order. These developments provide a basis for quantitative studies of parton dynamics in nuclei at the Electron-Ion Collider and for constraining cold nuclear matter effects in studies of the quark-gluon plasma.
\end{abstract}

\begin{keywords}
Cold nuclear matter \sep
Parton energy loss \sep
Transverse momentum broadening \sep
High-twist factorization \sep
Color Glass Condensate
\end{keywords}

\section{Introduction}
\label{sec:intro}
High-energy scattering off atomic nuclei provides a controlled setting for studying how energetic quarks and gluons interact with cold QCD matter. Nuclear sizes and density profiles can be constrained independently, while varying the nuclear species changes the amount of matter traversed by a fast parton. These features make nuclei useful for identifying the multiparton correlations that control momentum transfer and radiation, and for studying the transition from interactions with individual nucleons to coherent scattering extending across the nucleus.

These questions are central to the electron--nucleus ($e+A$) program of the Electron-Ion Colliders in the US (EIC)~\cite{Accardi:2012qut,AbdulKhalek:2021gbh} and China~\cite{Anderle:2021wcy}. The scattered lepton constrains the hard-scattering kinematics, enabling systematic studies of parton propagation as the collision energy, resolution scale, and nuclear target are varied. Complementary proton--nucleus ($p+A$) measurements probe initial- and final-state multiple scattering in processes with different color configurations, while forward production at the Relativistic Heavy-Ion Collider (RHIC) and the Large Hadron Collider (LHC) extends the reach to small target momentum fractions~\cite{Xing:2012ii,Arleo:2025oos}.

A quantitative understanding of these cold nuclear matter (CNM) effects is also essential for determining the properties of the hot quark-gluon plasma (QGP) created in heavy-ion collisions. Nuclear modifications of parton distributions and multiple parton scattering can alter the yields and transverse momentum distributions of hard probes even in the absence of a hot medium. Measurements in $e+A$ and $p+A$ collisions therefore help constrain the CNM contributions that must be disentangled from modifications induced by the QGP~\cite{Arleo:2025oos}.

The goal of this chapter is to review theoretical descriptions of multiple parton scattering in cold nuclear matter, focusing on its consequences for energy loss and transverse momentum broadening. Three questions guide the discussion: how multiple scattering generates these effects, to what extent the same nuclear correlations describe different hard processes, and how the description changes as coherent scattering becomes important at small $x$.

Both effects arise from successive soft interactions of a fast parton with the nuclear medium. The resulting momentum transfers broaden the parton's transverse momentum distribution and can also induce gluon radiation. In semi-inclusive deep inelastic scattering (SIDIS), final-state rescattering and radiation modify the fragmentation of the struck parton; in Drell--Yan (DY) production, initial-state energy loss modifies the momentum of the incoming projectile parton before the hard collision~\cite{Guo:2000nz,Wang:2001ifa,Xing:2011fb}. The strength of these interactions is commonly characterized by the jet transport coefficient $\hat q$, defined as the average transverse momentum squared acquired by a fast parton per unit path length. Whereas broadening reflects accumulated transverse momentum transfer, induced radiation also depends on the gluon formation time. When this time extends over successive scatterings, interference between radiation amplitudes suppresses emission relative to incoherent, independent scatterings. This Landau--Pomeranchuk--Migdal (LPM) effect connects the energy and nuclear-size dependence of radiation to the space-time structure of multiple scattering~\cite{Baier:1996sk,Zakharov:1996fv}.

The principal theoretical framework reviewed here is the high-twist (HT) approach, based on generalized collinear factorization~\cite{Qiu:1990xxa,Qiu:1990xy}. Additional parton scatterings appear as power corrections to leading-twist cross sections. Although suppressed by inverse powers of the hard scale, scatterings involving different nucleons can receive a nuclear-size enhancement proportional to $R_A\propto A^{1/3}$. For radiative corrections where the gluon formation time is comparable to the nuclear size, the LPM effect makes the enhancement quadratic in the nuclear radius, i.e., proportional to $R_A^2\propto A^{2/3}$. The leading nuclear-enhanced power corrections arise at twist 4 and involve quark--gluon and gluon--gluon correlation functions~\cite{Luo:1993ui,Luo:1994np}. For a large, loosely bound nucleus, the relevant diagonal multiparton correlations can be approximated by a parton distribution multiplied by a path integral of $\hat q$. This relates the underlying QCD matrix elements to the transport coefficient employed in phenomenology. Explicit next-to-leading-order (NLO) calculations of transverse-momentum-weighted SIDIS and DY cross sections establish factorization for these twist-4 observables, test the process independence of the relevant quark--gluon correlations, and derive their scale evolution, which in general involves additional multiparton correlations~\cite{Kang:2013raa,Kang:2014ela,Kang:2016ron}. Building on this framework, a global analysis of SIDIS, DY, and quarkonium broadening together with nuclear DIS data has extracted a kinematics-dependent $\hat q(x,Q^2)$ and examined whether the same transport coefficient can describe measurements from different processes~\cite{Ru:2019qvz}.

As the target momentum fraction $x$ decreases, the longitudinal coherence length grows, allowing scattering to extend across the nucleus. Within the HT framework, longitudinal phases in the multiparton correlation functions retain information about this coherence. At sufficiently small $x$, the increasing gluon density can also make nonlinear QCD dynamics important. The Color Glass Condensate (CGC) describes this regime in terms of color fields and Wilson lines that resum eikonal multiple scattering, with the saturation scale $Q_s$ characterizing the typical transverse momentum associated with the nuclear gluon field~\cite{McLerran:1993ni,McLerran:1993ka,Gelis:2010nm}. In the overlapping regime where the relevant hard scale satisfies $Q^2\gg Q_s^2$, a power expansion of CGC observables can be compared with the HT description. Recent studies of direct-photon production in $p+A$ collisions establish the correspondence between the HT and CGC descriptions through twist 4 at leading order~\cite{Fu:2023jqv,Fu:2024sba}. Recovering the full HT result requires going beyond the strict eikonal approximation to retain longitudinal momentum phases and include photon emission between successive scatterings. These ingredients preserve the formation-time information underlying LPM interference. This correspondence clarifies how the descriptions of incoherent multiple scattering and coherent small-$x$ dynamics meet in their common domain of validity.

The chapter is organized as follows. Section~\ref{sec:eloss} reviews medium-induced radiation and its applications to hadron attenuation in SIDIS and initial-state energy loss in DY. Section~\ref{sec:TMB} examines transverse momentum broadening as a test of nuclear correlations across processes, from NLO factorization to global extractions of $\hat q(x,Q^2)$. Section~\ref{sec:CGC} discusses the role of longitudinal coherence in the HT--CGC correspondence, using direct-photon production to establish matching through twist 4 at leading order. Section~\ref{sec:conclusions} summarizes the implications for understanding cold nuclear matter at the EIC and interpreting hard probes in heavy-ion collisions.

\section{Parton energy loss in cold nuclear matter}
\label{sec:eloss}

In proton-nucleus and nucleus-nucleus collisions,  initial partons from the proton or nucleus go through multiple scatterings with the nucleons inside the target nucleus before the hard processes of interest. Similarly, final-state partons from a hard process can also undergo final-state multiple scattering within the target nucleus. Such initial- or final-state multiple scattering can lead to transverse momentum broadening and parton energy loss. In these cases,  an energetic quark or gluon traversing the QCD matter interacts with the cold nuclear medium through elastic scattering and medium-induced radiation. Elastic collisions transfer energy and momentum directly to medium constituents, whereas inelastic processes modify the parton shower through induced gluon emissions, leading to parton energy loss. For sufficiently energetic light partons and extended media, induced gluon radiation generally provides the dominant contribution to the average energy loss which can be calculated in perturbative QCD.

Similar parton energy loss inside a hot quark-gluon plasma in high-energy nucleus-nucleus collisions was first investigated by several complementary approaches. The first calculation by Gyulassy and Wang employed a medium composed of static, screened scattering centers~\cite{Gyulassy:1993hr,Wang:1994fx}. The coherent interaction of the radiated gluon with multiple scattering centers, which is essential in the soft-radiation limit, was subsequently incorporated in the BDMPS framework~\cite{Baier:1994bd,Baier:1996sk,Baier:1996kr} and in Zakharov's path-integral formulation~\cite{Zakharov:1996fv}. The GLV~\cite{Gyulassy:1999zd,Gyulassy:2000fs} and Wiedemann~\cite{Wiedemann:2000za} approaches organize the calculation as an expansion in the medium opacity. The high-twist (HT) formalism instead uses a twist expansion within collinear factorization, encoding the medium in multi-parton correlation functions~\cite{Guo:2000nz,Wang:2001ifa,Zhang:2003yn,Schafer:2007xh,Majumder:2009ge}. More recently, the SCET$_{\rm G}$ framework has supplemented soft-collinear effective theory with Glauber gluons that describe transverse momentum exchange with the medium~\cite{Ovanesyan:2011xy,Ovanesyan:2011kn}. Though most of these approaches to parton energy loss in QGP can be applied to parton scattering inside a cold nuclear medium through some modeling of the static scattering centers or the sources of the Glauber gluon, the high-twist approach was formulated from the beginning for parton scattering inside a large nucleus. We will review the parton energy loss within this approach in the electron-nucleus and proton-nucleus collisions and discuss its connection to others at the end.

\subsection{High-twist approach to multiple parton scattering}

We begin with the vacuum radiation in semi-inclusive deeply inelastic scattering (SIDIS) off a large nucleus,
\begin{equation}
e(l_1)+A(p)\rightarrow e(l_2)+h(l_h)+\mathcal{X}, \qquad p=[p^+,0,0],
\qquad
q=l_1-l_2=
\left[-\frac{Q^2}{2q^-},q^-,0\right],
\end{equation}
where  \(Q^2=-q^2\) and \(p\) denotes the momentum per nucleon in the infinite-momentum frame for collinear factorization. The differential SIDIS cross section is
\begin{equation}
d\sigma=
\frac{e^4}{2s}\frac{1}{q^4} \sum_q e_q^2
\int\frac{d^4l_2}{(2\pi)^4}\,
2\pi\delta(l_2^2)\,
L_{\mu\nu}W_q^{\mu\nu},  \qquad L_{\mu\nu}=\frac{1}{2}\,{\rm Tr}
\left[ \slashed l_1\gamma_\mu \slashed l_2\gamma_\nu \right],
\end{equation}
where \(L_{\mu\nu}\) is the leptonic tensor and $e_q$ the fractional charge of the quark with flavor $q$. The semi-inclusive hadronic tensor is
\begin{equation}
E_{l_h}\frac{dW_q^{\mu\nu}}{d^3l_h} =
\int d^4y\,e^{-iq\cdot y}
\sum_{\mathcal{X}}
\langle A|J_q^\mu(y)|\mathcal{X},h\rangle
\langle h,\mathcal{X}|J_q^\nu(0)|A\rangle ,
\end{equation}
with \(J_q^\mu=\bar\psi_q\gamma^\mu\psi_q\) the vector current of the quark $q$.

At leading order and leading twist, the virtual photon strikes a quark carrying longitudinal momentum fraction \(x\). The outgoing quark subsequently fragments into measured hadrons. The factorized hadronic tensor is
\begin{equation}
\frac{dW^{\mu\nu}_{S(0)}}{dz_h}
=
\int dx\,
f_q^A(x)\,
H_{(0)}^{\mu\nu}(x,x_B)\,
D_{h/q}(z_h), \;\;  z_h=\frac{l_h^-}{q^-}=\frac{p\cdot l_h}{p\cdot q},\;\; x_{\rm B}=\frac{Q^2}{2p^+q^-}=\frac{Q^2}{2p\cdot q},
\label{eq:SIDIS-LO}
\end{equation}
where the nuclear quark distribution is
\begin{equation}
f_q^A(x) =\int\frac{dy^-}{2\pi} \, e^{-ixp^+y^-}
\frac{1}{2} \langle A|\bar\psi_q(y^-)\gamma^+\psi_q(0)|A\rangle ,
\label{eq:nuclear-pdf}
\end{equation}
and the lowest-order partonic hard part is
\begin{equation}
H_{(0)}^{\mu\nu}(x,x_B) =
\frac{1}{2}
{\rm Tr}
\left[
\slashed p\gamma^\mu
(\slashed q+x\slashed p)\gamma^\nu
\right]
\frac{2\pi}{2p^+q^-}
\delta(x-x_{\rm B}) .
\label{eq:SIDIS-hard-LO}
\end{equation}

\begin{figure}
\begin{center}
  \includegraphics[scale=0.25]{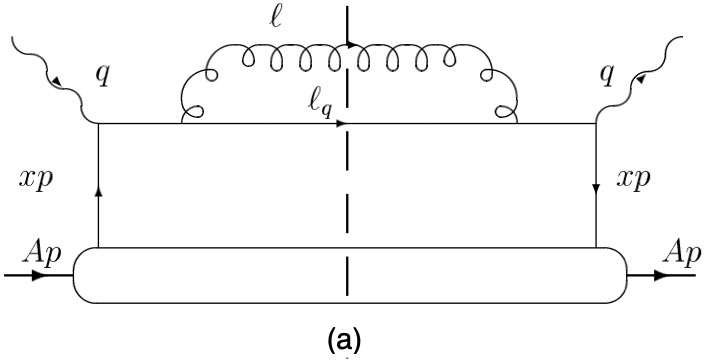}\hspace{0.5in}
   \includegraphics[scale=0.25]{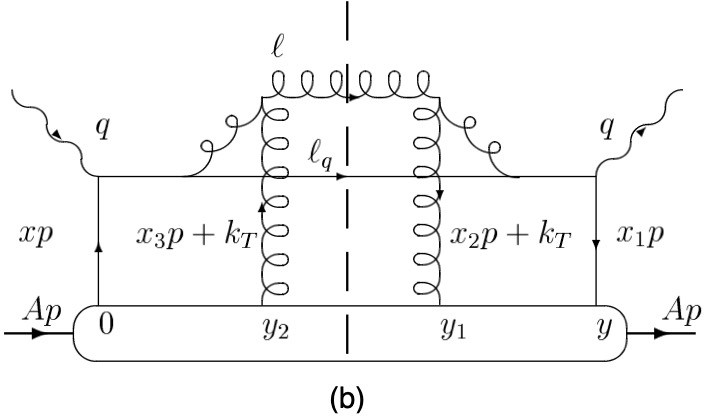}
  \caption{(a) The next-to-leading-order, leading-twist contribution to the hadronic tensor in the SIDIS process. (b) One Feynman diagram for gluon radiation induced by double scattering that contributes to the hadronic tensor in the SIDIS process.}
  \label{fig:dis}
  \end{center}
\end{figure}

At next-to-leading order, the outgoing quark may radiate a gluon as illustrated by the Feynman diagram in Fig.~\ref{fig:dis}(a). If the final quark carries the momentum fraction \(z=l_q^-/q^-\), the real-emission contribution to the hard part is
\begin{equation}
H_{(1)}^{\mu\nu}(x,x_B)=H_{(0)}^{\mu\nu}(x,x_B)\frac{\alpha_s}{2\pi}
\int\frac{dl_\perp^2}{l_\perp^2}\,P_{qq}(z),
\qquad
P_{qq}(z)=C_F\frac{1+z^2}{1-z}.
\label{eq:vacuum-splitting}
\end{equation}
Including hadron production from both daughter partons and adding the contribution from the virtual correction, which provides a subtraction of the infrared divergence in the real correction, one can absorb the collinear divergence from $1/l_\perp^2$ into an infrared-safe and scale-dependent fragmentation function,
\begin{equation}
\frac{dW_S^{\mu\nu}}{dz_h}=\int dx\,f_q^A(x)\,
H_{(0)}^{\mu\nu}(x,x_B)\,D_{h/q}(z_h,\mu^2),
\label{eq:SIDIS-renormalized}
\end{equation}
\begin{equation}
D_{h/q}(z_h,\mu^2)= D_{h/q}(z_h) +\frac{\alpha_s}{2\pi}
\int_0^{\mu^2}\frac{dl_\perp^2}{l_\perp^2}
\int_{z_h}^{1}\frac{dz}{z} \left[ P_{qq}^{+}(z)
D_{h/q}\left(\frac{z_h}{z}\right) +
P_{gq}(z) D_{h/g}\left(\frac{z_h}{z}\right) \right],
\label{eq:renormalized-FF}
\end{equation}
\begin{equation}
P_{qq}^{+}(z)=C_F \left[
\frac{1+z^2}{(1-z)_+}+\frac{3}{2}\delta(1-z) \right], \;\;P_{gq}(z)=C_F
\frac{1+(1-z)^2}{z},
\end{equation}
where the ``$+$"-function is defined as 
\begin{equation}
\int_0^1 dz\,\frac{F(z)}{(1-z)_+} =\int_0^1dz\,
\frac{F(z)-F(1)}{1-z}.
\end{equation}

Eq.~\eqref{eq:renormalized-FF} leads to the usual DGLAP evolution equation~\cite{Gribov:1972ri,Dokshitzer:1977sg,Altarelli:1977zs} for the fragmentation functions. Medium-induced radiation modifies this evolution by adding contributions generated by rescattering of the propagating partons inside the target nucleus.

In SIDIS off a large nucleus,  the outgoing quark can have a secondary scattering with a different nucleon inside the nucleus.  Let $\rho_A(y^-,\bm y_\perp)$ be the nucleon density, normalized as
$ \int dy^-d^2y_\perp\,\rho_A(y^-,\bm y_\perp)=A$. The hard photon-quark interaction occurs at \((y^-,\bm y_\perp)\), while the secondary scattering occurs at \((y_1^-,\bm y_\perp)\). We write \(\Delta y_1^-=y_1^--y^-\) for the longitudinal separation, and use \(z\) for the momentum fraction of the radiated gluon in the medium-induced spectrum below. The radiated gluon may be induced either by the initial hard photon-quark interaction or the secondary scattering, and both the quark and the radiated gluon can interact with the medium as shown in Fig.~\ref{fig:dis} (b) for one such secondary scattering in which the radiated gluon interacts with the second target nucleon and radiates another gluon. Interference among these amplitudes leads to the non-Abelian Landau--Pomeranchuk--Migdal (LPM) effect~\cite{Landau:1953gr,Migdal:1956tc}.

 Neglecting energy and longitudinal-momentum transfer from the medium, the induced spectrum can be written in the compact form
\begin{equation}
\frac{dN_g}{dl_\perp^2dz} = \int\frac{dy^-d^2y_\perp}{A}\,
\rho_A(y^-,\bm y_\perp)
\int_{y^-}^{\infty}dy_1^-\,\rho_A(y_1^-,\bm y_\perp)
\frac{2\pi^2\alpha_s}{N_c}
\int\frac{d^2k_\perp}{(2\pi)^2}\,
\frac{\phi(0,\bm k_\perp)}{k_\perp^2}\,
\mathcal{K}
(\bm l_\perp,\bm k_\perp,z;\Delta y_1^-)
\label{eq:general-induced-spectrum}
\end{equation}
where \(\phi(0,\bm k_\perp)\) is the small-\(x\) transverse-momentum-dependent (TMD) gluon distribution of the target nucleon inside the nucleus, and the medium-induced splitting kernel \(\mathcal{K}\) contains the color factors, gluon radiation amplitude, and LPM phases. Explicit expressions for this kernel, including nonzero energy--momentum transfer, can be found in Refs.~\cite{Zhang:2021tcc,Zhang:2019toi}.

The small-$x$ TMD gluon distribution in Eq.~\eqref{eq:general-induced-spectrum}  is related to the effective parton--nucleon cross section,
\begin{equation}
\sigma_{aN}=\frac{4\pi^2\alpha_s C_2(R)}{N_c^2-1}
\int\frac{d^2k_\perp}{(2\pi)^2}\,
\frac{\phi(0,\bm k_\perp)}{k_\perp^2},
\label{eq:qN-cross-section}
\end{equation}
for a parton $a$ in color representation \(R\), where \(C_2(R)\) is the 
quadratic Casimir color factor in SU$(N_c)$. For a quark in the fundamental representation, $C_F=(N_c^2-1)/2N_c$. Given \(\rho_A \sigma_{aN}\) as the inverse mean free path or the interaction rate of the parton in the nucleus, one can understand the radiative gluon spectrum in Eq.~\eqref{eq:general-induced-spectrum} as the normalized probability of finding the first nucleon in the photon-quark hard scattering times the quark-nucleon interaction rate \(\sigma_{aN}\rho_A\) and the medium-induced splitting kernel \(\mathcal{K}\). With the parton-nucleon cross section, one can also define the parton transport coefficient, 
\begin{equation}
\hat q_a(y)\equiv \int d^2k_\perp \rho_A(y) k_\perp^2\frac{d\sigma_{aN}}{d^2k_\perp}
=\frac{4\pi^2\alpha_s C_2(R)}{N_c^2-1}
\rho_A(y)
\int\frac{d^2k_\perp}{(2\pi)^2}\,
\phi(0,\bm k_\perp),
\label{eq:qhat-TMD}
\end{equation}
as the average transverse momentum squared transferred to the parton per unit path length.

The LPM effect suppresses gluon emissions whose formation time is larger than the separation between successive interactions. For an induced splitting where the gluon is radiated from the quark, the characteristic formation time is
\begin{equation}
\tau_{1f} = \frac{1}{x_{\rm L}p^+} =
\frac{2q^-z(1-z)}{l_\perp^2}.
\label{eq:formation-time}
\end{equation}
For an induced splitting where the gluon is emitted from the medium gluon as shown in Fig.~\ref{fig:dis} (b), the transverse momentum exchange \(\bm{k}_\perp\) enters the gluon splitting and the formation time is
\begin{equation}
\tau_{2f}=\frac{1}{x_{\rm LD}p^+}=\frac{2q^-z(1-z)}{(\bm l_\perp-\bm k_\perp)^2}.
\end{equation}
These two longitudinal momentum fractions, $x_{\rm L}$ and $x_{\rm LD}$, and formation times enter the interference phases, \(1-\cos(x_{\rm L}p^+\Delta y_1^-)\) and
\(1-\cos(x_{\rm LD}p^+\Delta y_1^-)\) in the medium-induced splitting kernel \(\mathcal{K}\),  suppressing coherent radiation when the formation time exceeds the distance between the two scattering points.

The structure of the splitting kernel \(\mathcal{K}\) simplifies considerably in the soft-gluon limit, \(z\ll1\), where the \( g+g*\rightarrow g\) splitting as shown in Fig.~\ref{fig:dis} (b) dominates~\cite{Zhang:2018kkn,Zhang:2019toi},
\begin{equation}
\mathcal{K}(\bm l_\perp,\bm k_\perp,z;\Delta y_1^-)=
C_A\frac{\alpha_s}{2\pi}\frac{1+(1-z)^2}{z}
\frac{2\bm k_\perp\cdot\bm l_\perp}
{l_\perp^2(\bm l_\perp-\bm k_\perp)^2}
\left[1-\cos(x_{\rm LD}p^+\Delta y_1^-)\right].
\label{eq:soft-GLV-spectrum}
\end{equation}

When \(\Delta y_1^-\ll\tau_f\), the phase factor \(1-\cos\left(\Delta y_1^-/\tau_f\right)\) is approximately \(\tfrac12(\Delta y_1^-/\tau_f)^2\). Medium-induced gluon radiation is then strongly suppressed. For separations larger than the formation time, the phase oscillates rapidly and the emission becomes effectively incoherent. The characteristic path-length dependence of radiative energy loss follows from integrating this phase over the medium profile,
\begin{equation}
\Delta E_{\rm rad}=E\int dz\,z\int dl_\perp^2\,
\frac{dN_g}{dl_\perp^2dz},
\label{eq:radiative-energy-loss}
\end{equation}
with the integration limits fixed by the kinematics. In a static, approximately uniform medium, the first-order result exhibits an approximately quadratic dependence on the path length when formation-time effects are important.

In the original HT calculation~\cite{Guo:2000nz,Wang:2001ifa,Zhang:2003yn,Schafer:2007xh}, one expands the splitting kernel \(\mathcal{K}(\bm l_\perp,\bm k_\perp,z;\Delta y_1^-)\) in the medium momentum $\bm k_\perp$. Keeping the terms that survive azimuthal averaging gives
\begin{equation}
\frac{2\bm k_\perp\cdot\bm l_\perp}
{l_\perp^2(\bm l_\perp-\bm k_\perp)^2}\left[1-\cos(x_{\rm LD}p^+\Delta y_1^-)\right]
\simeq 4\frac{(\bm k_\perp\cdot\bm l_\perp)^2}{l_\perp^6}
\left[1-\cos(x_{\rm L}p^+\Delta y_1^-)-x_{\rm L}p^+\Delta y_1^-
\sin(x_{\rm L}p^+\Delta y_1^-) \right] +\mathcal{O}(k_\perp^4),
\label{eq:collinear-expansion}
\end{equation}
in the limit $k_\perp\ll l_\perp$. The final sine term in the above equation originates from differentiating the phase during the collinear expansion.

The above collinear expansion is not uniform over the entire transverse momentum phase space. In particular, it misses the region $\bm k_\perp\simeq\bm l_\perp$ where \((\bm l_\perp-\bm k_\perp)^2\) is small and the corresponding formation time \(\tau_{2f}\) becomes large. This region can provide an important contribution despite violating the nominal assumption \(k_\perp\ll l_\perp\). One approximation to its effect is to drop the phase-derivative (sine) term in Eq.~\eqref{eq:collinear-expansion}. The \(k_\perp\) integration can then be factorized into the transport coefficient in Eq.~\eqref{eq:qhat-TMD} and yields the radiative gluon spectrum in the HT approach \cite{Guo:2000nz,Wang:2001ifa,Zhang:2003yn,Schafer:2007xh},
\begin{equation}
\frac{dN_g^{\rm HT}}{dl_\perp^2dz} =
\int\frac{dy^-d^2y_\perp}{A}\,\rho_A(y^-,\bm y_\perp)
\int_{y^-}^{\infty}dy_1^-\,\hat q_F(y_1^-,\bm y_\perp)
C_A \frac{\alpha_s}{\pi}\frac{1+(1-z)^2}{z}\frac{1}{l_\perp^4}
\left[1-\cos(x_{\rm L}p^+\Delta y_1^-)\right].
\label{eq:HT-induced-spectrum}
\end{equation}

Considering quark scattering with partons inside the target nucleon through a gluon exchange as given by a static potential screened by a saturation scale $Q_s$,
\begin{equation}
\frac{d\sigma}{dk_\perp^2}=\sum_a
\frac{C_F K_2(a)}{N_c^2-1}
\frac{4\pi\alpha_s^2}
{(k_\perp^2+Q_s^2)^2},
\label{eq:GW-cross-section}
\end{equation}
where \(K_2(a)\) is the color-factor-weighted effective parton density inside the target nucleon. Comparison with Eq.~\eqref{eq:qN-cross-section} gives
\begin{equation}
\frac{\phi(0,\bm k_\perp)}{k_\perp^2}=
\sum_a K_2(a)
\frac{4\alpha_s}{(k_\perp^2+Q_s^2)^2}\equiv \frac{4\alpha_s K_2}{(k_\perp^2+Q_s^2)^2}
\label{eq:GW-TMD}
\end{equation}
One can complete the integration over the medium $\bm k_\perp$ in Eqs.~(\ref{eq:general-induced-spectrum}) and (\ref{eq:soft-GLV-spectrum}) by the change of variables $\bm l_\perp-\bm k_\perp = \bm l^\prime_\perp$
\begin{equation}
\int d^2k_\perp \frac{k_\perp l^\prime_\perp\cos\phi+k_\perp^2}{{l^\prime}_\perp^2(k_\perp^2+{l^\prime}^2_\perp+2k_\perp l^\prime_\perp\cos\phi)   
(k_\perp^2+Q^2_s)^2}
=\frac{\pi}{{l^\prime}_\perp^2} \int_{{l^\prime}^2_\perp}^\infty \frac{dk_\perp^2}{(k_\perp^2+Q_s^2)^2}
=\frac{\pi}{{l^\prime}^2_\perp({l^\prime}^2_\perp+Q_s^2)}
\end{equation}
 The radiative gluon spectrum can be expressed as
\begin{equation}
\frac{dN_g}{dl_\perp^2dz}
\simeq 
\int\frac{dy^-d^2y_\perp}{A}\,\rho_A(y^-,\bm y_\perp)
\int_{y^-}^{\infty}dy_1^-\, \hat q^\prime_F(y_1^-,\bm y_\perp)
C_A\frac{\alpha_s}{\pi} \frac{1+(1-z)^2}{z} 
\frac{1}{l_\perp^2(l_\perp^2+Q_s^2)}
\left[1-\cos(x_{\rm L}p^+\Delta y_1^-)\right],
\label{eq:GLV-static-final}
\end{equation}
where the effective jet transport coefficient $\hat q^\prime_F$ is 
\begin{equation}
    \hat q^\prime _F(y^-,\bm y_\perp)=\frac{\pi\alpha_s^2K_2}{N_c}\rho_A(y^-,\bm y_\perp).
\end{equation}
For \(l_\perp^2\gg Q_s^2\), the transverse factor behaves as \(1/l_\perp^4\), and one recovers the original HT results in Eq.~(\ref{eq:HT-induced-spectrum}). However, the effective jet transport coefficient $\hat q^\prime_F$ differs from the quark transport coefficient  according to Eq.~(\ref{eq:qhat-TMD}) in this model of parton-nucleon scattering,
\begin{equation}
\hat q_F(y^-,\bm y_\perp,Q^2) =
\frac{2\pi\alpha_s^2}{N_c} K_2\rho_A(y^-,\bm y_\perp)
\left[
\ln\left(1+\frac{Q^2}{Q_s^2}\right)
-\frac{Q^2}{Q^2+Q_s^2}
\right],
\label{eq:qhat-GW}
\end{equation}
by a factor of $\sim 2\log(Q^2/Q_s^2)$. Thus, the generalized high-twist and the original HT spectra have the same leading transverse and phase structure once their definitions of \(\hat q_F\), integration limits, and treatment of the phase-derivative contribution are matched. Quantitative comparisons nevertheless require care because the collinear expansion integrates over a different region of transverse phase space than the full opacity kernel.

\subsection{Final-state parton energy loss}

In deep-inelastic scattering (DIS), parton energy loss reduces the yield of energetic hadrons from a nuclear target relative to that from a nucleon or deuterium target. This suppression can be calculated by solving the medium-modified DGLAP (mDGLAP) evolution equations for the fragmentation functions~\cite{Wang:2009qb,Chang:2014fba,Majumder:2011uk,Majumder:2013re}:
\begin{eqnarray}
 \label{eq: modified DGLAP1}
 \frac{\partial \tilde{D}_{h/q}(z_h,Q^2)}{\partial \ln Q^2}\hspace{-4pt} &=&\hspace{-4pt}\frac{\alpha_s(Q^2)}{2\pi}\hspace{-4pt} \int_{z_h}^1
 \frac{dz}{z}\left [ \tilde{P}_{qq}(z,Q^2)\tilde{D}_{h/q}(\frac{z_h}{z},Q^2)
 + \tilde{P}_{gq}(z,Q^2)\tilde{D}_{h/g}(\frac{z_h}{z},Q^2)\right ] ,\\
\label{eq: modified DGLAP2}
  \frac{\partial \tilde{D}_{h/g}(z_h,Q^2)}{\partial\ln Q^2}\hspace{-4pt}&=&\hspace{-4pt}\frac{\alpha_s(Q^2)}{2\pi} \hspace{-4pt} \int_{z_h}^1
 \frac{dz}{z}\left [ \tilde{P}_{gg}(z,Q^2)\tilde{D}_{h/g}(\frac{z_h}{z},Q^2)
 + \sum_{q=1}^{2n_f}\tilde{P}_{qg}(z,Q^2)\tilde{D}_{h/q}(\frac{z_h}{z},Q^2)\right ] ,
\end{eqnarray}
where the modified splitting functions $\tilde P _{ba}$ are given by the sum of the vacuum ones 
and the medium corrections \cite{Schafer:2007xh},
\begin{equation}
 \tilde P _{ba}(z,Q^2)=P_{ba}(z)+ \Delta P_{ba}(z,Q^2).
\end{equation}
For a quark jet according to Eqs.~(\ref{eq:HT-induced-spectrum}) and (\ref{eq:GLV-static-final}), for example,
  \begin{equation}  
 \Delta P_{qq}(z,\ell_T^2) \approx \frac{2 C_{A}}{\ell_T^2+Q_s^2} \frac{1+(1-z)^{2}}{z} \int_{y^-}^{\infty}dy_1^-\,\hat q_F(y^-_1,{\bf y}_\perp)\left[1-\cos\!\left(x_Lp^+\Delta y_1^-\right)\right],
 \label{delta_sp_qqg}
 \end{equation}
 where the hard interaction position $y^-$ enters the lower limit and the separation $\Delta y_1^-=y_1^--y^-$. The impact parameter is $\bm{y}_\perp$.

 Solving these evolution equations requires fragmentation functions at an initial scale \(Q_0\). There are several choices for these initial fragmentation functions. If vacuum fragmentation functions are imposed at \(Q_0\), medium interactions below that scale are excluded from the calculation. The resulting suppression must then be generated entirely by subsequent evolution inside the nuclear medium, affecting both its virtuality dependence and the value of the transport parameter \(\hat q_F\) inferred from data. Conversely, an initial condition obtained by applying medium-induced evolution below \(Q_0\) may overestimate the energy loss. A more reasonable choice is to use the induced single-gluon spectrum to construct the probability for multiple emissions at \(Q_0\) and then convolve the resulting parton and gluon energy distributions with vacuum fragmentation functions.
 
For the convoluted initial condition, we assume that the number of induced gluon emissions from a quark follows a Poisson distribution, with each emission governed by the medium-induced single-gluon spectrum. Their convolution defines a quenching weight,
\begin{equation}
  P_a(\epsilon,Q_0^2)=\sum_{n=0}^\infty \frac{1}{n!}
  \left[ \prod_{i=1}^n \int_0^{1} dz_i \frac{dN_g^a}{dz_i}(Q_0^2)
    \right]
    \delta\left(\epsilon-\sum_{i=1}^n z_i\right)
 \exp\left[ - \int_0^1 dz\frac{dN_g^a}{dz}(Q_0^2)\right]\, ,
   \label{eq:pdeltaeps}
\end{equation}
which is the probability that a parton \(a\) loses a fraction \(\epsilon\) of its initial energy. The induced gluon spectrum per emission is given by
 \begin{eqnarray}
 \frac{dN_g^a}{dz}(Q_0^2) &=& \frac{\alpha_s(Q^2_0)}{2\pi}\int_0^{Q_0^2} \frac{d\ell_T^2}{\ell_T^2} \Delta P_{ga}(z,\ell_T^2), 
  \label{eq:dngdz}
  \end{eqnarray}
One can also determine a distribution \(G_a(\epsilon,Q_0^2)\) for gluons, subject to the constraint that the total fractional energy loss in an event cannot exceed one, $\int [d\epsilon (1-\epsilon)P_a(\epsilon,Q_0^2)+\epsilon G_a(\epsilon,Q_0^2)]=1$. The modified initial fragmentation function is then the sum of two terms,
\begin{equation}
\widetilde D_{h/q}(z,Q_0^2)=
\int_0^1 d\epsilon\,
P_q(\epsilon,Q_0^2)
\frac{1}{1-\epsilon}
D_{h/q}\!\left(\frac{z}{1-\epsilon},Q_0^2\right) +
\int_0^1 d\epsilon\,
G_q(\epsilon,Q_0^2)
\frac{1}{\epsilon}
D_{h/g}\!\left(\frac{z}{\epsilon},Q_0^2\right).
\end{equation}
The first term describes fragmentation of the leading quark with its reduced energy; the second describes fragmentation of the radiated gluons. Including fragmentation of the emitted gluons ensures that the initial fragmentation functions satisfy the momentum sum rule $ \sum_h\int_0^1 dz\,z\,\widetilde D_{h/a}(z,Q_0^2)=1$.

In semi-inclusive DIS, the struck quark is produced at a position $(y^-,\boldsymbol y_\perp)$ within the nucleus and propagates through the remaining nuclear matter. Both the mDGLAP evolution equations and the initial conditions implicitly depend on the production position. One should average its medium-modified fragmentation function over the production positions and propagation paths according to Eqs.~(\ref{eq:HT-induced-spectrum}) and (\ref{eq:GLV-static-final}). The jet transport coefficient is proportional to the nuclear density, which can be represented by a Woods--Saxon distribution. We define $\hat q_{F0} = \hat q_F(0,\boldsymbol 0_\perp)$ as the jet transport coefficient at the center of the nucleus.

The experimental observable for the suppression of hadron yields is the multiplicity ratio \(R_A^h\),
\begin{equation}
R_A^h(z,\nu) = \left(\frac{N^h(z,\nu)}{N^e(\nu)}|_A\right) / \left(\frac{N^h(z,\nu)}{N^e(\nu)}|_D\right) =\left(\frac{\Sigma e_q^2q(x_B)\widetilde D_q^h(z)}{\Sigma e_q^2q(x_B)}|_A\right) / \left(\frac{\Sigma e_q^2q(x_B) D_q^h(z)}{\Sigma e_q^2q(x_B)}|_D\right),
\end{equation}
where $\nu=E$ is the energy of the virtual photon transferred to the struck quark, $z=p_h/E$ is the energy fraction carried by the final hadron, and $x_B=Q^2/(2p^+q^-)=Q^2/(2M_N\nu)$ is the Bjorken variable for the fractional light-cone momentum carried by the initial quark. The sum runs over all quark and antiquark flavors, and $q(x_B)$ denotes the corresponding parton distribution inside the nucleus. 

\begin{center}
\begin{figure}[htb]
  \centering
     \includegraphics[width=3.2in]{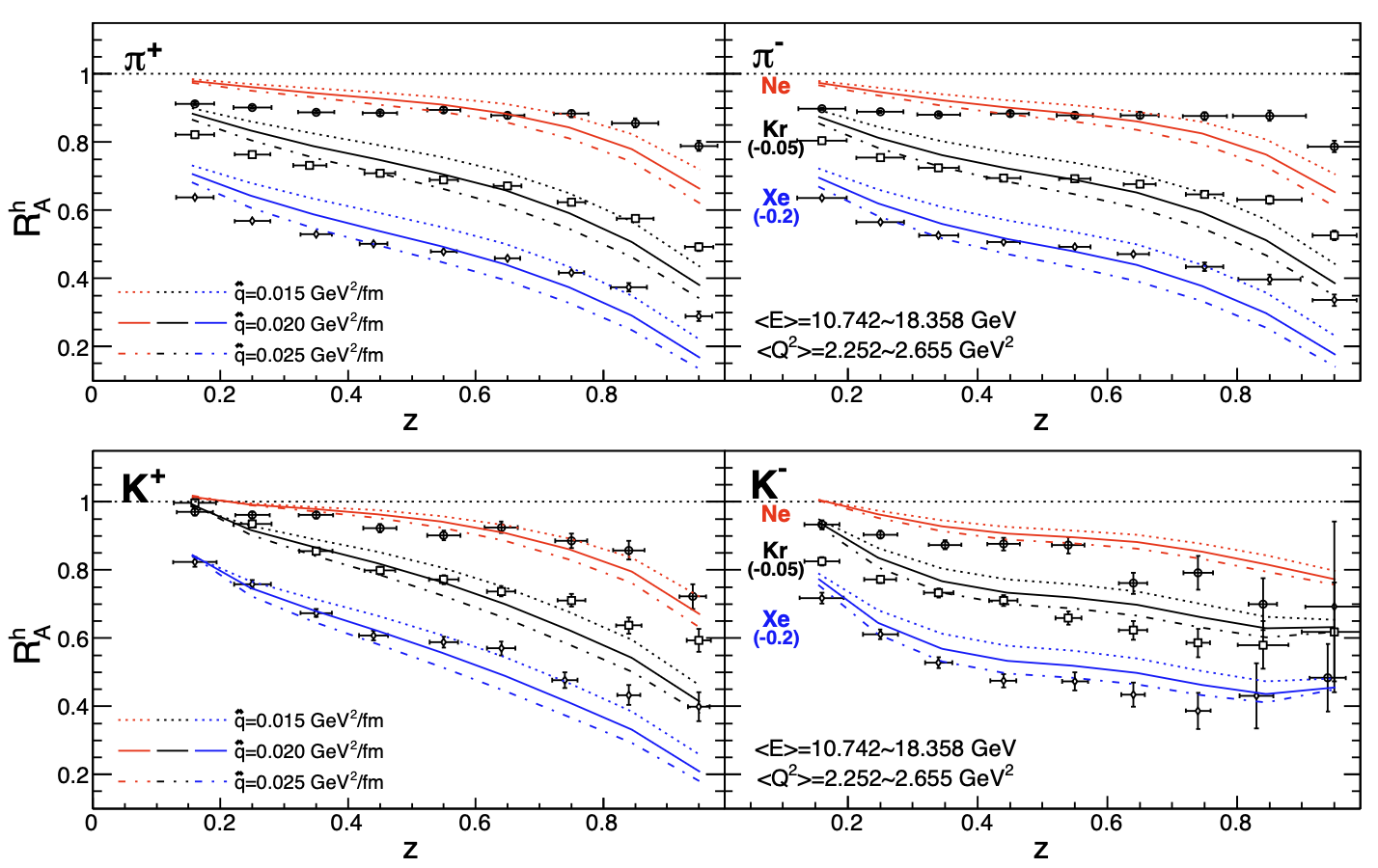}
     \includegraphics[width=3.2in]{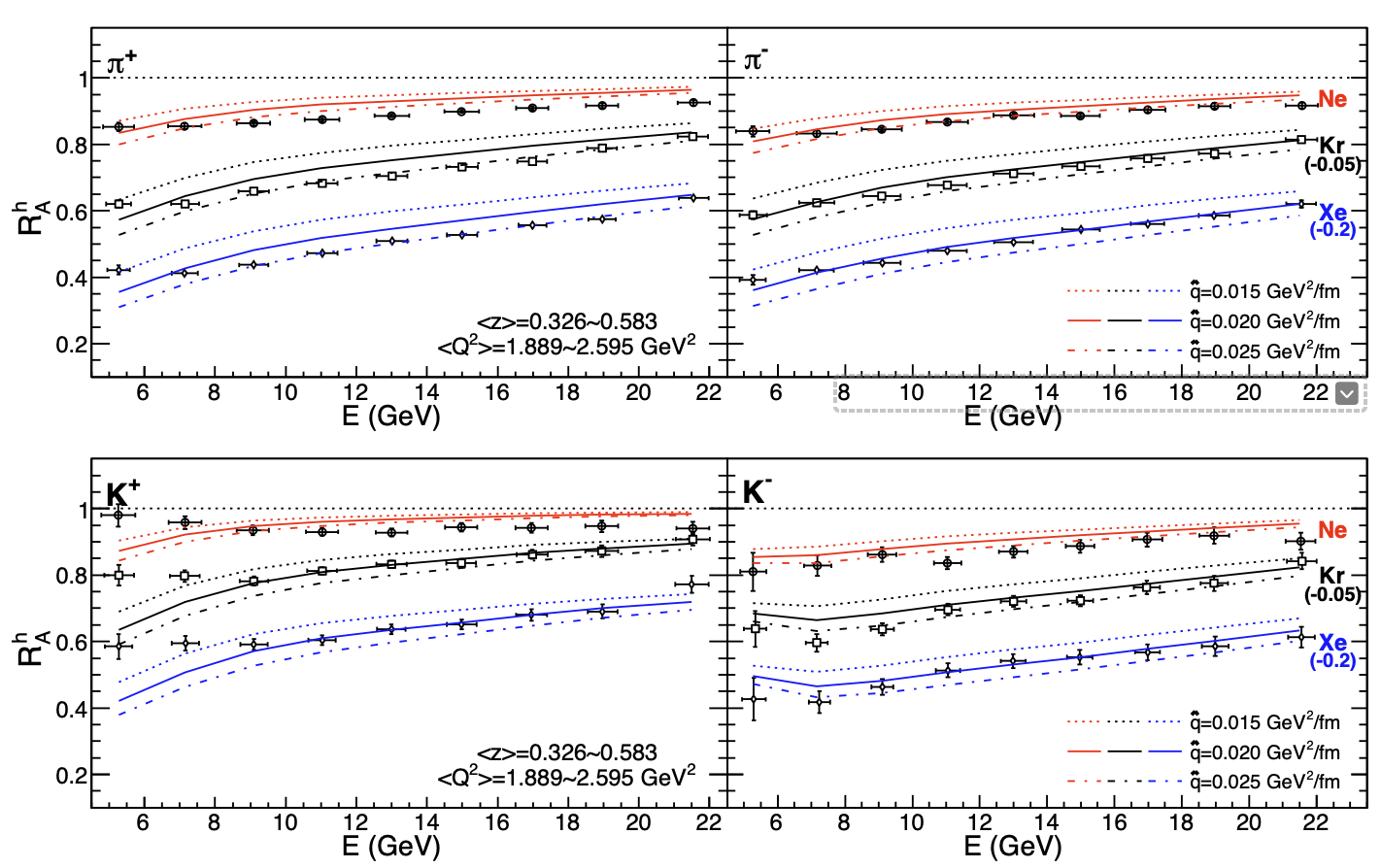}
   \caption{The $z$ (left) and $E$ (right) dependence of the suppression factor $R^h_A$ for pions (upper panels) and kaons (lower panels) with the convoluted initial condition for different values of $\hat{q}_{F0}$ compared with HERMES data \cite{HERMES:2007plz} for Ne, Kr and Xe targets. For clarity, values of $R_A^h$ for Kr and Xe targets are displaced by -0.05 and -0.2, respectively.}
  \label{fig-rhze}
\end{figure}
\end{center}

The calculated suppression factors for neon, krypton, and xenon targets as a function of the hadron momentum fraction \(z\) and struck-quark energy \(E\), as shown in Fig.~\ref{fig-rhze}, describe the HERMES data~\cite{HERMES:2007plz} for pions and kaons reasonably well over much of the measured region, particularly at small and intermediate \(z\). The calculation also reproduces the observed weak \(Q^2\) dependence. Agreement deteriorates at large \(z\) and especially for protons and antiprotons, where effects absent from the calculation, including hadronic interactions and nonperturbative baryon transport mechanisms \cite{Kharzeev:1996sq}, may become important. With these initial conditions for the mDGLAP evolution, fits to HERMES data on the \(z\), \(E\), and \(Q^2\) dependence with the lowest overall \(\chi^2\) per degree of freedom give the preferred jet transport coefficient,
\[
\hat q_{F0}\simeq0.020\pm0.005\ \mathrm{GeV^2/fm}.
\]

These analyses remain subject to the assumptions of the leading-order, twist-four treatment and the multiple-emission model. In particular, additional processes appear necessary to describe baryons and the large-\(z\) region of the fragmentation function. Nevertheless, the higher-twist evolution of the modified fragmentation functions with the convoluted initial conditions can describe the SIDIS experimental data well and provides a quantitative extraction of the jet transport coefficient in cold nuclear matter. This approach has recently been implemented in the eHIJING Monte Carlo model~\cite{Ke:2023xeo} for jet production in electron--ion collisions. The model could be extended to include final-state hadronic interactions, which may be needed to describe proton and antiproton yields and the data at large $z$.

\subsection{Initial-state parton energy loss}

In DIS off a nucleus, the struck quark after the hard photon-quark scattering goes through multiple scattering and loses energy, leading to suppression of final hadrons from the hadronization of the final quark. Similarly, the initial parton from a beam proton can also go through multiple scattering before the hard processes in p+A collisions, such as the Drell-Yan (DY) lepton-pair production \cite{Johnson:2001xfa,Arleo:2002ph,Garvey:2002sn,Fries:1999jj,Fries:2000da}. In this case, a quark from the incoming proton propagates through nuclear matter before annihilating with a target antiquark into a lepton pair through a virtual photon,

\[
q+\bar q\longrightarrow\gamma^*\longrightarrow\ell^+\ell^-.
\]

Let \(Q\) denote the dilepton invariant mass, \(x'\) the beam-parton momentum fraction, and \(x\) the target-parton momentum fraction. At leading order, the virtual-photon production cross section has the factorized form,

\begin{equation}
\frac{d\sigma^{(0)}_{hA\to\gamma^*}}{dQ^2\,dx'}=
\sum_q\int dx\,f_{\bar q/A}(x,Q^2)\,f_{q/p}(x',Q^2)\,H_0(x,p,q),
\label{DY}
\end{equation}
where \(H_0\) is the hard part for quark–antiquark annihilation. In leading-order kinematics, $xx's=Q^2$ and $s$ is the center-of-mass energy squared of p+A per nucleon. At next-to-leading order (NLO), one has to consider both radiative corrections to the quark-antiquark annihilation and quark-gluon Compton scattering. Together, they lead to the NLO correction of the hard part for DY production as well as the scale dependence of the parton distributions that is governed by the DGLAP evolution equations for parton distribution functions in Eq.~(\ref{DY}).

As in SIDIS off a nuclear target, the initial beam partons in DY production in p+A collisions go through multiple scattering with induced gluon radiation (quark-antiquark pair production in the Compton scattering process) inside the target nucleus. These radiative processes interfere with the gluon radiation ($q\bar q$ production) associated with the quark-antiquark annihilation. Such destructive LPM interference suppresses medium-induced gluon radiation when the gluon formation time $\tau_f\sim x^\prime \sqrt{s}/q_\perp^2$ is comparable to or larger than the nucleus size. Here $q_\perp$ is the transverse momentum of the DY lepton pair.  Considering the leading-logarithmic contributions from single and double scattering, one can combine the medium-induced gluon radiation into an effective beam-quark distribution \cite{Xing:2011fb},
\begin{equation}
\widetilde f_{q/p}(x',\mu^2,A) =f_{q/p}(x',\mu^2)+\frac{\alpha_s}{2\pi}
\int_0^{\mu^2}\frac{dq_T^2}{q_T^2}\int_{x'}^1\frac{d\xi}{\xi}
\left[f_{q/p}(\xi)\,\Delta P_{qq}\!\left(\frac{x'}{\xi},q_T^2\right)
+f_{g/p}(\xi)\,\Delta P_{qg}\!\left(\frac{x'}{\xi},q_T^2\right)
\right],
\label{eq:modpdf}
\end{equation}
where \(f_{q/p}\) and \(f_{g/p}\) are the parton distributions inside the beam proton in the vacuum, while \(\Delta P_{qq}\) and \(\Delta P_{qg}\) are the medium corrections to the splitting functions associated with quark-nucleon interactions in the nuclear medium. This effective distribution describes the momentum distribution of a beam quark after its passage through the target nucleus before the DY production. 

If induced emissions from further scattering are resummed in analogy with vacuum radiation, the beam parton distribution will follow the evolution equation which retains a DGLAP-like form with splitting functions $\widetilde P_{ba}=P_{ba}+\Delta P_{ba}$ \cite{Xing:2011fb}. Medium-induced radiation shifts the beam parton distribution away from the high-\(x'\) region. Since production at large \(x'\) requires an incoming quark with a large fraction of the beam momentum, this shift suppresses the DY cross section in the large \(x'\) region.

\begin{center}
\begin{figure}[htb]
  \centering
     \includegraphics[width=5.0in]{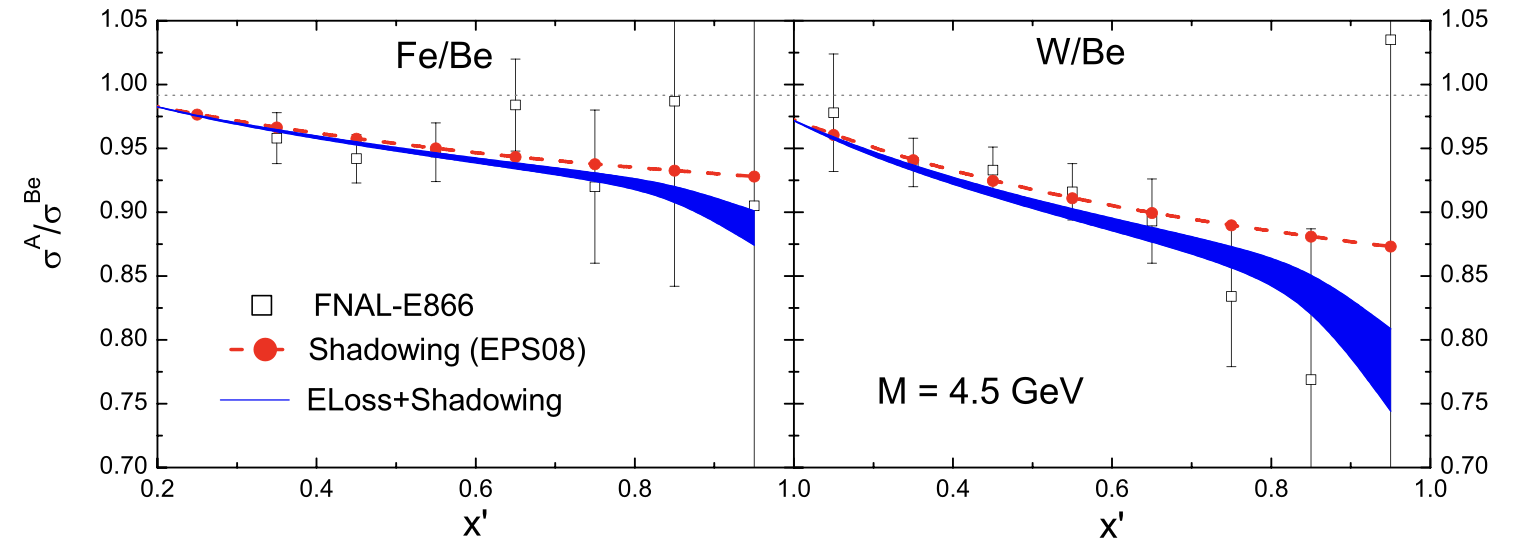}
   \caption{Nuclear modification of the DY cross section in p+Fe and p+W collisions as a function of $x'$ at $E_{lab}=800$ GeV and invariant dilepton mass $M=4.5$ GeV as compared to the Fermilab E866 \cite{NuSea:1999egr} experimental data on the ratio of p+A and p+Be cross sections.  The solid (dashed) lines are theoretical results with (without) initial beam parton energy loss. The shaded bands correspond to $\hat{q}_{F0}=0.024\pm0.008~\rm GeV^2/fm$ \cite{Wang:2009qb}.}
  \label{fig-dy}
\end{figure}
\end{center}

In Eq.~\eqref{eq:modpdf}, the medium dependence enters through the twist-four corrections \(\Delta P_{ba}\). These corrections have an LPM interference pattern analogous to that in Eq.~\eqref{delta_sp_qqg} and depend on the nuclear medium through the quark transport coefficient $\hat q_F$. Since the initial multiple scatterings occur before the hard process of DY production at $(y^-, {\bf y}_\perp)$, one should integrate over their location $(y^-_1 ,{\bf y}_\perp)$ from $y^-_1=-\infty$ to $y^-_1=y^-$ before averaging over the position of the DY production and the impact-parameter ${\bf y}_\perp$. In addition, one should also take into account the nuclear modification of the parton distribution in the target nucleus. One can do this by using the parameterized nuclear parton distribution functions (nPDF) such as the EPS08 parametrization \cite{Eskola:2008ca}.  Using this modified beam quark distribution, one can calculate the nuclear modification factor for the DY cross section,
\begin{eqnarray}
\frac{ \sigma^{pA}}{A \sigma^{pp}}=\frac{\sum_q \int dx
f_{\bar{q}/A}(x,\mu^2)\tilde{f}_{q/p}(x',\mu^2,A)H_0(x,p,q)}{A \sum_q
\int dx f_{\bar{q}/p}(x,\mu^2) f_{q/p}(x',\mu^2)H_0(x,p,q)}.
\label{eq-RDY}
\end{eqnarray}
Shown in Fig.~{\ref{fig-dy}} are the calculated nuclear modification factors for 
iron (Fe) and tungsten (W) targets from Ref.~\cite{Xing:2011fb}. In order to compare to the experimental data from the Fermilab E866 \cite{NuSea:1999egr} experiment which used the beryllium (Be) target to approximate a proton target, one replaces the parton distribution of a proton target with an isospin-corrected one in the denominator of Eq.~(\ref{eq-RDY}). The vacuum PDFs of the nucleons are given by the CTEQ6L \cite{Pumplin:2002vw} parameterization and the nuclear modification of parton distribution functions inside a nuclear target is given by the EPS08 \cite{Eskola:2008ca} parameterization without impact-parameter dependence. The value of the quark transport coefficient used, $\hat q_{F0}=0.024\pm0.008\ \mathrm{GeV^2/fm}$, is consistent with the value extracted from SIDIS measurements of hadron suppression due to final-state parton energy loss.

 The calculation with EPS08 nPDF for the target parton distributions alone already produces most of the nuclear suppression over the measured region. Including the modified effective beam distribution due to initial-state parton energy loss gives an additional suppression, but its magnitude is small over most of this kinematic range with \(\hat q_{F0}\) extracted from SIDIS data.  The effect of the initial-state parton energy loss becomes more visible toward large \(x'\), particularly for the heavier tungsten target. At lower proton-beam energy, $E_{lab}=120$ GeV, for example, the same dilepton mass and a moderately large \(x'\) correspond to a larger target momentum fraction \(x\). In the kinematic region considered here, the EPS08 shadowing effect is small. The energy loss of the beam parton, however, leads to significant suppression of the DY cross section, much stronger than at $E_{lab}=800$ GeV \cite{Xing:2011fb}. Another conclusion of this study is that one can safely neglect the effect of initial parton energy loss in p+A and A+A collisions at high RHIC and LHC energies, especially for jet production in the central rapidity region.


\section{Transverse momentum broadening in cold nuclear matter}
\label{sec:TMB}

When an energetic parton propagates through cold nuclear matter (CNM), either prior to a hard scattering in proton-nucleus ($p+A$) collisions or following a hard interaction in lepton-nucleus ($e+A$) and $p+A$ collisions, it experiences multiple elastic and inelastic scatterings with the medium constituents~\cite{Baier:1996sk,Luo:1993ui,Luo:1994np,Qiu:2003pm}. While medium-induced radiative processes govern parton energy loss and lead to the attenuation of leading hadron spectra as reviewed in Sec.~\ref{sec:eloss}, elastic scatterings exchange soft transverse momenta between the propagating parton and the background color fields of the target nucleus. The cumulative effect of these successive transverse momentum kicks leads to a broadening of the transverse momentum distribution of the observed final-state particles relative to their corresponding elementary proton-proton ($p+p$) or electron-proton ($e+p$) reactions. This phenomenon, known as transverse momentum broadening (TMB), is one of the most direct and sensitive probes of multiple parton scattering dynamics and the fundamental transport properties of cold nuclear media~\cite{Arleo:2025oos,Kang:2013raa,Kang:2014ela,Kang:2016ron,Ru:2019qvz}.

For single-inclusive particle production, such as semi-inclusive deep inelastic scattering (SIDIS) or Drell-Yan (DY) dilepton and heavy quarkonium hadroproduction, transverse momentum broadening is quantified by the shift in the average squared transverse momentum of the measured particle~\cite{Guo:1998rd,Fries:2002mu,Kang:2008us}:
\begin{equation}
\Delta\langle p_\perp^2 \rangle = \langle p_\perp^2 \rangle_{hA} - \langle p_\perp^2 \rangle_{hp}\,, \qquad
\langle p_\perp^2 \rangle \equiv \left. \int d^2\bm{p}_\perp\, p_\perp^2 \frac{d\sigma}{d\mathcal{PS}\,d^2\bm{p}_\perp} \right/ \frac{d\sigma}{d\mathcal{PS}}\,,
\label{eq:tmb-def-general}
\end{equation}
where $h=e,p$ denotes an electron or proton projectile, $A$ denotes the target nucleus and its mass number, and $d\mathcal{PS}$ represents the appropriate final-state phase space. For processes involving two final-state particles or jets produced nearly back-to-back in the transverse plane, multiple parton scatterings induce an acoplanarity, which is characterized by the enhancement $\Delta\langle q_\perp^2 \rangle=\langle q_\perp^2 \rangle_{hA}-\langle q_\perp^2 \rangle_{hp}$. Here $\bm{q}_\perp=\bm{p}_{1\perp}+\bm{p}_{2\perp}$ is the sum of the transverse momenta of the two observed particles~\cite{Luo:1993ui,Qiu:2003vd,Kang:2011bp,Xing:2012ii}. 

Describing the shape of the differential spectrum at $p_\perp\ll Q$ requires transverse-momentum-dependent (TMD) factorization and Collins-Soper-Sterman (CSS) resummation~\cite{Collins:1984kg,Boussarie:2023izj}. However, once integrated over $p_\perp^2$ as in Eq.~\eqref{eq:tmb-def-general}, the weighted cross section becomes an inclusive observable controlled by a single hard scale $Q^2$~\cite{Kang:2008us,Kang:2013raa,Kang:2014ela}. Within the generalized collinear factorization framework developed by Qiu and Sterman~\cite{Qiu:1990xxa,Qiu:1990xy,Luo:1992fz,Luo:1993ui,Luo:1994np}, nuclear-enhanced multiple scatterings are systematically organized as higher-twist power corrections to the cross section. Although higher-twist contributions are nominally suppressed by inverse powers of the hard scale $1/Q^2$, scatterings off different nucleons along the parton trajectory yield a nuclear size enhancement proportional to the nuclear radius $R_A \propto A^{1/3}$. Such a nuclear enhancement compensates the $1/Q^2$ power suppression of the double-scattering amplitude, making the twist-4 parton-parton correlation functions the leading contribution to $\Delta\langle p_\perp^2 \rangle$ and $\Delta\langle q_\perp^2 \rangle$. Therefore, we focus on the discussion of twist-4 contributions in this section.

In this section, we review the theory and phenomenology of transverse momentum broadening and imbalance in cold nuclear matter within the high-twist (HT) expansion formalism. We first present the collinear expansion method and next-to-leading order (NLO) perturbative QCD factorization at twist-4 for SIDIS in $e+A$ collisions in Sec.~\ref{sec:TMB-eA}. In Sec.~\ref{sec:TMB-pA}, we extend the discussion to initial-state and final-state broadening of Drell-Yan, heavy quarkonia, and electroweak vector bosons in $p+A$ collisions, highlighting the universality of twist-4 nuclear correlation functions and the role of color dynamics. In Sec.~\ref{sec:TMB-imbalance}, we examine the transverse momentum imbalance of back-to-back dijet, dihadron, photon-jet, and heavy-flavor pair production in $e+A$ and $p+A$ collisions. Finally, in Sec.~\ref{sec:TMB-global}, we review the global extraction of the jet transport coefficient $\hat{q}$ for CNM from world data.

\subsection{Transverse momentum broadening in electron-nucleus collisions}
\label{sec:TMB-eA}

\begin{figure}[htb]
\centering
\includegraphics[width=2.5in]{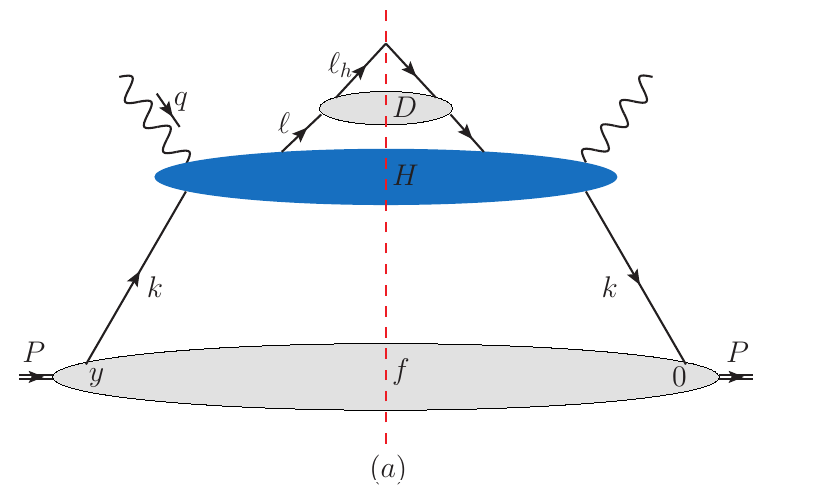}
\hskip 0.2in
\includegraphics[width=2.5in]{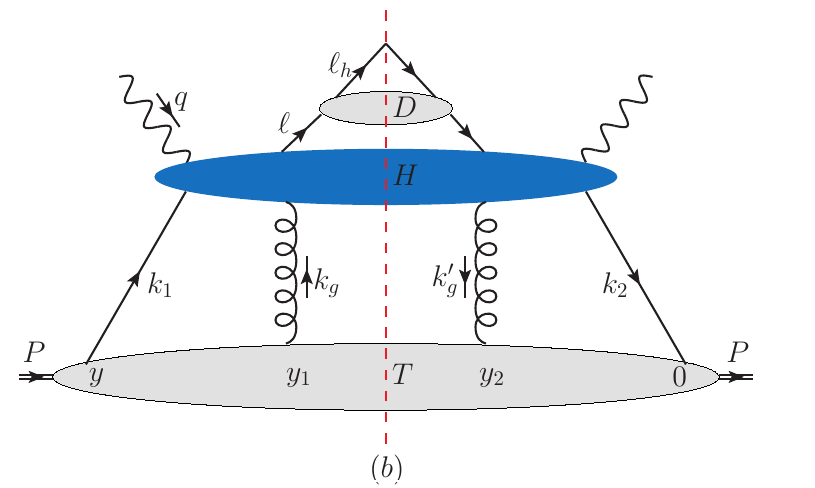}
\caption{The general diagrams for SIDIS off a nuclear medium: (a) single scattering; (b) quark-gluon double scattering. }
\label{fig-DIS}
\end{figure}

Semi-inclusive deep inelastic scattering off a nuclear target, $e(L_1) + A(p) \to e(L_2) + h(\ell_h) + X$, provides a clean environment to investigate final-state parton rescattering in cold nuclear matter, because the virtual photon interacts electromagnetically and only the struck parton propagates through the nuclear medium~\cite{Guo:1998rd,Kang:2013raa,Kang:2014ela}. In a large nucleus, the differential cross section for SIDIS can be expanded in terms of the number of partonic scatterings~\cite{Luo:1994np,Kang:2014ela}:
\begin{equation}
d\sigma = d\sigma^S + d\sigma^D + \cdots\,,
\label{eq:sidis-expansion}
\end{equation}
where the superscripts ``$S$'' and ``$D$'' denote single and double parton scatterings, as shown in Fig.~\ref{fig-DIS}(a) and (b), respectively. Single scattering is localized within a single nucleon and scales linearly with the mass number $A$ up to leading-twist nuclear parton distribution function (nPDF) modifications. In the hadron frame (where the virtual photon and target nucleon move along the longitudinal direction)~\cite{Meng:1991da,Kang:2014ela}, the transverse momentum $\bm{\ell}_{h\perp}$ of the observed hadron is measured with respect to the photon-nucleon collision axis. While single scattering contributes to the average transverse momentum squared $\langle \ell_{h\perp}^2 \rangle$ in both $e+A$ and $e+p$ collisions, these ``vacuum" contributions can be neglected in the difference $\Delta\langle \ell_{h\perp}^2 \rangle = \langle \ell_{h\perp}^2 \rangle_{eA} - \langle \ell_{h\perp}^2 \rangle_{ep}$. Consequently, the leading nuclear-size-enhanced ($\propto A^{1/3}$) contribution to the broadening arises from the double-scattering cross section $d\sigma^D$~\cite{Kang:2013raa,Kang:2014ela}:
\begin{equation}
\Delta\langle \ell_{h\perp}^2 \rangle \approx \left. \frac{d\langle \ell_{h\perp}^2 \sigma^D \rangle}{d\mathcal{PS}} \right/ \frac{d\sigma^S}{d\mathcal{PS}}\,, \qquad
\frac{d\langle \ell_{h\perp}^2 \sigma^D \rangle}{d\mathcal{PS}} \equiv \int d^2\bm{\ell}_{h\perp}\, \ell_{h\perp}^2 \frac{d\sigma^D}{d\mathcal{PS}\,d^2\bm{\ell}_{h\perp}}\,,
\label{eq:sidis-tmb-def}
\end{equation}
where the longitudinal phase space element is given by $d\mathcal{PS} = dx_B\,dy\,dz_h$, with the standard Lorentz-invariant SIDIS variables defined as
\begin{equation}
x_B = \frac{Q^2}{2p\cdot q}\,, \qquad y = \frac{p\cdot q}{p\cdot L_1}\,, \qquad z_h = \frac{p\cdot \ell_h}{p\cdot q}\,.
\label{eq:sidis-kinematics}
\end{equation}
$p^\mu = [p^+, 0, 0]$ is the average momentum per nucleon. In Eq. (\ref{eq:sidis-tmb-def}), we neglect $A$-independent higher-twist contributions, such as twist-4 fragmentation correlations or target-mass corrections, which lack the $A^{1/3}$ geometric enhancement in large nuclei ($A \gg 1$).

\subsubsection{High-twist approach for transverse momentum broadening}

To isolate the physical double-scattering contribution to $d\langle \ell_{h\perp}^2 \sigma^D \rangle / d\mathcal{PS}$, we employ the generalized high-twist factorization formalism~\cite{Qiu:1990xxa,Qiu:1990xy,Luo:1994np}. In this approach, the outgoing parton of momentum $\ell$, fragmenting into the hadron $h$ with $\bm{\ell}_{h\perp} = z_h \bm{\ell}_\perp$ (assuming collinear fragmentation at leading power in the fragmentation sector), undergoes secondary scattering with gluons from the nuclear target. Working in a covariant gauge, the dominant contribution to the hadronic tensor comes from the $A^+$ components of the exchanged gluons, which are enhanced by a factor of $p^+$ under Lorentz boosts compared to transverse components $\omega_\alpha^{\ \alpha'} A_{\alpha'}$~\cite{Luo:1994np}.

Let $k_1 = x_1 p$, $k_2 = x_2 p + k_{2\perp}$, $k_3 = (x_2 - x_3)p + k_{3\perp}$, and $k_4 = (x_1 + x_3)p + k_{2\perp} - k_{3\perp}$ denote the momenta of the two quark and two gluon lines connecting the hard scattering part $\hat{H}_{\mu\nu}^{\alpha\beta}$ to the nuclear matrix element. The double-scattering hadronic tensor takes the form~\cite{Luo:1994np,Kang:2014ela}
\begin{align}
W_{\mu\nu}^D &= \int \frac{dy^-}{2\pi} \frac{dy_1^-}{2\pi} \frac{dy_2^-}{2\pi} \int d^2\bm{y}_{1\perp} d^2\bm{y}_{2\perp} \int dx_1 dx_2 dx_3 \int \frac{d^2\bm{k}_{2\perp}}{(2\pi)^2} \frac{d^2\bm{k}_{3\perp}}{(2\pi)^2}\,
e^{ix_1 p^+ y^-} e^{ix_2 p^+ (y_1^- - y_2^-)} e^{ix_3 p^+ y_2^-}
\nonumber\\
&\quad \times e^{-i\bm{k}_{2\perp}\cdot \bm{y}_{1\perp}} e^{i\bm{k}_{3\perp}\cdot \bm{y}_{2\perp}}
{\rm Tr}\left[ \hat{H}_{\mu\nu}^{\alpha\beta}(\{x_i\}, \bm{k}_{2\perp}, \bm{k}_{3\perp}, p, q, \ell)\, \frac{\slashed{p}}{4}\, \frac{p_\alpha p_\beta}{(p\cdot n)^2} \langle A | \bar{\psi}_q(0) \gamma^+ n\cdot A(y_2) n\cdot A(y_1) \psi_q(y^-) | A \rangle \right]\,,
\label{eq:sidis-WD-raw}
\end{align}
where $n^\mu = [0, 1, 0]$ is the conjugate light-cone vector with $p\cdot n = p^+$. Because the exchanged transverse momenta $\bm{k}_{2\perp}, \bm{k}_{3\perp}$ are soft compared to the hard scale $Q$, one performs a Taylor expansion (collinear expansion) of the partonic hard part around $\bm{k}_{2\perp} = \bm{k}_{3\perp} = 0$~\cite{Luo:1994np}:
\begin{align}
\hat{H}_{\mu\nu}^{\alpha\beta}(\{x_i\}, \bm{k}_{2\perp}, \bm{k}_{3\perp}) &= \hat{H}_{\mu\nu}^{\alpha\beta}(\{x_i\}, 0, 0)
+ \left. \frac{\partial \hat{H}_{\mu\nu}^{\alpha\beta}}{\partial k_{2\perp}^\rho} \right|_{\bm{k}_{i\perp}=0} k_{2\perp}^\rho
+ \left. \frac{\partial \hat{H}_{\mu\nu}^{\alpha\beta}}{\partial k_{3\perp}^\sigma} \right|_{\bm{k}_{i\perp}=0} k_{3\perp}^\sigma
 + \left. \frac{\partial^2 \hat{H}_{\mu\nu}^{\alpha\beta}}{\partial k_{2\perp}^\rho \partial k_{3\perp}^\sigma} \right|_{\bm{k}_{i\perp}=0} k_{2\perp}^\rho k_{3\perp}^\sigma + \cdots\,.
\label{eq:collinear-exp-general}
\end{align}
In unpolarized collisions, the linear terms vanish upon azimuthal integration. The zeroth-order term contributes to the gauge-invariant Wilson line of the twist-2 quark distribution, leaving no nuclear-size enhancement due to unitarity cancellations among different final-state cuts~\cite{Luo:1994np}. In the second-order term, the factors of $k_{2\perp}^\rho k_{3\perp}^\sigma$ are converted into transverse spatial derivatives $\partial_\perp^\rho A^+(y_2) \partial_\perp^\sigma A^+(y_1)$ via integration by parts over $\bm{y}_{1\perp}$ and $\bm{y}_{2\perp}$. Up to higher-twist corrections, these derivatives form the gauge-covariant gluon field strength tensors $F_\sigma^{\ +}(y_2^-) F^{\sigma+}(y_1^-)$~\cite{Luo:1994np}.

\begin{figure}[htb]
\centering
\includegraphics[width=2.0in]{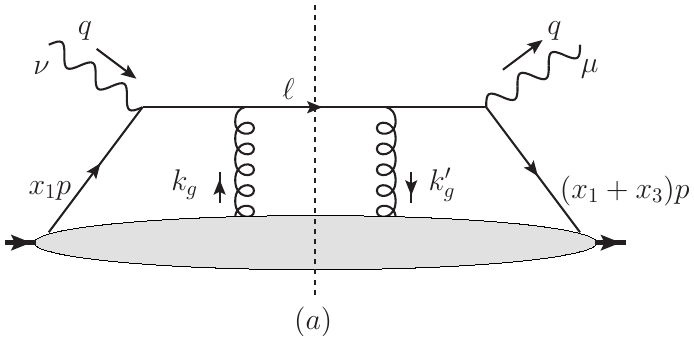}
\hskip 0.1in
\includegraphics[width=2.0in]{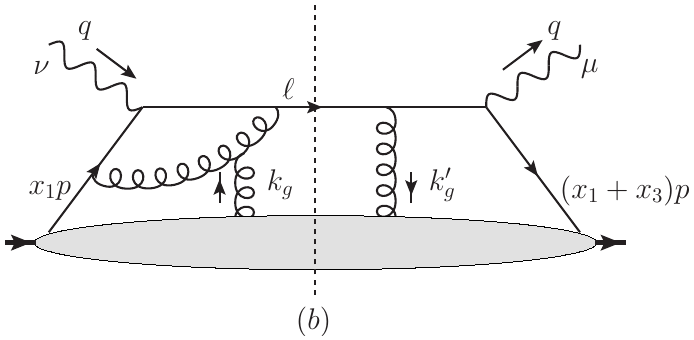}
\hskip 0.1in
\includegraphics[width=2.0in]{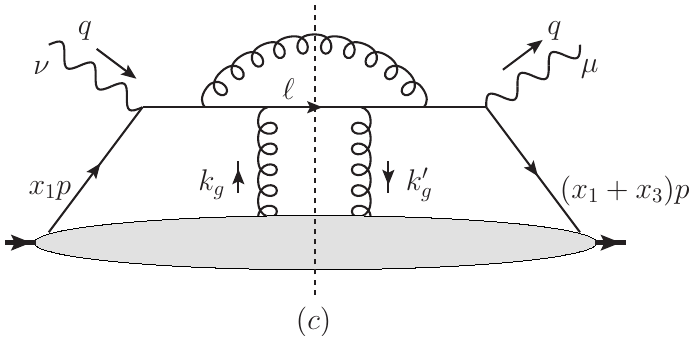}
\caption{Representative Feynman diagrams for double-scattering contributions to the SIDIS differential cross section at twist-4: (a) leading-order process, (b) NLO virtual correction, and (c) NLO real gluon emission. The exchanged gluon momenta are $k_g = x_2 p + \bm{k}_\perp$ and $k'_g = (x_2 - x_3)p + \bm{k}_\perp$.}
\label{fig:sidis-tmb-feyn}
\end{figure}

At leading order (LO) in $\alpha_s$, the double-scattering process is represented by the central-cut diagram in Fig.~\ref{fig:sidis-tmb-feyn}(a), where the struck quark scatters off a soft gluon with momentum $k_g = x_2 p + \bm{k}_\perp$ in the amplitude and $k'_g = (x_2 - x_3)p + \bm{k}_\perp$ in the conjugate amplitude (setting $\bm{k}_{2\perp} = \bm{k}_{3\perp} = \bm{k}_\perp$). At this order, the transverse momentum of the outgoing quark is directly given by the soft gluon kick, $\bm{\ell}_\perp = \bm{k}_\perp$, reflected by the phase space delta function $\delta^2(\bm{\ell}_\perp - \bm{k}_\perp)$. In the $\ell_{h\perp}^2$-weighted cross section, integrating the delta function replaces the weight $z_h^2\ell_\perp^2$ by $z_h^2 k_\perp^2$. The weight therefore supplies the full quadratic dependence needed for the twist-4 contribution, so the remaining partonic hard part can be evaluated at $\bm{k}_\perp=0$~\cite{Guo:1998rd,Kang:2011bp,Kang:2014ela}.

The longitudinal momentum fractions $x_2$ and $x_3$ are fixed by contour integrations around the unpinched poles of the quark propagators adjacent to the photon-quark vertices~\cite{Luo:1994np,Kang:2014ela}:
\begin{equation}
\frac{1}{(x_1 p + q)^2 + i\epsilon} = \frac{1}{2p\cdot q} \frac{1}{x_1 - x_B + i\epsilon}\,, \qquad
\frac{1}{[(x_1 + x_3)p + q]^2 - i\epsilon} = \frac{1}{2p\cdot q} \frac{1}{x_1 + x_3 - x_B - i\epsilon}\,,
\label{eq:sidis-LO-poles}
\end{equation}
while the final-state cut on the outgoing quark line $\delta_+(\ell^2) = \delta_+[(x_1 + x_2)p + q + k_\perp]^2$ fixes $x_1 + x_2 = x_B$ in the collinear limit $\bm{k}_\perp \to 0$. Closing the integration contours of $x_2$ and $x_3$ in the complex plane generates the light-cone ordering step functions $\theta(y_2^-)\theta(y_1^- - y^-)$ and sets $x_1 = x_B, x_2 = x_3 = 0$. Note that asymmetric-cut diagrams (interference between single and triple scattering) yield step functions $\theta(y_1^- - y_2^-)\theta(y_2^- - y^-)$ or $\theta(y_2^- - y_1^-)\theta(-y_2^-)$; when combined with the central cut if the momentum kick did not enter the observed hadron, they would restrict the integral to $|y^-| > |y_1^-| > |y_2^-|$ (a localized contact term)~\cite{Luo:1994np}. However, because the final-state cut fixes the observed hadron momentum $\bm{\ell}_{h\perp}$, the central cut is kinematically distinct and survives the cancellation~\cite{Luo:1994np,Guo:1998rd}.

In $n = 4 - 2\epsilon$ dimensions (used for dimensional regularization at NLO), the LO double-scattering contribution to the $\ell_{h\perp}^2$-weighted cross section reads~\cite{Kang:2013raa,Kang:2014ela}:
\begin{equation}
\frac{d\langle \ell_{h\perp}^2 \sigma^D \rangle^{\rm LO}}{d\mathcal{PS}} = \sigma_h \sum_q e_q^2 \int \frac{dz}{z} D_{h/q}(z) \int \frac{dx}{x} T_{qg}(x, 0, 0)\, \delta(1 - \hat{x})\delta(1 - \hat{z})\,,
\label{eq:sidis-weighted-LO}
\end{equation}
where $\hat{x} = x_B / x$, $\hat{z} = z_h / z$, and the prefactor $\sigma_h$ is related to the Born cross section factor $\sigma_0$ by
\begin{equation}
\sigma_h = \frac{4\pi^2 \alpha_s z_h^2}{N_c}\, \sigma_0\,, \qquad
\sigma_0 = \frac{2\pi \alpha_{\rm em}^2}{Q^2} \frac{1 + (1-y)^2}{y} (1 - \epsilon)\,.
\label{eq:sidis-sigma0}
\end{equation}
Here $D_{h/q}(z)$ is the standard leading-twist quark fragmentation function, and $T_{qg}(x_1, x_2, x_3)$ is the universal twist-4 quark-gluon correlation function in the nucleus, defined as~\cite{Wang:2001ifa,Osborne:2002st,Kang:2008us,Kang:2013raa}
\begin{align}
T_{qg}(x_1, x_2, x_3) &= \int \frac{dy^-}{2\pi} e^{ix_1 p^+ y^-} \int \frac{dy_1^- dy_2^-}{4\pi} e^{ix_2 p^+ (y_1^- - y_2^-)} e^{ix_3 p^+ y_2^-} \theta(y_2^-)\theta(y_1^- - y^-)
\langle A | \bar{\psi}_q(0) \gamma^+ F_\sigma^{\ +}(y_2^-) F^{\sigma+}(y_1^-) \psi_q(y^-) | A \rangle\,,
\label{eq:Tqg-def}
\end{align}
where gauge links along the light-cone minus direction between the field operators are implied in covariant gauges. Dividing Eq.~\eqref{eq:sidis-weighted-LO} by the LO single-scattering SIDIS cross section $d\sigma^S / d\mathcal{PS} = \sigma_0 \sum_q e_q^2 f_{q/A}(x_B) D_{h/q}(z_h)$, one obtains the well-known LO expression for transverse momentum broadening in SIDIS~\cite{Guo:1998rd,Ru:2019qvz,Kang:2014ela}:
\begin{equation}
\Delta\langle \ell_{h\perp}^2 \rangle^{\rm LO} = \frac{4\pi^2 \alpha_s z_h^2}{N_c} \frac{\sum_q e_q^2 T_{qg}(x_B, 0, 0) D_{h/q}(z_h)}{\sum_q e_q^2 f_{q/A}(x_B) D_{h/q}(z_h)}\,.
\label{eq:sidis-tmb-LO-ratio}
\end{equation}
Notice that for a single quark flavor (or when $T_{qg}(x_B, 0, 0)$ is proportional to $f_{q/A}(x_B)$), the fragmentation function $D_{h/q}(z_h)$ cancels between the numerator and denominator at LO, predicting a clean quadratic scaling $\Delta\langle \ell_{h\perp}^2 \rangle \propto z_h^2$.

\subsubsection{Next-to-leading-order radiative corrections and QCD factorization at twist-4}

To establish the validity of the high-twist factorization formalism beyond LO and determine the scale evolution of $T_{qg}$, a complete NLO perturbative QCD calculation of $d\langle \ell_{h\perp}^2 \sigma^D \rangle / d\mathcal{PS}$ was carried out by Kang, Wang, Wang, and Xing~\cite{Kang:2013raa,Kang:2014ela}. At NLO, one must evaluate both one-loop virtual corrections to the $\gamma^* + q$ double-scattering process, illustrated in Fig.~\ref{fig:sidis-tmb-feyn}(b), and real radiative corrections from $\gamma^* + q \to q + g$ and $\gamma^* + g \to q + \bar{q}$ accompanied by secondary rescattering, illustrated in Fig.~\ref{fig:sidis-tmb-feyn}(c).

In the virtual diagrams, loop integrations are performed in $n = 4 - 2\epsilon$ dimensions. After tensor reduction and contour integration over the longitudinal momentum fractions, the complete NLO virtual correction is proportional to the LO cross section in Eq.~\eqref{eq:sidis-weighted-LO}, multiplied by the following factor~\cite{Kang:2013raa,Kang:2014ela}
\begin{equation}
\mathcal{V}_{\rm SIDIS}^{\rm NLO} = \frac{\alpha_s}{2\pi} C_F \left( \frac{4\pi\mu^2}{Q^2} \right)^\epsilon \frac{1}{\Gamma(1-\epsilon)} \left( -\frac{2}{\epsilon^2} - \frac{3}{\epsilon} - 8 \right)\,,
\label{eq:sidis-virtual}
\end{equation}
which exhibits both soft-collinear double poles ($1/\epsilon^2$) and collinear/soft single poles ($1/\epsilon$).

For the real emission contribution in the $\gamma^* + q$ channel (such as Fig.~\ref{fig:sidis-tmb-feyn}(c)), an outgoing quark of momentum $\ell$ and a radiated gluon of momentum $\ell_g = x_1 p + k_g + q - \ell$ are produced in the final state. 
The on-shell condition for the unobserved radiated gluon, $\delta_+(\ell_g^2)$, fixes one relation among the momentum fractions, $x_1 + x_2 = x + x_c$, where $x = (Q^2 + 2q\cdot \ell) / [2p\cdot(q-\ell)] = x_B / \hat{x}$ and $x_c$ encodes the $\bm{k}_\perp$ dependence. The remaining two fractions $x_2$ and $x_3$ are fixed by contour integrations around propagator poles on either side of the final-state cut. Depending on which propagator goes on shell in the contour integration, the real corrections naturally separate into four distinct physical mechanisms~\cite{Wang:2001ifa,Kang:2013raa,Kang:2014ela}:
\begin{enumerate}
    \item \textbf{Soft-soft ($ss$) rescattering}: The quark propagator immediately after the hard $\gamma^* q \to qg$ subdiagram is put on shell on both sides of the cut. In the collinear limit $\bm{k}_\perp \to 0$, the fixed momentum fractions are $x_1 = x$, $x_2 = 0$, and $x_3 = 0$. Since both exchanged gluons carry vanishing longitudinal momentum ($k_g, k'_g \to 0$), this corresponds to a hard $\gamma^* q \to qg$ scattering followed by soft rescattering of the outgoing parton. Because $x_1$ and $x_2$ depend on $\bm{k}_\perp$ prior to taking the limit, the collinear expansion generates not only the diagonal matrix element $T_{qg}(x, 0, 0)$ but also its first and second derivatives, $x\,dT_{qg}(x, 0, 0)/dx$ and $x^2\,d^2T_{qg}(x, 0, 0)/dx^2$, which can be reorganized via integration by parts.
    \item \textbf{Hard-hard ($hh$) rescattering}: The quark propagator immediately adjacent to the $\gamma^* q$ vertex is put on shell, while the gluon is radiated during the secondary quark-gluon scattering subprocess $q + g \to q + g$. In the limit $\bm{k}_\perp \to 0$, the initial quark carries $x_1 = x_B$, while the exchanged gluons carry finite longitudinal momentum fractions $x_2 = x - x_B$ and $x_3 = 0$. This probes the off-diagonal correlation function $T_{qg}(x_B, x - x_B, 0)$. Due to gauge invariance of the secondary $2\to 2$ scattering amplitude, the hard part vanishes at $\bm{k}_\perp = 0$, so no derivative terms of $T_{qg}$ appear in the $hh$ contribution.
    \item \textbf{Soft-hard ($sh$) and hard-soft ($hs$) interferences}: The gluon emission occurs before the secondary scattering in the amplitude (soft pole) and is induced by the secondary scattering in the conjugate amplitude (hard pole), or vice versa. These quantum interference terms probe the off-diagonal matrix elements $T_{qg}(x, 0, x_B - x)$ and $T_{qg}(x_B, x - x_B, x - x_B)$. Physically, the destructive interference between the soft-radiation amplitude and the medium-induced hard-radiation amplitude is the manifestation of the Landau-Pomeranchuk-Migdal (LPM) effect in the twist-4 NLO cross section~\cite{Wang:2001ifa,Kang:2013raa}.
    \item \textbf{Single-triple scattering interferences (asymmetric cuts)}: In addition to central-cut diagrams, asymmetric left-cut ($L$) and right-cut ($R$) diagrams (where both exchanged gluons are on the same side of the cut) must be included. While $ss$ asymmetric cuts reduce to contact terms without nuclear enhancement, $sh$ and $hs$ asymmetric cuts yield finite NLO contributions involving $T_{qg}^L$ and $T_{qg}^R$~\cite{Kang:2014ela,Kang:2016ron}.
\end{enumerate}

Evaluating all 57 real-emission diagrams in the $\gamma^* + q$ channel together with the photon-gluon fusion channel $\gamma^* + g \to q + \bar{q}$ (which receives only soft-soft rescattering contributions and probes the twist-4 gluon-gluon correlation function $T_{gg}$), and expanding in powers of $\epsilon$, one finds that all soft-collinear double poles ($1/\epsilon^2$) and soft single poles cancel exactly against the virtual correction in Eq.~\eqref{eq:sidis-virtual}~\cite{Kang:2013raa,Kang:2014ela}. The remaining collinear divergences ($1/\epsilon$) take the factorized form
\begin{align}
\frac{d\langle \ell_{h\perp}^2 \sigma^D \rangle}{d\mathcal{PS}} &= \sigma_h \sum_q e_q^2 \int_{z_h}^1 \frac{dz}{z} D_{h/q}(z) \int_{x_B}^1 \frac{dx}{x} \Bigg\{ \delta(1-\hat{x})\delta(1-\hat{z}) T_{qg}(x, 0, 0)
+ \frac{\alpha_s}{2\pi} \left( -\frac{1}{\hat{\epsilon}} + \ln\frac{Q^2}{\mu_f^2} \right)
\nonumber\\
&\quad \times \Big[ \delta(1-\hat{x}) P_{qq}(\hat{z}) T_{qg}(x, 0, 0)
+ \delta(1-\hat{z}) \left( \mathcal{P}_{qg\to qg} \otimes T_{qg} + P_{qg}(\hat{x}) T_{gg}(x, 0, 0) \right) \Big] + \cdots \Bigg\}\,,
\label{eq:sidis-divergent}
\end{align}
where $1/\hat{\epsilon} = 1/\epsilon - \gamma_E + \ln(4\pi)+\ln(\mu^2/\mu_f^2)$ in the $\overline{\rm MS}$ scheme, $\mu_f$ is the factorization scale, and the ellipsis denotes finite NLO hard coefficient functions. Here $P_{qq}(\hat{z}) = C_F \left[ \frac{1+\hat{z}^2}{(1-\hat{z})_+} + \frac{3}{2}\delta(1-\hat{z}) \right]$ and $P_{qg}(\hat{x}) = T_R \left[ \hat{x}^2 + (1-\hat{x})^2 \right]$ are the standard leading-twist DGLAP splitting functions, while the twist-4 gluon-gluon correlation function $T_{gg}$ is defined as~\cite{Kang:2013ufa,Kang:2013raa}
\begin{align}
T_{gg}(x_1, x_2, x_3) &= \frac{1}{x_1 p^+} \int \frac{dy^-}{2\pi} e^{ix_1 p^+ y^-} \int \frac{dy_1^- dy_2^-}{2\pi} e^{ix_2 p^+ (y_1^- - y_2^-)} e^{ix_3 p^+ y_2^-} \theta(y_2^-)\theta(y_1^- - y^-)
 \langle A | F_\alpha^{\ +}(0) F^{\sigma+}(y_2^-) F_\sigma^{\ +}(y_1^-) F^{+\alpha}(y^-) | A \rangle\,.
\label{eq:Tgg-def}
\end{align}
The twist-4 evolution kernel $\mathcal{P}_{qg\to qg} \otimes T_{qg}$ in Eq.~\eqref{eq:sidis-divergent} combines the collinear divergences from soft-soft, hard-hard, and soft-hard/hard-soft interference diagrams~\cite{Kang:2013raa,Kang:2014ela}:
\begin{align}
\mathcal{P}_{qg\to qg} \otimes T_{qg} &\equiv P_{qq}(\hat{x}) T_{qg}(x, 0, 0) + \frac{C_A}{2} \Bigg\{ \frac{4}{(1-\hat{x})_+} T_{qg}(x_B, x - x_B, 0)
\nonumber\\
&\quad - \frac{1+\hat{x}}{(1-\hat{x})_+} \Big[ T_{qg}(x, 0, x_B - x) + T_{qg}(x_B, x - x_B, x - x_B) \Big] \Bigg\} + 2C_A \delta(1-\hat{x}) T_{qg}(x, 0, 0)\,.
\label{eq:Pqg-kernel}
\end{align}
Note that in the soft-radiation limit $\hat{x} \to 1$ ($x \to x_B$), where the gluon formation time $\tau_f = 1/[(x - x_B)p^+] \gg R_A$ exceeds the nuclear size, the off-diagonal matrix elements approach the diagonal one, $T_{qg}(x_B, 0, 0)$, and the terms proportional to $C_A / (1-\hat{x})_+$ in Eq.~\eqref{eq:Pqg-kernel} cancel exactly between the hard-hard and soft-hard/hard-soft interference channels~\cite{Kang:2013raa}. This demonstrates how the LPM interference regularizes the soft divergence in medium-induced radiation.

Following the standard collinear factorization procedure, the collinear divergence proportional to $\delta(1-\hat{x}) P_{qq}(\hat{z})$ in Eq.~\eqref{eq:sidis-divergent} is absorbed into the renormalized fragmentation function $D_{h/q}(z, \mu_f^2)$, which obeys the vacuum DGLAP evolution equation. Simultaneously, the initial-state collinear divergence proportional to $\delta(1-\hat{z})$ is absorbed into the renormalized twist-4 quark-gluon correlation function~\cite{Kang:2013raa},
leading to the QCD evolution equation for the twist-4 quark-gluon correlation function~\cite{Kang:2013raa,Kang:2014ela}:
\begin{equation}
\frac{\partial}{\partial \ln\mu_f^2} T_{qg}(x_B, 0, 0, \mu_f^2) = \frac{\alpha_s}{2\pi} \int_{x_B}^1 \frac{dx}{x} \Big[ \mathcal{P}_{qg\to qg} \otimes T_{qg} + P_{qg}(\hat{x}) T_{gg}(x, 0, 0, \mu_f^2) \Big]\,.
\label{eq:Tqg-evolution}
\end{equation}
In terms of the renormalized distributions, the $\ell_{h\perp}^2$-weighted differential cross section at twist-4 is completely infrared-safe at NLO~\cite{Kang:2013raa,Kang:2014ela}:
\begin{align}
\frac{d\langle \ell_{h\perp}^2 \sigma^D \rangle}{d\mathcal{PS}} &= \sigma_h \sum_q e_q^2 \int_{z_h}^1 \frac{dz}{z} D_{h/q}(z, \mu_f^2) \int_{x_B}^1 \frac{dx}{x} \Bigg\{ \delta(1-\hat{x})\delta(1-\hat{z}) T_{qg}(x, 0, 0, \mu_f^2)
+ \frac{\alpha_s}{2\pi} \ln\left(\frac{Q^2}{\mu_f^2}\right) \Big[ \delta(1-\hat{x}) P_{qq}(\hat{z}) T_{qg}(x, 0, 0, \mu_f^2)
\nonumber\\
&\quad 
+ \delta(1-\hat{z}) \left( \mathcal{P}_{qg\to qg} \otimes T_{qg} + P_{qg}(\hat{x}) T_{gg}(x, 0, 0, \mu_f^2) \right) \Big]
+ \frac{\alpha_s}{2\pi} \Big[ H_q^{\rm NLO}(\hat{x}, \hat{z}) \otimes T_{qg} + H_g^{\rm NLO}(\hat{x}, \hat{z}) \otimes T_{gg} \Big] \Bigg\}\,,
\label{eq:sidis-NLO-final}
\end{align}
where $H_{q,g}^{\rm NLO}$ are the finite NLO hard coefficient functions (explicitly given in Ref.~\cite{Kang:2014ela}, along with gluon fragmentation channels). Eq.~\eqref{eq:sidis-NLO-final} establishes, for the first time at one-loop order, the collinear factorization theorem for twist-4 observables in nuclear media~\cite{Kang:2013raa}. Note that Eq.~\eqref{eq:Tqg-evolution} is not closed because the evolution of the diagonal correlator $T_{qg}(x_B, 0, 0, \mu_f^2)$ couples to off-diagonal correlators $T_{qg}(x_1, x_2, x_3, \mu_f^2)$, a generic feature of higher-twist evolution equations~\cite{Braun:2009mi,Vogelsang:2009pj,Kang:2012em}. Under effective models for the nuclear geometry and two-nucleon correlations, Eq.~\eqref{eq:Tqg-evolution} directly governs the QCD scale dependence of the jet transport parameter $\hat{q}(\mu_f^2)$~\cite{Kang:2013raa}, complementing small-$x$ and double-logarithmic resummations of radiative corrections to $\hat{q}$~\cite{Liou:2013qya,Blaizot:2014bha,Iancu:2014kga}.

\subsection{Transverse momentum broadening in proton-nucleus collisions}
\label{sec:TMB-pA}

Hadron-nucleus ($p+A$ and $\pi+A$) collisions offer a golden process to test the universality of twist-4 nuclear correlation functions and to contrast initial-state versus final-state multiple parton scatterings~\cite{Guo:1998rd,Fries:2002mu,Kang:2008us,Kang:2012am,Kang:2016ron,Arleo:2025oos}. Depending on the color charge of the produced hard probe, multiple scatterings occur at different stages of the collision: colorless electroweak probes such as Drell-Yan lepton pairs and $W^\pm/Z^0$ bosons experience purely initial-state rescattering~\cite{Guo:1998rd,Kang:2008us,Kang:2016ron}, whereas heavy quarkonia ($J/\psi, \Upsilon$) experience both initial-state parton rescattering and final-state rescattering of the heavy-quark pair ($Q\bar{Q}$) as it traverses the nuclear medium~\cite{Kang:2008us,Kang:2012am}.

\subsubsection{Transverse momentum broadening of Drell-Yan dilepton production}

In the Drell-Yan process, $p(p') + A(p) \to [\gamma^* \to]\,\ell^+\ell^-(q) + \mathcal{X}$, an incoming quark (or antiquark) of momentum $x'p'$ from the projectile proton propagates through the target nucleus and undergoes initial-state multiple scatterings before annihilating with a target antiquark (or quark) into a virtual photon of invariant mass $Q^2 = q^2$ and transverse momentum $q_\perp$~\cite{Guo:1998rd,Fries:2002mu,Kang:2016ron}. The transverse momentum broadening is given by~\cite{Kang:2016ron}
\begin{equation}
\Delta\langle q_\perp^2 \rangle = \langle q_\perp^2 \rangle_{pA} - \langle q_\perp^2 \rangle_{pp} \approx \left. \frac{d\langle q_\perp^2 \sigma^D \rangle}{dQ^2} \right/ \frac{d\sigma^S}{dQ^2}\,, \qquad
\frac{d\langle q_\perp^2 \sigma^D \rangle}{dQ^2} \equiv \int dq_\perp^2\, q_\perp^2 \frac{d\sigma^D}{dQ^2\,dq_\perp^2}\,.
\label{eq:dy-tmb-def}
\end{equation}

\begin{figure}[htb]
\centering
\includegraphics[width=2.0in]{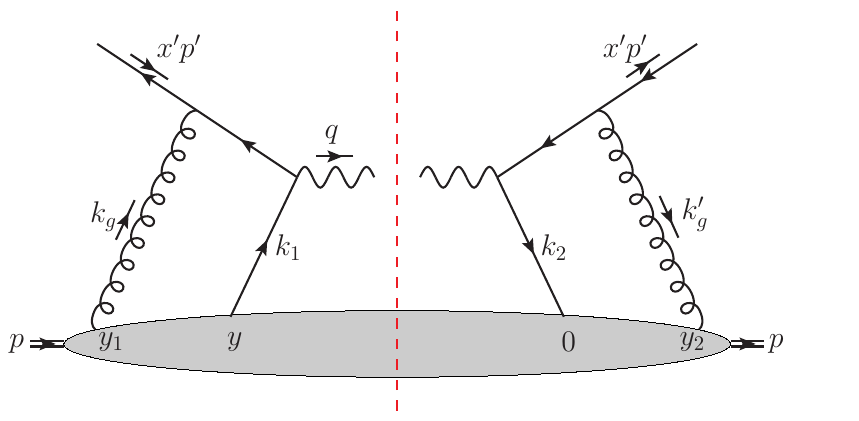}
\hskip 0.1in
\includegraphics[width=2.0in]{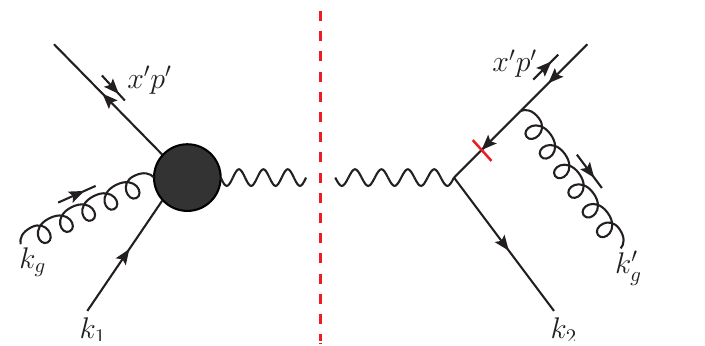}
\hskip 0.1in
\includegraphics[width=2.0in]{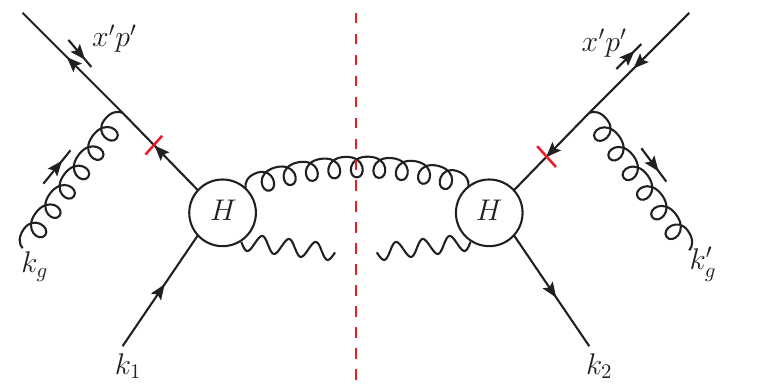}
\hskip 0.1in
\caption{Schematic Feynman diagrams for initial-state double scattering in Drell-Yan production in $p+A$ collisions. From left to right, the panels show the leading-order contribution, the virtual NLO correction, and a real-emission NLO contribution.}
\label{fig:dy-tmb-feyn}
\end{figure}

At leading order, depicted in the first diagram in Fig.~\ref{fig:dy-tmb-feyn}, the incoming antiquark exchanges a soft gluon with the nucleus prior to the $\bar{q} q \to \gamma^*$ annihilation. Following the collinear expansion procedure in $n = 4 - 2\epsilon$ dimensions, the LO $q_\perp^2$-weighted double-scattering cross section reads~\cite{Kang:2016ron}
\begin{equation}
\frac{d\langle q_\perp^2 \sigma^D \rangle^{\rm LO}}{dQ^2} = \sigma_\ell \sum_q \int \frac{dx'}{x'} f_{\bar{q}/p}(x') \int \frac{dx}{x} T_{gq}(x, 0, 0)\, \delta(1 - z)\,, 
\label{eq:dy-weighted-LO}
\end{equation}
where $\sigma_\ell = \frac{4\pi^2 \alpha_s}{N_c} \sigma_0^{\rm DY}$ with $\sigma_0^{\rm DY} = \frac{4\pi\alpha_{\rm em}^2 e_q^2}{9Q^2 s}(1-\epsilon)$, $f_{\bar{q}/p}(x')$ is the proton antiquark PDF, $z = x_B / x = Q^2 / \hat{s}$ with $x_B = Q^2 / (2p\cdot q)$ and $\hat{s} = (x'p' + xp)^2$, and the sum $\sum_q$ runs over all quark and antiquark flavors. The twist-4 nuclear gluon-quark correlation function for initial-state rescattering is defined by~\cite{Guo:1998rd,Kang:2016ron}
\begin{align}
T_{gq}(x_1, x_2, x_3) &= \int \frac{dy^-}{2\pi} e^{ix_1 p^+ y^-} \int \frac{dy_1^- dy_2^-}{4\pi} e^{ix_2 p^+ (y_1^- - y_2^-)} e^{ix_3 p^+ y_2^-} \theta(-y_2^-)\theta(y^- - y_1^-)
\langle A | F_\sigma^{\ +}(y_2^-) \bar{\psi}_q(0) \gamma^+ \psi_q(y^-) F^{\sigma+}(y_1^-) | A \rangle\,.
\label{eq:Tgq-def}
\end{align}
The DY and SIDIS operators in Eqs.~\eqref{eq:Tgq-def} and~\eqref{eq:Tqg-def} differ only in the light-cone ordering step functions: $\theta(-y_2^-)\theta(y^- - y_1^-)$ in DY reflects that the soft gluon scatterings at $y_1^-, y_2^-$ occur prior to the hard annihilation at $0, y^-$, whereas $\theta(y_2^-)\theta(y_1^- - y^-)$ in SIDIS reflects final-state rescattering after the hard vertex~\cite{Kang:2016ron}. By translational invariance and time-reversal symmetry in a large loosely bound nucleus, their diagonal limits are identical, $T_{gq}(x, 0, 0) = T_{qg}(x, 0, 0)$~\cite{Guo:1998rd,Kang:2011bp,Ru:2019qvz}. Dividing Eq.~\eqref{eq:dy-weighted-LO} by the LO DY cross section yields~\cite{Guo:1998rd,Kang:2008us,Kang:2016ron}
\begin{equation}
\Delta\langle q_\perp^2 \rangle_{\rm DY}^{\rm LO} = \frac{4\pi^2 \alpha_s}{N_c} \frac{\sum_q e_q^2 \int \frac{dx'}{x'} f_{\bar{q}/p}(x') T_{gq}(x_B, 0, 0)}{\sum_q e_q^2 \int \frac{dx'}{x'} f_{\bar{q}/p}(x') f_{q/A}(x_B)}\,.
\label{eq:dy-tmb-LO-ratio}
\end{equation}

To verify whether QCD factorization for initial state multiple scattering holds under quantum radiative corrections, Kang, Qiu, Wang, and Xing computed the complete NLO perturbative QCD corrections to $d\langle q_\perp^2 \sigma^D \rangle / dQ^2$ in the Drell-Yan process~\cite{Kang:2013raa,Kang:2016ron}. As illustrated in Fig.~\ref{fig:dy-tmb-feyn}, the NLO calculation encompasses the virtual correction and real-emission diagrams, which include soft-soft, hard-hard, and soft-hard/hard-soft interferences in the $\bar{q} + q \to \gamma^* + g$ channel; asymmetric-cut single-triple interference diagrams; and the Compton scattering channel $\bar{q} + g \to \gamma^* + \bar{q}$ accompanied by soft gluon rescattering.


In the two-body final-state phase space of DY at NLO, the partonic scattering angle $\theta$ is parameterized by $v = (1 + \cos\theta)/2$. Consequently, the limit $z \to 1$ corresponds to soft gluon emission, $v \to 1$ ($\theta \to 0$) corresponds to radiation collinear to the projectile proton, and $v \to 0$ ($\theta \to \pi$) corresponds to radiation collinear to the target nucleus~\cite{Kang:2016ron}. When the real-emission contributions are combined with the virtual correction, all $1/\epsilon^2$ double poles and soft $1/\epsilon$ poles cancel completely. The remaining collinear divergences separate cleanly into the projectile ($v \to 1$) and target ($v \to 0$) regions~\cite{Kang:2016ron}:
\begin{align}
\frac{d\langle q_\perp^2 \sigma^D \rangle}{dQ^2} &= \sigma_\ell \frac{\alpha_s}{2\pi} \sum_q \int \frac{dx'}{x'} f_{\bar{q}/p}(x') \int_{x_B}^1 \frac{dx}{x} \int_0^1 dv \Bigg\{ \left( -\frac{1}{\hat{\epsilon}} + \ln\frac{Q^2}{\mu_f^2} \right) \Big[ \delta(1-v) P_{qq}(z) T_{gq}(x, 0, 0)
\nonumber\\
&\quad + \delta(v) \left( \mathcal{P}_{qg\to qg} \otimes T_{gq} + P_{qg}(z) T_{gg}(x, 0, 0) \right) \Big] + H_{\rm DY}^{\rm NLO} \Bigg\}\,.
\label{eq:dy-divergent}
\end{align}
The collinear divergence at $v=1$ is absorbed into the DGLAP-evolved antiquark PDF $f_{\bar{q}/p}(x', \mu_f^2)$ of the projectile proton, while the collinear divergence at $v=0$ is absorbed into the renormalized twist-4 nuclear gluon-quark correlation function $T_{gq}(x_B, 0, 0, \mu_f^2)$~\cite{Kang:2016ron}. Strikingly, the evolution kernel $\mathcal{P}_{qg\to qg} \otimes T_{gq} + P_{qg}(z) T_{gg}$ in Eq.~\eqref{eq:dy-divergent} is identical to that found in SIDIS, leading to exactly the same evolution equation as in SIDIS, Eq.~\eqref{eq:Tqg-evolution}~\cite{Kang:2013raa,Kang:2016ron}.
After subtracting the collinear divergences in the $\overline{\rm MS}$ scheme, the factorized NLO $q_\perp^2$-weighted Drell-Yan cross section is given by~\cite{Kang:2016ron}
\begin{align}
\frac{d\langle q_\perp^2 \sigma^D \rangle}{dQ^2} &= \sigma_\ell \sum_q \int \frac{dx'}{x'} f_{\bar{q}/p}(x', \mu_f^2) \Bigg\{ \int_{x_B}^1 \frac{dx}{x} T_{gq}(x, 0, 0, \mu_f^2)\, \delta(1-z)
+ \frac{\alpha_s}{2\pi} \int_{x_B}^1 \frac{dx}{x} \bigg[ \ln\left(\frac{Q^2}{\mu_f^2}\right) P_{qq}(z) T_{gq}(x, 0, 0, \mu_f^2)
\nonumber\\
&\quad + \ln\left(\frac{Q^2}{\mu_f^2}\right) \Big( \mathcal{P}_{qg\to qg} \otimes T_{gq} + P_{qg}(z) T_{gg}(x, 0, 0, \mu_f^2) \Big) + H_{\rm DY}^{\rm NLO} \bigg] \Bigg\}\,,
\label{eq:dy-NLO-final}
\end{align}
where the complete analytical expressions for $H_{\rm DY}^{\rm NLO}$ are presented in Ref.~\cite{Kang:2016ron}. The exact agreement between the twist-4 evolution equations in SIDIS and Drell-Yan demonstrates the process-independence (universality) of the twist-4 parton correlation functions and confirms the predictive power of the high-twist factorization approach across $e+A$ and $p+A$ collisions~\cite{Kang:2013raa,Kang:2016ron}.

\subsubsection{Transverse momentum broadening for heavy quarkonium and electroweak vector boson production}

Comparing the transverse momentum broadening of colorless electroweak bosons ($\gamma^*, W^\pm, Z^0$) with that of heavy quarkonia ($J/\psi, \Upsilon$) in $p+A$ collisions provides a powerful diagnostic tool for color dynamics and the quarkonium hadronization mechanism~\cite{Kang:2008us,Kang:2012am}. While $W^\pm$ and $Z^0$ bosons are color singlets produced directly at the hard vertex and experience only initial-state parton rescattering, heavy quarkonia are formed from a heavy-quark pair $Q\bar{Q}$ created at short distances $\sim 1/(2m_Q)$. As the $Q\bar{Q}$ pair propagates through the nuclear medium prior to hadronizing into a physical quarkonium state $H$, it undergoes final-state multiple scatterings that are directly sensitive to the color configuration of the pair~\cite{Kang:2008us,Kang:2012am}.

Within the high-twist approach, Kang and Qiu derived the transverse momentum broadening for heavy quarkonium production in both the Color Evaporation Model (CEM) and Nonrelativistic QCD (NRQCD) factorization~\cite{Kang:2008us,Kang:2012am}. To evaluate the broadening quantitatively, the diagonal twist-4 quark-gluon and gluon-gluon correlation functions are parameterized in terms of the leading-twist nuclear PDFs~\cite{Luo:1993ui,Kang:2008us,Xing:2012ii}:
\begin{equation}
T_{q/A}^{(I)}(x) = T_{q/A}^{(F)}(x) \equiv T_{gq}(x, 0, 0) = \lambda^2 A^{1/3} f_{q/A}(x)\,, \qquad
T_{g/A}^{(I)}(x) = T_{g/A}^{(F)}(x) \equiv T_{gg}(x, 0, 0) = \lambda^2 A^{1/3} f_{g/A}(x)\,,
\label{eq:T4-lambda-model}
\end{equation}
where $\lambda^2$ sets the characteristic scale of the twist-4 nuclear matrix element per nucleon.

In the CEM, the cross section for producing a heavy quarkonium state $H$ is obtained by integrating the perturbative $Q\bar{Q}$ pair production cross section over the invariant mass interval $4m_Q^2 < Q^2 < 4M_Q^2$ (where $m_Q$ is the heavy-quark mass and $M_Q$ is the mass of the open-heavy-flavor meson that sets the pair threshold) and multiplying the result by a universal transition probability $F_{Q\bar{Q}\to H}$~\cite{Kang:2008us,Kang:2012am}. Taking into account initial-state double scattering of the incoming quark or gluon, final-state double scattering of the outgoing $Q\bar{Q}$ pair, and their color interference, the transverse momentum broadening in $p+A$ collisions reads~\cite{Kang:2008us,Xing:2012ii,Kang:2012am}:
\begin{equation}
\Delta\langle q_\perp^2 \rangle_{\rm HQ}^{\rm CEM} = \left( \frac{8\pi^2 \alpha_s}{N_c^2 - 1} \lambda^2 A^{1/3} \right) \frac{(C_F + C_A)\sigma_{q\bar{q}} + 2C_A \sigma_{gg} + \Delta\sigma_{gg}}{\sigma_{q\bar{q}} + \sigma_{gg}}\,,
\label{eq:tmb-HQ-CEM}
\end{equation}
where $\sigma_{q\bar{q}}$ and $\sigma_{gg}$ are the integrated leading-order $q\bar{q}\to Q\bar{Q}$ and $gg\to Q\bar{Q}$ cross sections:
\begin{equation}
\sigma_{ab} = F_{Q\bar{Q}\to H} \int_{4m_Q^2}^{4M_Q^2} dQ^2 \int dx'\, f_{a/p}(x') \int dx\, f_{b/A}(x)\, \frac{d\hat{\sigma}_{ab\to Q\bar{Q}}}{dQ^2}\,,
\label{eq:sigma-ab-CEM}
\end{equation}
and $\Delta\sigma_{gg}$ is a small color-suppressed correction to the final-state $gg\to Q\bar{Q}$ channel~\cite{Xing:2012ii,Kang:2012am}. The partonic cross sections in Eqs.~\eqref{eq:tmb-HQ-CEM} and~\eqref{eq:sigma-ab-CEM} are given in Refs.~\cite{Kang:2008us,Kang:2012am}.
The color factors in the numerator of Eq.~\eqref{eq:tmb-HQ-CEM} have a transparent physical interpretation~\cite{Kang:2008us,Kang:2012am}. In the $q\bar{q}\to Q\bar{Q}$ channel, the incoming quark carries color factor $C_F$ in initial-state rescattering, while the $s$-channel virtual gluon forces the outgoing $Q\bar{Q}$ pair into a pure color-octet state, which rescatters in the final state with color factor $C_A$, yielding the sum $C_F + C_A$. In the $gg\to Q\bar{Q}$ channel, the incoming gluon rescatters in the initial state with color factor $C_A$, while the outgoing $Q\bar{Q}$ pair (summed over colors in the CEM) rescatters in the final state with an effective color-octet factor $C_A$ up to the small correction $\Delta\sigma_{gg}$, yielding $2C_A$.

In contrast, for Drell-Yan and $W^\pm/Z^0$ production at LO, only initial-state quark rescattering occurs with color factor $C_F$:
\begin{equation}
\Delta\langle q_\perp^2 \rangle_{\rm DY,\,W/Z}^{\rm LO} = C_F \left( \frac{8\pi^2 \alpha_s}{N_c^2 - 1} \lambda^2 A^{1/3} \right)\,.
\label{eq:tmb-DY-CF}
\end{equation}
At high collision energies (such as at Fermilab, RHIC, and the LHC), heavy quarkonium production is overwhelmingly dominated by gluon-gluon fusion ($\sigma_{gg} \gg \sigma_{q\bar{q}}, \vert{}\Delta\sigma_{gg}\vert{}$), so Eq.~\eqref{eq:tmb-HQ-CEM} simplifies to~\cite{Kang:2008us,Kang:2012am}
\begin{equation}
\Delta\langle q_\perp^2 \rangle_{\rm HQ}^{\rm CEM} \approx 2C_A \left( \frac{8\pi^2 \alpha_s}{N_c^2 - 1} \lambda^2 A^{1/3} \right) \approx \frac{2C_A}{C_F}\, \Delta\langle q_\perp^2 \rangle_{\rm DY,\,W/Z}^{\rm LO} = \frac{9}{2}\, \Delta\langle q_\perp^2 \rangle_{\rm DY,\,W/Z}^{\rm LO}\,.
\label{eq:tmb-ratio-9half}
\end{equation}
Similarly, in NRQCD factorization, where the $Q\bar{Q}[^{2S+1}L_J^{(1,8)}]$ pair is produced in specific color-singlet or color-octet states, a color-octet pair rescatters as a net color octet ($C_A$) at leading order in the small spatial separation of the pair, whereas a color-singlet final-state rescattering is strongly suppressed~\cite{Kang:2008us,Kang:2012am}. The dramatic enhancement factor of $2C_A / C_F = 9/2$ for heavy quarkonium broadening relative to Drell-Yan and $W/Z$ broadening is a signature prediction of pQCD multiple scattering~\cite{Kang:2008us,Kang:2012am}. It successfully explains the Fermilab E772/E866 measurements of $J/\psi$, $\Upsilon$, and DY broadening as well as RHIC and LHC data~\cite{Kang:2008us,Kang:2012am,Ru:2019qvz}, and indicates that the heavy-quark pair remains in a colored configuration while traversing the nuclear medium.

\begin{figure}[htb]
\centering
\includegraphics[width=2.0in]{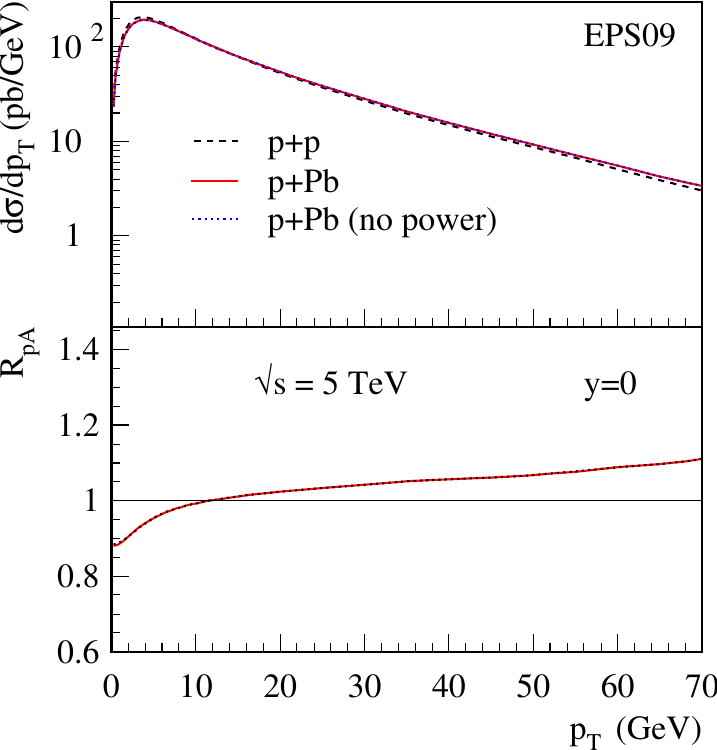}
\hskip 0.5 in
\includegraphics[width=2.5in]{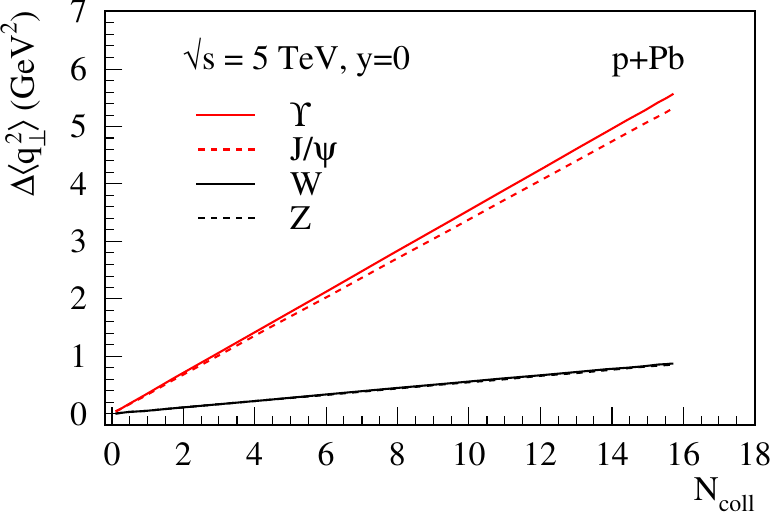}
\caption{Left: Nuclear modification factor $R_{pA}(p_\perp)$ for $Z^0$ boson production at midrapidity ($y=0$) in $p+\text{Pb}$ collisions at $\sqrt{s_{NN}}=5$~TeV calculated within the CSS resummation formalism using EPS09 nuclear PDFs ~\cite{Kang:2012am} (with and without $A^{1/3}$-enhanced dynamical power corrections). Right: Predictions for the transverse momentum broadening $\Delta\langle q_\perp^2 \rangle$ of $\Upsilon$, $J/\psi$, $W$, and $Z^0$ bosons as a function of the number of binary collisions $N_{\rm coll}$ in $p+\text{Pb}$ collisions at $\sqrt{s_{NN}}=5$~TeV~\cite{Kang:2012am}.}
\label{fig:pA-WZ-HQ}
\end{figure}

Furthermore, for massive $Z^0$ boson production in $p+\text{Pb}$ collisions at the LHC, the full $p_\perp$ spectrum can be evaluated using the CSS resummation formalism~\cite{Collins:1984kg,Qiu:2000ga,Qiu:2000hf,Kang:2012am},
\begin{equation}
\frac{d\sigma_{AB\to Z^0+\mathcal{X}}}{dy\,dp_\perp^2} = \frac{1}{(2\pi)^2} \int d^2\bm{b}\, e^{i\bm{p}_\perp\cdot \bm{b}} \widetilde{W}(b, M_Z, x_A, x_B) + Y(p_\perp, M_Z, x_A, x_B)\,,
\label{eq:CSS-Z0}
\end{equation}
where nuclear-size-enhanced multiple scatterings enter the nonperturbative Sudakov factor at $b > b_{\rm max}$ via the dynamical power-correction parameter $g_2 \to g_2 A^{1/3}$~\cite{Kang:2012am}. As shown in Fig.~\ref{fig:pA-WZ-HQ} (left), because the hard scale $M_Z$ is very large, the $A^{1/3}$-enhanced power correction has a negligible impact on the $Z^0$ spectrum at LHC energies, while gluon-initiated subprocesses dominate $Z^0$ production at $p_\perp > 20$~GeV~\cite{Kang:2012am}. This makes the nuclear modification factor $R_{pA}(p_\perp)$ of $Z^0$ bosons a clean probe of leading-twist nuclear gluon antishadowing, whereas the integrated broadening $\Delta\langle q_\perp^2 \rangle$ in Fig.~\ref{fig:pA-WZ-HQ} (right) clearly separates $J/\psi$ and $\Upsilon$ from $W/Z^0$ bosons as a function of $N_{\rm coll}$~\cite{Kang:2012am}.

\subsection{Transverse momentum imbalance for two-particle production in electron-nucleus and proton-nucleus collisions}
\label{sec:TMB-imbalance}

Beyond single-inclusive observables, two-particle (or dijet) angular and momentum correlations in $\gamma/e+A$ and $p+A$ collisions provide detailed multi-dimensional access to both initial-state and final-state multiple parton scatterings~\cite{Luo:1993ui,Qiu:2003vd,Kang:2011bp,Xing:2012ii,Arleo:2025oos}. Consider the inclusive production of two particles or jets $h_1$ and $h_2$ with rapidities $y_1, y_2$ and transverse momenta $\bm{p}_{1\perp}, \bm{p}_{2\perp}$:
\begin{equation}
h(P') + A(p) \to h_1(p_1) + h_2(p_2) + \mathcal{X}\,,
\label{eq:2particle-process}
\end{equation}
where $h = \gamma, \gamma^*, p$. At leading order in pQCD, the two partons in the $2\to 2$ hard scattering are produced strictly back-to-back in the transverse plane, $\bm{p}_{1\perp} = -\bm{p}_{2\perp}$. In the presence of a nuclear target, multiple parton scatterings before and/or after the hard $2\to 2$ scattering deflect the partons from the back-to-back configuration, generating a net transverse momentum imbalance $\bm{q}_\perp \equiv \bm{p}_{1\perp} + \bm{p}_{2\perp}$ (with average transverse momentum $\bm{P}_\perp \equiv (\bm{p}_{1\perp} - \bm{p}_{2\perp})/2$ and $q_\perp \ll P_\perp$). 

\subsubsection{Dijet and heavy-quark pair imbalance in $\gamma+A$ and $e+A$ collisions}

Historically, the first extraction of twist-4 nuclear parton correlation functions was performed by Luo, Qiu, and Sterman in their analysis of dijet momentum imbalance in photon-nucleus ($\gamma+A$) photoproduction~\cite{Luo:1993ui,Luo:1994np}. In direct photoproduction, $\gamma(P') + A(p) \to J_1(p_1) + J_2(p_2) + \mathcal{X}$, the real photon enters the hard scattering directly without initial-state strong interactions, and the two outgoing partons undergo final-state double scattering inside the nucleus via the QCD Compton ($\gamma q \to qg$) and photon-gluon fusion ($\gamma g \to q\bar{q}$) channels.

Extending this framework to virtual photoproduction in $e+A$ DIS ($\gamma^* + A \to J_1 + J_2 + \mathcal{X}$) and heavy-quark pair production ($\gamma^* + A \to Q + \bar{Q} + \mathcal{X}$), Xing, Kang, Vitev, and Wang derived the exact color and kinematic structures of the hard parts~\cite{Xing:2012ii}. In the virtual photon-nucleon center-of-mass frame, the LO dijet cross section in $e+p$ collisions is
\begin{equation}
\frac{d\sigma}{dy_1 dy_2 dP_\perp^2} = \frac{\pi \alpha_{\rm em}\alpha_s}{s^2} \sum_b \frac{f_{b/p}(x)}{x} H_{\gamma^* b\to cd}^U(\hat{s}, \hat{t}, \hat{u})\,,
\label{eq:eA-dijet-LO}
\end{equation}
where $x = P_\perp(e^{-y_1} + e^{-y_2})/\sqrt{s}$ for massless jets, and $H_{\gamma^* b\to cd}^U$ are the unpolarized Born hard functions for $b=q,\bar{q},g$. When the two final-state partons $c$ and $d$ rescatter off soft gluons in the nucleus, four final-state double-scattering diagrams contribute (rescattering on parton $c$ in both amplitude and conjugate amplitude, on parton $d$ in both, and interferences between $c$ and $d$)~\cite{Luo:1993ui,Kang:2011bp,Xing:2012ii}. Summing over all diagrams and performing the collinear expansion on $\delta^2(\bm{q}_\perp - \bm{k}_\perp)$, the nuclear broadening of the dijet transverse momentum imbalance in $e+A$ collisions takes the compact factorized form~\cite{Luo:1993ui,Xing:2012ii}:
\begin{equation}
\Delta\langle q_\perp^2 \rangle_{eA}^{\rm dijet} = \left( \frac{8\pi^2 \alpha_s}{N_c^2 - 1} \right) \frac{\sum_b \frac{1}{x} T_{b/A}^{(F)}(x) H_{\gamma^* b\to cd}^F(\hat{s}, \hat{t}, \hat{u})}{\sum_b \frac{1}{x} f_{b/A}(x) H_{\gamma^* b\to cd}^U(\hat{s}, \hat{t}, \hat{u})}\,,
\label{eq:eA-dijet-imbalance}
\end{equation}
where $T_{b/A}^{(F)}(x)$ ($b=q,g$) are the final-state twist-4 correlation functions defined in Eqs.~\eqref{eq:Tqg-def} and~\eqref{eq:Tgg-def} with $x_2=x_3=0$. The final-state rescattering hard functions $H_{\gamma^* b\to cd}^F$ are related to the Born hard functions $H_{\gamma^* b\to cd}^U$ by specific color and kinematic factors~\cite{Luo:1993ui,Xing:2012ii}:
\begin{align}
H_{\gamma^* q\to qg}^F &= \left( C_F \frac{\hat{s}}{-\hat{t}} + C_A \frac{-\hat{t}}{\hat{s}} \right) \frac{\hat{s}\hat{t}}{\hat{s}^2 + \hat{t}^2}\, H_{\gamma^* q\to qg}^U\,,
\label{eq:H-gamma-q-F}\\
H_{\gamma^* g\to q\bar{q}}^F &= \left( C_F + C_A \frac{\hat{t}\hat{u}}{\hat{t}^2 + \hat{u}^2} \right) H_{\gamma^* g\to q\bar{q}}^U\,.
\label{eq:H-gamma-g-F}
\end{align}
In Eq.~\eqref{eq:H-gamma-q-F}, the factor $C_F \hat{s}/(-\hat{t})$ reflects soft gluon rescattering on the outgoing quark when the quark is produced at a forward angle, whereas $C_A (-\hat{t})/\hat{s}$ reflects rescattering on the outgoing gluon~\cite{Luo:1993ui}. For heavy-quark pair ($Q\bar{Q}$) or heavy-meson pair ($D\bar{D}, B\bar{B}$) production in $e+A$ collisions, only the photon-gluon fusion channel $\gamma^* g \to Q\bar{Q}$ contributes at LO, probing the twist-4 gluon-gluon correlator $T_{g/A}^{(F)}(x)$ cleanly~\cite{Xing:2012ii}. Comparing Eq.~\eqref{eq:eA-dijet-imbalance} with Fermilab E683 dijet photoproduction data yielded the first empirical estimate $\lambda^2 \approx 0.05\text{--}0.1~\text{GeV}^2$~\cite{Luo:1993ui,Luo:1994np}.

\subsubsection{Dijet, dihadron, photon-jet, and heavy-flavor imbalance in $p+A$ collisions}

\begin{figure}[htb]
\centering
\includegraphics[width=2.0in]{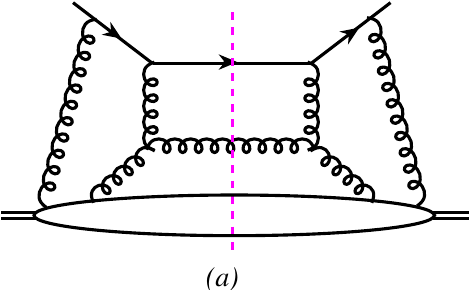}
\hskip 0.5 in
\includegraphics[width=2.0in]{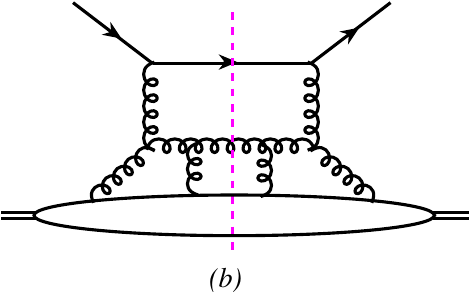}
\caption{Representative double-scattering Feynman diagrams contributing to the dijet and dihadron transverse momentum imbalance $\Delta\langle q_\perp^2 \rangle$ in $p+A$ collisions~\cite{Kang:2011bp}: (a) initial-state double scattering and (b) final-state double scattering.}
\label{fig:pA-imbalance-feyn}
\end{figure}

In proton--nucleus and deuteron--nucleus collisions, $h(P')+A(p)\to h_1(p_1)+h_2(p_2)+\mathcal{X}$ with $h=p,d$, two-particle production receives double-scattering contributions from both the incoming projectile parton (initial-state rescattering, Fig.~\ref{fig:pA-imbalance-feyn}(a)) and the two outgoing partons after the hard $2\to 2$ interaction (final-state rescattering, Fig.~\ref{fig:pA-imbalance-feyn}(b))~\cite{Qiu:2003vd,Kang:2011bp,Xing:2012ii}.

For dijet production in $p+A$ collisions, let $x' = P_\perp(e^{y_1} + e^{y_2})/\sqrt{s}$ and $x = P_\perp(e^{-y_1} + e^{-y_2})/\sqrt{s}$ be the collinear momentum fractions of the active partons $a$ (from the proton) and $b$ (from the nucleus) in the $ab \to cd$ subprocess. Evaluating both initial-state and final-state double-scattering diagrams across all partonic channels, the general formula for the nuclear enhancement of the dijet transverse momentum imbalance can be written as~\cite{Kang:2011bp}:
\begin{equation}
\Delta\langle q_\perp^2 \rangle_{pA}^{\rm dijet} = \left( \frac{8\pi^2 \alpha_s}{N_c^2 - 1} \right) \frac{\sum_{a,b} \frac{f_{a/p}(x')}{x'x} \Big[ T_{b/A}^{(I)}(x) H_{ab\to cd}^I(\hat{s}, \hat{t}, \hat{u}) + T_{b/A}^{(F)}(x) H_{ab\to cd}^F(\hat{s}, \hat{t}, \hat{u}) \Big]}{\sum_{a,b} \frac{f_{a/p}(x') f_{b/A}(x)}{x'x} H_{ab\to cd}^U(\hat{s}, \hat{t}, \hat{u})}\,.
\label{eq:pA-dijet-imbalance}
\end{equation}
Here $T_{b/A}^{(I)}(x)$ and $T_{b/A}^{(F)}(x)$ ($b=q,\bar{q},g$) are the initial-state and final-state twist-4 quark-gluon and gluon-gluon correlation functions [Eqs.~\eqref{eq:Tgq-def}, \eqref{eq:Tqg-def}, and \eqref{eq:Tgg-def}]. Because initial-state rescattering occurs on the single incoming parton $a$ from the projectile proton, the initial-state hard functions $H_{ab\to cd}^I$ are simply proportional to the unpolarized Born hard functions $H_{ab\to cd}^U$ weighted by the color Casimir of parton $a$~\cite{Kang:2011bp,Xing:2012ii}:
\begin{equation}
H_{ab\to cd}^I(\hat{s}, \hat{t}, \hat{u}) = \begin{cases}
C_F\, H_{qb\to cd}^U(\hat{s}, \hat{t}, \hat{u})\,, & a = q, \bar{q}\,, \\[4pt]
C_A\, H_{gb\to cd}^U(\hat{s}, \hat{t}, \hat{u})\,, & a = g\,.
\end{cases}
\label{eq:Hab-I}
\end{equation}
On the other hand, the final-state hard functions $H_{ab\to cd}^F$ depend on the color flow of the $2\to 2$ scattering amplitude and the interference between rescatterings of the two outgoing partons $c$ and $d$. Their explicit forms for all eight massless partonic channels are~\cite{Kang:2011bp}:
\begin{align}
H_{qq'\to qq'}^F &= \frac{(N_c^2 - 3)(N_c^2 - 1)}{2N_c^3} \left( \frac{\hat{s}^2 + \hat{u}^2}{\hat{t}^2} \right)\,,
\label{eq:HF-qqp}\\
H_{qq\to qq}^F &= \frac{(N_c^2 - 3)(N_c^2 - 1)}{2N_c^3} \left( \frac{\hat{s}^2 + \hat{u}^2}{\hat{t}^2} + \frac{\hat{s}^2 + \hat{t}^2}{\hat{u}^2} \right) + \frac{2(N_c^2 - 1)}{N_c^4} \left( \frac{\hat{s}^2}{\hat{t}\hat{u}} \right)\,,
\label{eq:HF-qq}\\
H_{q\bar{q}\to q'\bar{q}'}^F &= C_A\, H_{q\bar{q}\to q'\bar{q}'}^U\,, \qquad
H_{q\bar{q}\to q\bar{q}}^F = C_A\, H_{q\bar{q}\to q\bar{q}}^U - \frac{(N_c^2 - 1)^2}{2N_c^3} \left( \frac{\hat{s}^2 + \hat{u}^2}{\hat{t}^2} \right)\,,
\label{eq:HF-qqbar}\\
H_{qg\to qg}^F &= C_F\, H_{qg\to qg}^U - \frac{N_c}{2} \left[ \frac{\hat{s}(\hat{s}^2 + \hat{u}^2)}{\hat{t}^2 \hat{u}} \right]\,,
\label{eq:HF-qg}\\
H_{q\bar{q}\to gg}^F &= C_A\, H_{q\bar{q}\to gg}^U - \frac{N_c^2 - 1}{2N_c^2} \left( \frac{\hat{t}}{\hat{u}} + \frac{\hat{u}}{\hat{t}} \right)\,, \qquad
H_{gg\to q\bar{q}}^F = C_A\, H_{gg\to q\bar{q}}^U - \frac{1}{2(N_c^2 - 1)} \left( \frac{\hat{t}}{\hat{u}} + \frac{\hat{u}}{\hat{t}} \right)\,,
\label{eq:HF-qqgg}\\
H_{gg\to gg}^F &= C_A\, H_{gg\to gg}^U + \frac{2N_c^3}{(N_c^2 - 1)^2} \left( \frac{\hat{t}}{\hat{u}} + \frac{\hat{u}}{\hat{t}} + 1 \right)^2\,,
\label{eq:HF-gggg}
\end{align}
where the standard unpolarized Born hard functions $H_{ab\to cd}^U(\hat{s}, \hat{t}, \hat{u})$ can be found in Refs.~\cite{Owens:1986mp,Kang:2011bp}.

This formalism has been systematically generalized to several key experimental channels in $p+A$ collisions~\cite{Kang:2011bp,Xing:2012ii}:
\begin{enumerate}
    \item \textbf{Dihadron production ($p+A \to h_1 + h_2 + \mathcal{X}$)}: When the two outgoing partons $c$ and $d$ fragment into hadrons $h_1$ and $h_2$ with fragmentation fractions $z_1$ and $z_2 = z_1 p_{2\perp}/p_{1\perp}$, let $\pm\bm{P}_{\perp,\mathrm{parton}}$ denote the transverse momenta before fragmentation. Then $\bm{p}_{1\perp}=z_1\bm{P}_{\perp,\mathrm{parton}}$ and $\bm{p}_{2\perp}=-z_2\bm{P}_{\perp,\mathrm{parton}}$, so the dihadron momentum imbalance in $p+p$ collisions is $\bm{q}_\perp=(z_1-z_2)\bm{P}_{\perp,\mathrm{parton}}$. In $p+A$ collisions, because the random soft momentum kick $\bm{k}_\perp$ from the nuclear medium is uncorrelated with the vacuum fragmentation direction, the nuclear broadening of the dihadron imbalance is given by convoluting Eq.~\eqref{eq:pA-dijet-imbalance} with the fragmentation functions $D_{h_1/c}(z_1)$ and $D_{h_2/d}(z_2)$~\cite{Kang:2011bp}:
    \begin{equation}
    \Delta\langle q_\perp^2 \rangle_{pA}^{h_1 h_2} = \left( \frac{8\pi^2 \alpha_s}{N_c^2 - 1} \right) \frac{\sum_{abcd} \int \frac{dz_1}{z_1} D_{h_1/c}(z_1) D_{h_2/d}(z_2) \frac{f_{a/p}(x')}{x'x} \Big[ T_{b/A}^{(I)}(x) H_{ab\to cd}^I + T_{b/A}^{(F)}(x) H_{ab\to cd}^F \Big]}{\sum_{abcd} \int \frac{dz_1}{z_1} D_{h_1/c}(z_1) D_{h_2/d}(z_2) \frac{f_{a/p}(x') f_{b/A}(x)}{x'x} H_{ab\to cd}^U}\,,
    \label{eq:pA-dihadron-imbalance}
    \end{equation}
    where $x' = \frac{p_{1\perp}}{z_1\sqrt{s}}(e^{y_1} + e^{y_2})$ and $x = \frac{p_{1\perp}}{z_1\sqrt{s}}(e^{-y_1} + e^{-y_2})$.
    \item \textbf{Direct photon-jet ($\gamma+\text{jet}$) and photon-hadron ($\gamma+h$) production}: For $p+A \to \gamma(p_1) + J(p_2) + \mathcal{X}$, the produced photon does not interact strongly in the final state, so only the recoil parton $d$ experiences final-state rescattering~\cite{Xing:2012ii}:
    \begin{equation}
    H_{q\bar{q}\to \gamma g}^I = C_F\, H_{q\bar{q}\to \gamma g}^U\,, \quad H_{q\bar{q}\to \gamma g}^F = C_A\, H_{q\bar{q}\to \gamma g}^U\,, \qquad
    H_{qg\to \gamma q}^I = C_F\, H_{qg\to \gamma q}^U\,, \quad H_{gq\to \gamma q}^I = C_A\, H_{gq\to \gamma q}^U\,, \quad H_{qg\to \gamma q}^F = C_F\, H_{qg\to \gamma q}^U\,.
    \label{eq:photon-jet-H}
    \end{equation}
    For $\gamma+h$ production, where the recoil parton fragments with fraction $z = p_{2\perp}/p_{1\perp}$, the broadening of the hadronic imbalance gets weighted by $z^2$ in the final-state rescattering term: $H_{ab\to \gamma d}^I + z^2 H_{ab\to \gamma d}^F$~\cite{Xing:2012ii}.
    \item \textbf{Heavy-quark pair ($Q\bar{Q}$) and heavy-meson pair ($D\bar{D}, B\bar{B}$) production}: In $p+A \to Q(p_1) + \bar{Q}(p_2) + \mathcal{X}$ via $q\bar{q}\to Q\bar{Q}$ and $gg\to Q\bar{Q}$, the massive hard functions $H_{ab\to Q\bar{Q}}^{U,I,F}(\hat{s}, \hat{t}, \hat{u}, m_Q)$ incorporate finite heavy-quark mass effects, providing a sensitive probe of the gluon twist-4 correlator $T_{g/A}^{(I,F)}(x)$~\cite{Xing:2012ii}.
\end{enumerate}

\subsubsection{Phenomenology of dihadron azimuthal correlations in $d+\text{Au}$ collisions}

At RHIC, back-to-back dihadron azimuthal correlations in $d+\text{Au}$ collisions at $\sqrt{s_{NN}}=200$~GeV have been extensively measured by the PHENIX and STAR collaborations across mid-mid, mid-forward, and forward-forward rapidities~\cite{PHENIX:2011puq,STAR:2021fgw}. Two prominent cold nuclear matter effects are observed in the away-side peak ($\Delta\phi \approx \pi$): a broadening of the azimuthal width $\sigma_F$ and a strong suppression of the correlated yield (quantified by the nuclear modification factor $R_{dA}^{(2)}$) at forward rapidities.

In Refs. \cite{Qiu:2003vd,Qiu:2004da,Kang:2011bp}, within the high-twist formalism, the twist-4 correlation functions are parameterized as~
\begin{equation}
\frac{4\pi^2 \alpha_s}{N_c} T_{q,g/A}^{(I)}(x) = \frac{4\pi^2 \alpha_s}{N_c} T_{q,g/A}^{(F)}(x) = \xi^2 (A^{1/3} - 1) f_{q,g/A}(x)\,,
\label{eq:T4-xi2-model}
\end{equation}
where $\xi^2 = 0.09\text{--}0.12~\text{GeV}^2$ is the characteristic scale of high-twist power corrections per nucleon extracted from nuclear shadowing in DIS~\cite{Qiu:2003vd}. Substituting Eq.~\eqref{eq:T4-xi2-model} into Eq.~\eqref{eq:pA-dihadron-imbalance} gives the dihadron imbalance broadening $\Delta\langle q_\perp^2 \rangle$. The away-side azimuthal width $\sigma_F$ is related to the total imbalance $\langle q_\perp^2 \rangle_{dA} = \langle q_\perp^2 \rangle_{pp} + \Delta\langle q_\perp^2 \rangle$ and the vacuum fragmentation transverse momentum $\langle |j_{\perp y}| \rangle \approx 0.4$~GeV via kinematic geometry~\cite{Qiu:2003vd,Kang:2011bp}:
\begin{equation}
\cos^2\left( \sqrt{\frac{2}{\pi}}\,\sigma_F \right) = \frac{1 - \langle |j_{\perp y}| \rangle^2/p_{\perp,{\rm assoc}}^2}{1 + \langle |j_{\perp y}| \rangle^2/p_{\perp,{\rm trig}}^2 + \frac{1}{\pi}\langle q_\perp^2 \rangle/p_{\perp,{\rm trig}}^2}\,.
\label{eq:sigmaF-imbalance}
\end{equation}

\begin{figure}[htb]
\centering
\includegraphics[width=1.55in]{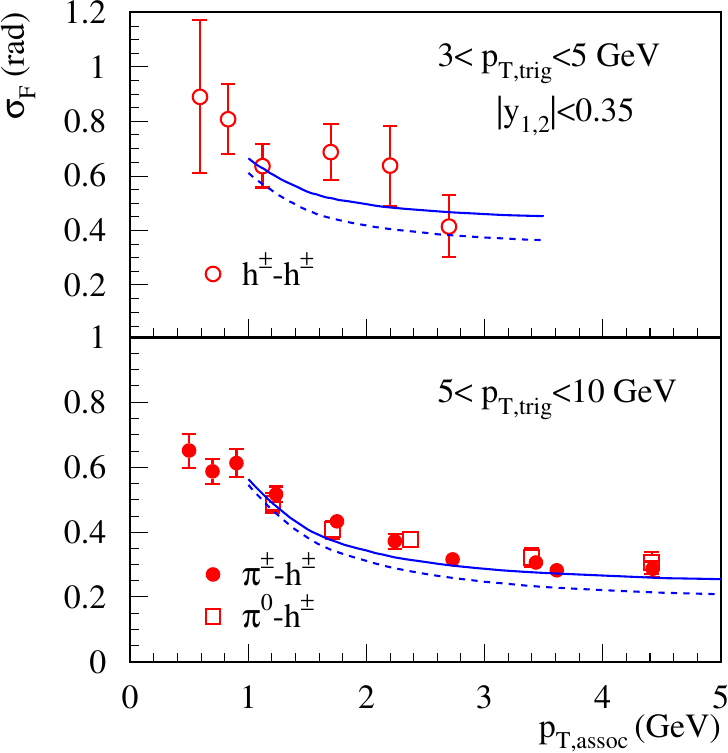}
\hskip 0.2 in
\includegraphics[width=2.35in]{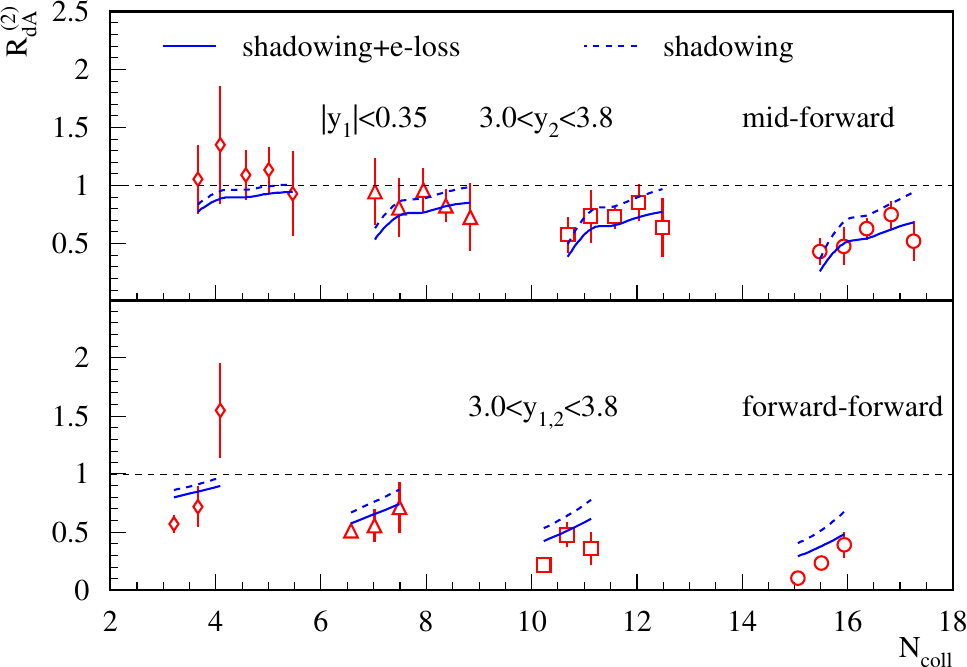}
\hskip 0.2 in
\includegraphics[width=1.8in]{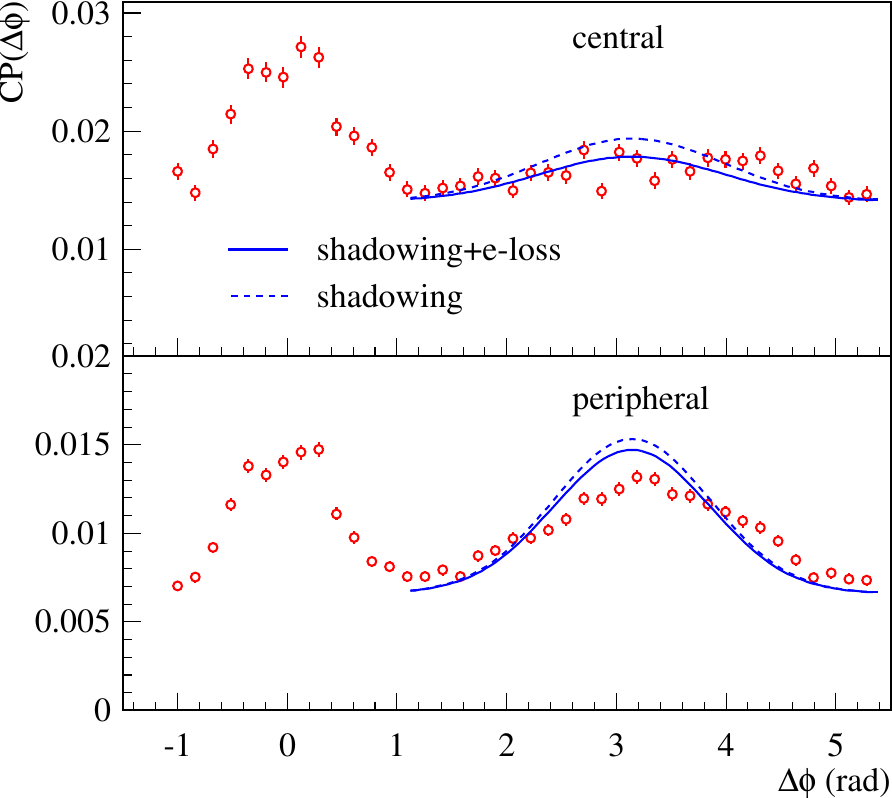}
\caption{Left: Width $\sigma_F$ of the away-side dihadron azimuthal correlation at midrapidity as a function of $p_{\perp,{\rm assoc}}$ in $d+\text{Au}$ (solid curves, including multiple-scattering broadening $\Delta\langle q_\perp^2 \rangle$) and $p+p$ (dashed curves) collisions compared with PHENIX data~\cite{PHENIX:2005veb,Kang:2011bp}. Middle: Nuclear modification factor $R_{dA}^{(2)}$ for mid-forward and forward-forward dihadron production as a function of $N_{\rm coll}$ compared with PHENIX data~\cite{PHENIX:2011puq,Kang:2011bp}. Right: Forward dihadron azimuthal correlation function $CP(\Delta\phi)$ ($\langle y_{1,2}\rangle = 3.2$) in central and peripheral $d+\text{Au}$ collisions compared with STAR data~\cite{Braidot:2010ig,Kang:2011bp}.}
\label{fig:dAu-dihadron}
\end{figure}

At the same time, the suppression of the dihadron production rate $R_{dA}^{(2)}$ at forward rapidity is governed by the interplay of two high-twist mechanisms~\cite{Qiu:2004da,Vitev:2006bi,Kang:2011bp}:
\begin{enumerate}
    \item \textbf{Coherent power corrections (dynamical shadowing)}: At forward rapidities, the target parton momentum fraction $x$ is small, so the coherence length exceeds the nuclear size $(\lambda \gg L)$ and the incoming parton scatters coherently off multiple nucleons~\cite{Qiu:2003vd,Qiu:2004da}. Resumming the all-twist series of maximally nuclear-enhanced power corrections $(\xi^2 A^{1/3} / |\hat{t}|)^n$ leads to a shift in the target momentum fraction $x$ in the $t$-channel (and analogously for $\hat{u}$ and $\hat{s}$ channels)~\cite{Qiu:2004da,Vitev:2006bi,Kang:2011bp}:
    \begin{equation}
    x \to x \left( 1 + C_d \frac{\xi^2 (A^{1/3} - 1)}{-\hat{t}} \right)\,,
    \label{eq:dynamical-shadowing-shift}
    \end{equation}
    where $C_d = 1$ ($C_A / C_F$) if parton $d$ is a quark (gluon). Notice that both the dynamical shadowing shift in Eq.~\eqref{eq:dynamical-shadowing-shift} and the imbalance broadening $\Delta\langle q_\perp^2 \rangle$ originate from the same twist-4 scale $\xi^2$~\cite{Kang:2011bp}.
    \item \textbf{Initial-state cold nuclear matter energy loss}: Medium-induced gluon bremsstrahlung from the incoming projectile parton shifts its momentum fraction $x' \to x' / (1-\epsilon)$, convoluted with the energy-loss probability distribution $P_{q,g}(\epsilon)$~\cite{Vitev:2007ve,Kang:2011bp}:
    \begin{equation}
    f_{a/p}(x') \to \int_0^1 d\epsilon\, P_a(\epsilon)\, f_{a/p}\!\left( \frac{x'}{1-\epsilon} \right)\,.
    \label{eq:initial-eloss-conv}
    \end{equation}
\end{enumerate}
Combining the away-side width broadening $\sigma_F$ from Eq.~\eqref{eq:sigmaF-imbalance} and the area suppression $A_F^{d{\rm Au}} = R_{dA}^{(2)} A_F^{pp}$ inside the Gaussian correlation function $CP(\Delta\phi) = B + \frac{A_N}{\sqrt{2\pi}\sigma_N} e^{-\Delta\phi^2/(2\sigma_N^2)} + \frac{A_F}{\sqrt{2\pi}\sigma_F} e^{-(\Delta\phi-\pi)^2/(2\sigma_F^2)}$, the high-twist formalism provides a consistent description of the PHENIX away-side width, the PHENIX mid-forward and forward-forward $R_{dA}^{(2)}$ across centralities, and the STAR forward dihadron azimuthal correlation $CP(\Delta\phi)$, as shown in Fig.~\ref{fig:dAu-dihadron}~\cite{Kang:2011bp}.

\subsection{Global determination of the jet transport parameter in cold nuclear matter}
\label{sec:TMB-global}

The jet transport coefficient $\hat{q}$, defined as the average squared transverse momentum broadening per unit path length of a fast parton traversing a QCD medium, is a fundamental parameter characterizing the transverse properties of both hot quark-gluon plasma (QGP)~\cite{JET:2013cls,Gross:2022hyw,Niida:2021wut} and cold nuclear matter~\cite{Baier:1996sk,Ru:2019qvz}. Within the high-twist collinear factorization framework, this quantity is related to the twist-4 matrix element.

Under the approximation of a large and loosely bound nucleus, where momentum and spatial correlations between two different nucleons can be neglected, the four-parton nuclear matrix element factorizes into the product of a single-nucleon twist-2 parton distribution and the two-gluon field strength expectation value integrated along the parton path~\cite{Luo:1994np,Osborne:2002st,Kang:2013raa}:
\begin{equation}
T_{qg}(x, 0, 0, \mu^2) \approx \frac{N_c}{4\pi^2 \alpha_s} f_{q/A}(x, \mu^2) \int dy^-\, \hat{q}(x, \mu^2, y^-) = \frac{9 R_A}{8\pi^2 \alpha_s} f_{q/A}(x, \mu^2)\, \hat{q}(x, \mu^2)\,,
\label{eq:Tqg-qhat-relation}
\end{equation}
where $N_c = 3$, $R_A = r_0 A^{1/3}$ ($r_0 \approx 1.12\text{--}1.2$~fm) is the nuclear radius, and $\hat{q}(x, \mu^2)$ denotes the nuclear-geometry-averaged quark jet transport coefficient~\cite{Ru:2019qvz,Ru:2023ars}. Similarly, for the gluon-gluon correlation function $T_{gg}(x, 0, 0, \mu^2)$, the same factorization holds with $f_{q/A}(x, \mu^2)$ replaced by $f_{g/A}(x, \mu^2)$ (and the corresponding gluon transport coefficient $\hat{q}_A = \frac{C_A}{C_F} \hat{q}$ accounted for in the hard color factors)~\cite{Kang:2008us,Ru:2019qvz,Ru:2023ars}. Comparing Eq.~\eqref{eq:Tqg-qhat-relation} with Eqs.~\eqref{eq:T4-lambda-model} and~\eqref{eq:T4-xi2-model} immediately connects the different high-twist notations used in the literature:
\begin{equation}
\frac{4\pi^2 \alpha_s}{N_c} \lambda^2 (A^{1/3}-1) = \xi^2 (A^{1/3}-1) \approx \frac{3}{2} R_A\, \hat{q}(x, \mu^2),\,\, {\rm for}\,\, A\gg 1
\label{eq:qhat-lambda-xi-bridge}
\end{equation}
Substituting Eq.~\eqref{eq:Tqg-qhat-relation} into Eqs.~\eqref{eq:sidis-tmb-LO-ratio}, \eqref{eq:dy-tmb-LO-ratio}, and \eqref{eq:tmb-HQ-CEM} yields simple, direct relations between experimental broadening observables and $\hat{q}$ at LO~\cite{Ru:2019qvz,Ru:2023ars}:
\begin{equation}
\Delta\langle \ell_{h\perp}^2 \rangle_{\rm SIDIS}^{\rm LO} = \frac{3}{2} R_A z_h^2\, \hat{q}\,, \qquad
\Delta\langle q_\perp^2 \rangle_{\rm DY}^{\rm LO} = \frac{3}{2} R_A\, \hat{q}\,, \qquad
\Delta\langle q_\perp^2 \rangle_{\rm HQ}^{\rm CEM,\,LO} \approx \frac{2C_A}{C_F} \left( \frac{3}{2} R_A\, \hat{q} \right)\,.
\label{eq:tmb-qhat-summary}
\end{equation}
For centrality-dependent measurements in $d+\text{Au}$ and $p+\text{Pb}$ collisions at RHIC and the LHC, the nuclear path length scales with the mean number of binary nucleon-nucleon collisions, replacing $A^{1/3} \to A^{1/3} \langle N_{\rm coll}(b) \rangle / \langle N_{\rm coll}(b_{\rm min.\,bias}) \rangle$~\cite{Kang:2011bp,Kang:2012am,Ru:2019qvz}.


In early phenomenological studies, $\hat{q}$ (or equivalently $\lambda^2$ and $\xi^2$) was treated as a constant independent of the probing scale $\mu^2$ and momentum fraction $x$~\cite{Luo:1993ui,Qiu:2003vd,Kang:2008us}. However, the derivation of the NLO twist-4 evolution equation in Eq.~\eqref{eq:Tqg-evolution}~\cite{Kang:2013raa,Kang:2014ela,Kang:2016ron}, as well as small-$x$ renormalization group analyses~\cite{Liou:2013qya,Blaizot:2014bha,Iancu:2014kga}, demonstrated that $\hat{q}(x, \mu^2)$ possesses nontrivial dependence on both the momentum fraction $x$ and the hard scale $\mu^2 = Q^2$. To test the universality of $\hat{q}$ and extract its kinematic dependence in a data-driven manner, the first global QCD analysis of the jet transport coefficient in cold nuclear matter was performed~\cite{Ru:2019qvz,Ru:2023ars}. The global analysis incorporates $215$ experimental data points from $8$ world datasets spanning a wide range of collision systems and energies ($Q^2 > 1~\text{GeV}^2$)~\cite{Ru:2019qvz}: Hadron ($\pi^\pm, K^\pm$) transverse momentum broadening in $e+A$ SIDIS from HERMES (probing $\nu$, $x_B$, $Q^2$, and $A$ dependences)~\cite{HERMES:2009uge}; Drell-Yan dilepton broadening in $p+A$ collisions from Fermilab E772~\cite{McGaughey:1999mq} and in $\pi+A$ collisions from CERN SPS NA10~\cite{NA10:1987hho}; Heavy quarkonium broadening in $p/d+A$ collisions including $\Upsilon$ from E772~\cite{Alde:1991sw}, $J/\psi$ from E866~\cite{Leitch:1995yc}, $J/\psi$ from PHENIX at RHIC $\sqrt{s_{NN}}=200$~GeV~\cite{PHENIX:2012czk}, and $J/\psi$ from ALICE at the LHC $\sqrt{s_{NN}}=5.02$~TeV ~\cite{ALICE:2015sru}; Nuclear structure function ratios $F_2^A(x_B, Q^2) / F_2^D(x_B, Q^2)$ in $e/\mu+A$ DIS from Fermilab E665 (C, Ca, Xe, and Pb targets)~\cite{E665:1992ivw,E665:1995xur}. 

When fitting each dataset individually under the traditional assumption of a constant $\hat{q}(x, \mu^2) = \hat{q}_0$, the extracted values of $\hat{q}_0$ vary by more than a factor of three across different experiments: from $\hat{q}_0 \approx 0.01\text{--}0.015~\text{GeV}^2/\text{fm}$ in HERMES SIDIS, E772 DY, E866 $J/\psi$, and backward-rapidity RHIC/LHC $J/\psi$ (which probe moderate $x \sim 10^{-2}\text{--}10^{-1}$), up to $\hat{q}_0 \approx 0.038~\text{GeV}^2/\text{fm}$ in forward-rapidity ALICE $J/\psi$ production (which probes small $x \sim 10^{-4}$)~\cite{Ru:2019qvz}. A simultaneous fit to all 215 data points with a constant $\hat{q}_0$ yields a poor $\chi^2 = 387.9$ ($\chi^2/\text{d.o.f.} \approx 1.81$), failing particularly in the small-$x$ LHC forward region~\cite{Ru:2019qvz}.

\begin{figure}[htb]
\centering
\includegraphics[width=3.0in]{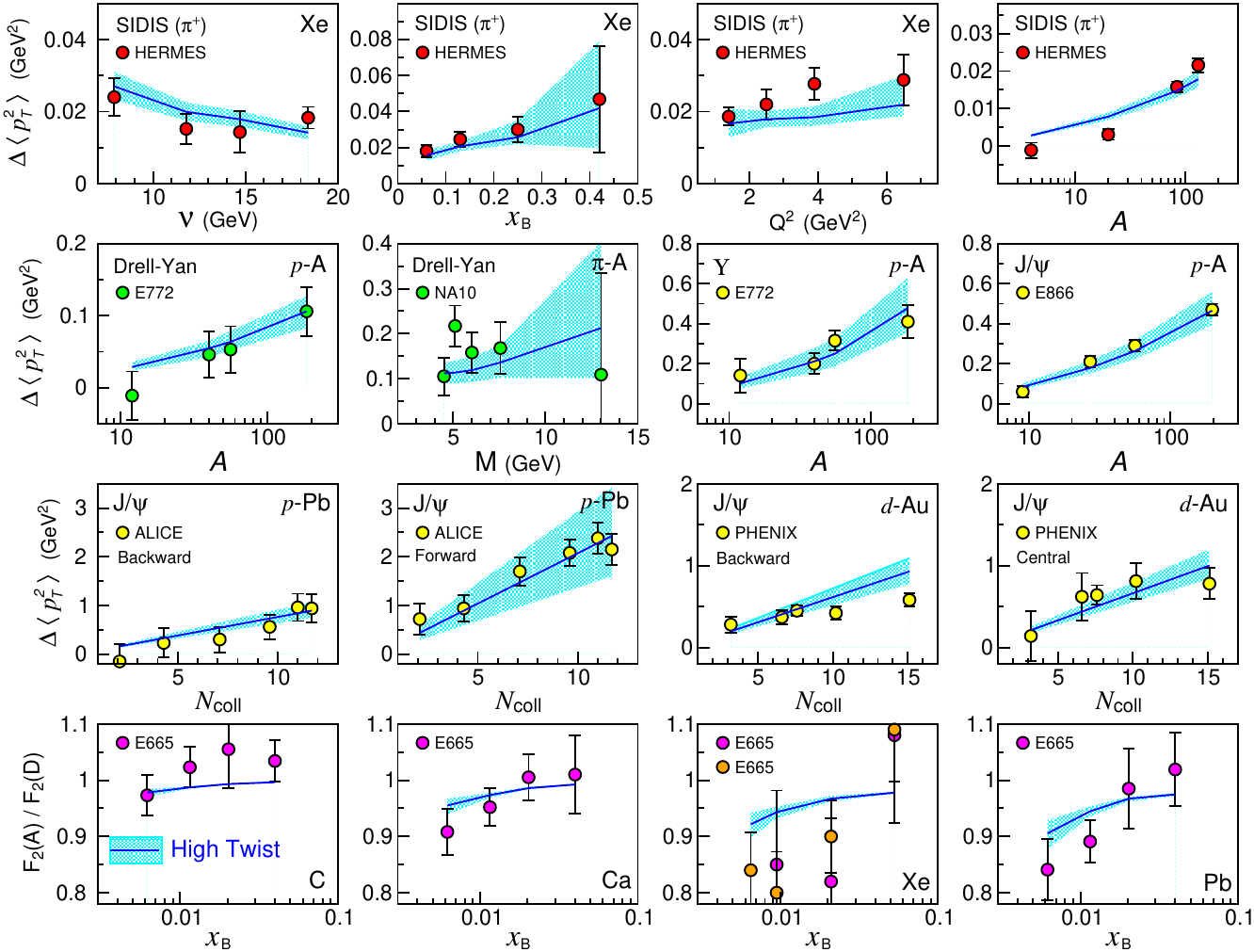}
\hskip 0.7 in
\includegraphics[width=2.5in]{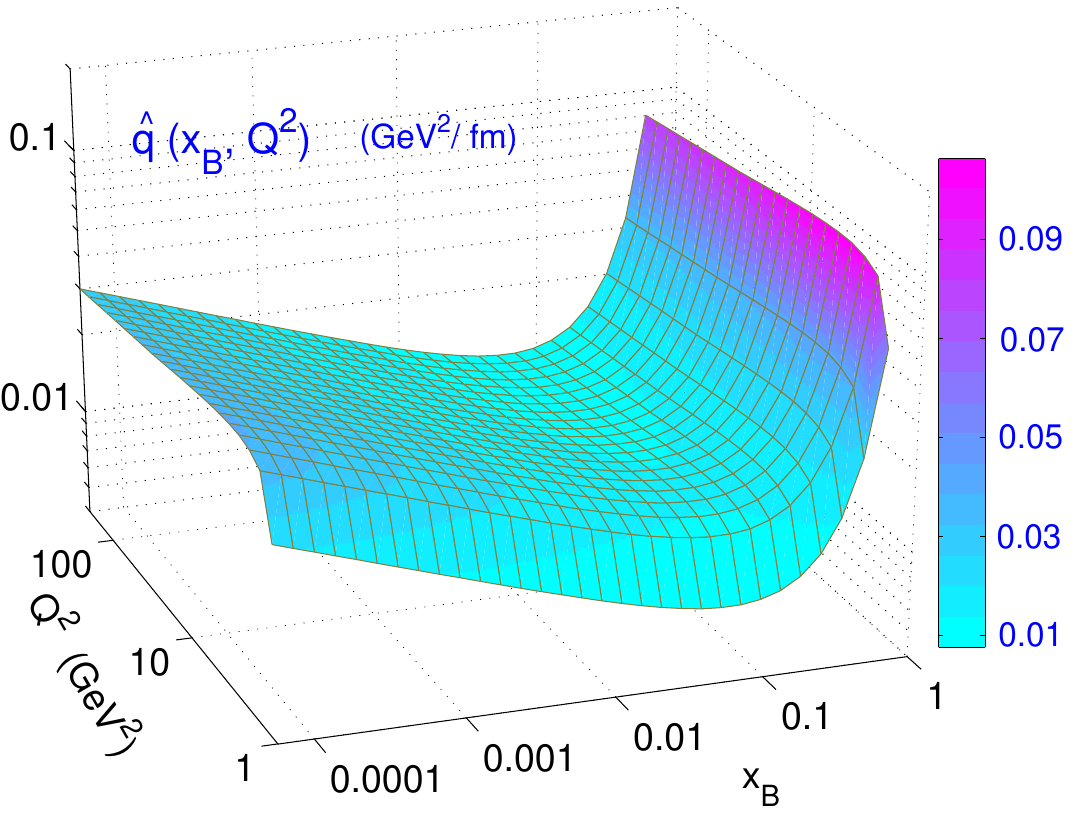}
\caption{Left: Comparison of the global high-twist theoretical fit (solid blue curves with shaded 90\% confidence bands) against world data on transverse momentum broadening $\Delta\langle p_T^2 \rangle$ in SIDIS (HERMES), Drell-Yan (E772, NA10), $\Upsilon$ (E772), $J/\psi$ (E866, PHENIX, ALICE), and on DIS structure-function ratios $F_2(A)/F_2(D)$ (E665)~\cite{Ru:2019qvz}. Right: Extracted cold nuclear matter jet transport coefficient $\hat{q}(x_B, Q^2)$ as a function of Bjorken $x_B$ and scale $Q^2$~\cite{Ru:2019qvz}.}
\label{fig:global-qhat}
\end{figure}

To capture the underlying QCD dynamics, Ref.~\cite{Ru:2019qvz} introduced a flexible 4-parameter kinematic parametrization for $\hat{q}(x, \mu^2)$:
\begin{equation}
\hat{q}(x, \mu^2) = \hat{q}_0\, \alpha_s(\mu^2)\, x^\alpha (1 - x)^\beta \ln^\gamma(\mu^2 / \mu_0^2)\,,
\label{eq:qhat-4param}
\end{equation}
where $\mu_0 = 1$~GeV. In Eq.~\eqref{eq:qhat-4param}, the explicit factor of $\alpha_s(\mu^2)$ compensates the $\alpha_s$ in the denominator of Eq.~\eqref{eq:Tqg-qhat-relation}, $\ln^\gamma(\mu^2/\mu_0^2)$ parameterizes the logarithmic scale evolution difference between $T_{qg}(x, 0, 0, \mu^2)$ and $f_{q/A}(x, \mu^2)$, $x^\alpha$ captures the small-$x$ growth associated with rising gluon densities and saturation scales ($Q_s^2 \propto x^{-\lambda}$)~\cite{Golec-Biernat:1998zce}, and $(1-x)^\beta$ accounts for higher-twist enhancements at large $x$~\cite{Dokshitzer:1995qm}. Performing a global $\chi^2$ minimization (with Lagrange multiplier and Hessian uncertainty analyses at the 90\% confidence level) yields a remarkably improved fit with $\chi^2 = 260.2$ ($\chi^2/\text{d.o.f.} = 1.21$) and the optimal parameters~\cite{Ru:2019qvz} (see also the Hessian analysis in Ref.~\cite{Ru:2023ars}):
\begin{equation}
\hat{q}_0 = 0.0195^{+0.0085}_{-0.0065}~\text{GeV}^2/\text{fm}\,, \qquad
\alpha = -0.174^{+0.077}_{-0.069}\,, \qquad
\beta = -2.79^{+1.99}_{-2.04}\,, \qquad
\gamma = 0.254^{+0.174}_{-0.166}\,.
\label{eq:qhat-fit-values}
\end{equation}
As demonstrated in Fig.~\ref{fig:global-qhat} (left), the global fit simultaneously describes all world datasets across $e+A$ and $p+A$ collisions, providing strong quantitative confirmation of the universality of $\hat{q}(x, \mu^2)$ within the generalized high-twist factorization formalism~\cite{Ru:2019qvz}.

The extracted 3D profile of $\hat{q}(x_B, Q^2)$ in Fig.~\ref{fig:global-qhat} (right) reveals several key physical features of cold nuclear matter~\cite{Ru:2019qvz,Ru:2023ars}. $\hat{q}$ increases significantly as $x_B$ decreases toward $10^{-4}$, consistent with the growth of the small-$x$ gluon density and saturation scale in heavy nuclei. Notice also that in the target rest frame of DIS ($x_B = Q^2 / (2M_N \nu)$), a negative $\alpha$ implies that at fixed $Q^2$, $\hat{q}$ increases with the initial jet energy $\nu$, an important feature for jet quenching phenomenology.
The negative power $\beta$, driven primarily by large-$x_B$ HERMES SIDIS data, suggests a faster-than-twist-2 rise of relative power corrections as $x \to 1$, though uncertainties remain sizable due to limited data at $x_B > 0.3$.
A mild positive logarithmic growth with $Q^2$ is favored by the data.
In the intermediate kinematic region, $\hat{q} \sim 0.02\text{--}0.05~\text{GeV}^2/\text{fm}$ in cold nuclear matter (corresponding to $\hat{q}_0 \approx 0.05~\text{GeV}^2/\text{fm}$ when defined at $x=10^{-2}$),  which is an order of magnitude smaller than $\hat{q}_{\rm QGP} \sim 1\text{--}2~\text{GeV}^2/\text{fm}$ extracted in relativistic heavy-ion collisions at RHIC and the LHC~\cite{JET:2013cls,Ru:2019qvz,Gross:2022hyw,Niida:2021wut}. 

Recently, comprehensive impact studies for the future Electron-Ion Collider (EIC) and fixed-target programs (such as Jefferson Lab 12~GeV) have shown that precision measurements of hadron and jet transverse momentum broadening across wide $(x_B, Q^2, z_h, A)$ bins in $e+A$ collisions will dramatically reduce the parameter uncertainties of $\hat{q}(x, \mu^2)$ in both the small-$x$ saturation regime and the large-$x$ region~\cite{Ru:2023ars,Arleo:2025oos}. Incorporating these constraints together with full NLO twist-4 hard parts and evolution equations will enable high-precision three-dimensional jet tomography of cold nuclear matter.

\section{Correspondence between high-twist and CGC}
\label{sec:CGC}

In Secs.~\ref{sec:eloss} and~\ref{sec:TMB} multiple scattering in a large nucleus was organized as a power expansion: each additional soft scattering brings a suppression factor of $1/Q^2$ and a nuclear-size enhancement proportional to $A^{1/3}$, so that the leading nuclear effects reside in twist-4 multi-parton correlation functions. The color glass condensate (CGC) effective theory describes the same nuclear medium from the opposite end of the kinematics~\cite{McLerran:1993ni,McLerran:1993ka,Gelis:2010nm,Morreale:2021pnn}. At small momentum fraction $x$ the gluon density grows until nonlinear recombination tames it~\cite{Gribov:1984tu,Mueller:1985wy}, and the resulting saturated state is characterized by an emergent transverse momentum scale $Q_s$, which depends on the energy and on the nuclear size. In the CGC the large-$x$ partons are treated as static color sources and the small-$x$ gluons as classical fields. The interaction of an energetic parton with the nucleus is encoded in a light-like Wilson line that resums eikonal scatterings to all orders. The medium is then described by correlators of Wilson lines, which obey the Balitsky--Kovchegov (BK) and Jalilian-Marian--Iancu--McLerran--Weigert--Leonidov--Kovner (JIMWLK) evolution equations~\cite{Balitsky:1995ub,Kovchegov:1999yj,Jalilian-Marian:1997qno,Iancu:2000hn}.

The two frameworks therefore differ in three respects: i) the factorization they rely on, collinear factorization generalized to higher twists~\cite{Qiu:1990xxa,Luo:1993ui} versus leading-order transverse-momentum-dependent (TMD) factorization formulae at small $x$, ii) the evolution of the nonperturbative objects, of the Dokshitzer--Gribov--Lipatov--Altarelli--Parisi (DGLAP) type for the multi-parton correlators~\cite{Kang:2013raa} versus the BK/JIMWLK type for the Wilson line correlators, and iii) the treatment of multiple scattering, order by order in the high-twist (HT) expansion versus exponentiated into the Wilson line in the CGC. Both frameworks nevertheless carry the same $A^{1/3}$ enhancement, and both should describe the region $x\ll 1$ with $Q^2\gtrsim Q_s^2$, where the expansion in $Q_s^2/Q^2$ converges and the eikonal approximation is accurate. In that overlap region the two must agree.

Establishing this correspondence has attracted considerable effort~\cite{Arleo:2025oos}. Subeikonal corrections to the parton propagators in the CGC have been computed in a number of settings~\cite{Altinoluk:2014oxa,Chirilli:2018kkw,Altinoluk:2021lvu}, most recently for forward parton--nucleus scattering, for dijets in DIS and for the DIS structure functions themselves~\cite{Altinoluk:2024zom,Altinoluk:2025ang}, and they are known to be required for spin physics at small $x$~\cite{Kovchegov:2015pbl}. The rapidity evolution of unintegrated gluon distributions from small to moderate $x$, and its interplay with DGLAP evolution, has been studied in~\cite{Balitsky:2015qba,Caucal:2025mth}. Semi-classical formulations that interpolate between Regge and Bjorken kinematics have been proposed in~\cite{Boussarie:2021wkn}, a background-field factorization valid at arbitrary $x$ has been constructed for back-to-back dijets~\cite{Mukherjee:2026cte}, and collinear parton distributions have been defined and matched directly in the shockwave formalism~\cite{Bhattacharya:2025fnz}. Most of these studies are formulated at the level of operators or of parton propagators. In this section we review the correspondence at the level of a physical observable, following~\cite{Fu:2023jqv,Fu:2024sba}, where direct photon production in proton--nucleus collisions was computed in both frameworks and compared order by order in $Q_s^2/p_{\gamma\perp}^2$. Direct photon production is the simplest process for this purpose, because only the quark line undergoes multiple scattering, while the photon's transverse momentum $p_{\gamma\perp}$ directly measures the hard scale. We first describe the HT result and its behavior in the incoherent (large-$x$) and coherent (small-$x$) regimes. We then turn to the CGC result and its expansion at leading twist and at twist 4, and we close with the ingredients that must be added to the CGC to reproduce the HT expansion at finite $x$. The hard factors are evaluated at tree level on both sides, and the detailed derivations can be found in~\cite{Fu:2024sba}.

\begin{figure}[htb]
    \centering
\includegraphics[width=0.55\textwidth]{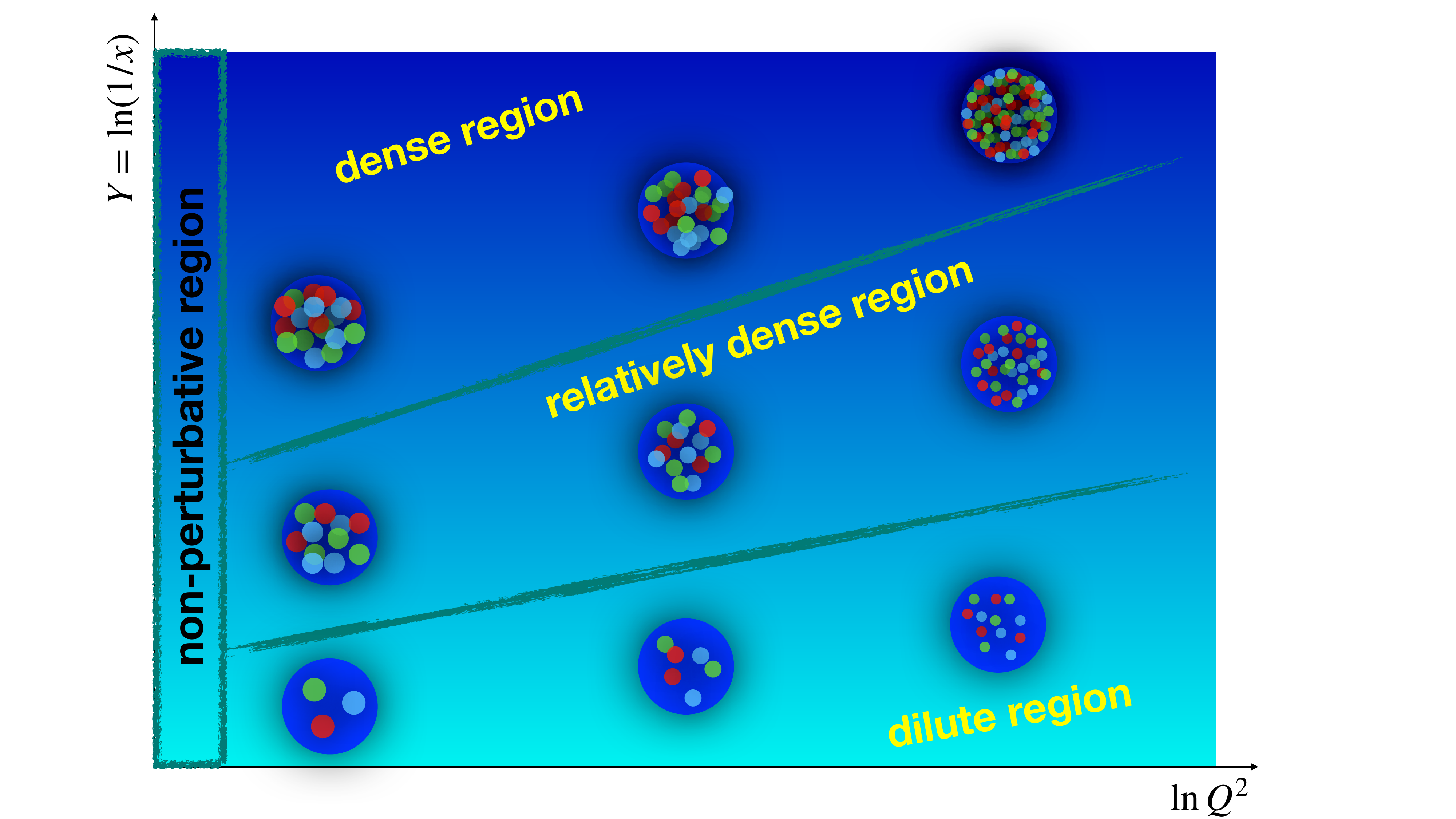}
\caption{Schematic phase diagram of parton density in a nucleus as a function of the gluon momentum fraction $x$ and of the hard scale $Q^2$. Leading-twist collinear factorization describes the dilute region $x\sim\mathcal{O}(1)$, the high-twist expansion describes the incoherent multiple scattering regime $x\lesssim\mathcal{O}(1)$, and the CGC describes the coherent, dense regime $x\to 0$. The two formalisms share the region $x\ll 1$ with $Q^2\gtrsim Q_s^2$. Figure taken from Ref.~\cite{Fu:2023jqv}.}
\label{fig:phase-diagram}
\end{figure}

Throughout this section we consider
\begin{equation}
p(P)+A(p)\rightarrow \gamma(p_\gamma)+\mathcal{X}\,,
\qquad
P=[0,P^-,0]\,, \qquad p=[p^+,0,0]\,,
\label{eq:cgc-process}
\end{equation}
with $p$ the momentum per nucleon as in Sec.~\ref{sec:eloss}. We focus on photons produced in the proton-going direction, where the dominant partonic channel is a quark from the proton scattering off gluons from the nucleus. We denote by $x_q$ the momentum fraction of the incoming quark, by $\xi=p_\gamma^-/(x_qP^-)$ the fraction of its light-cone momentum carried by the photon, and by
\begin{equation}
x=\frac{p_{\gamma\perp}^2}{\xi(1-\xi)x_q s}\,, \qquad s=(P+p)^2\,,
\label{eq:cgc-x}
\end{equation}
the momentum fraction of the gluon from the nucleus fixed by the $2\to2$ kinematics. The variable $x$ and the hard scale $p_{\gamma\perp}^2$ are the two coordinates of the ``phase diagram'' of parton density in Fig.~\ref{fig:phase-diagram}, and the regimes discussed below are regions of this plane.

\subsection{High-twist expansion at large $x$: incoherent multiple scattering}

\label{sec:HT-largex}

\begin{figure}[h!tb] \centering \includegraphics[width=0.78\textwidth]{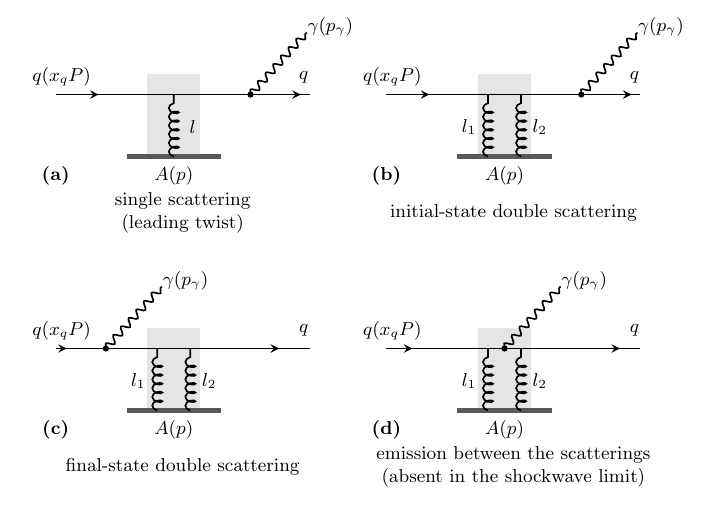} \caption{Representative amplitude-level diagrams for direct photon production in proton--nucleus collisions. The thick line represents the nucleus, the shaded band the region in which the quark crosses its color field, and the curly lines are gluon exchanges with transverse momenta $\bm{l}_{i\perp}$. (a) Single scattering, which gives the leading-twist contribution. (b) Initial-state and (c) final-state double scattering, which contribute at twist 4 together with the single--triple interference (not shown). (d) Photon emission between two scatterings, which is omitted in the shockwave approximation and carries the LPM phase built from the formation time of Eq.~(\ref{eq:cgc-tau}).} \label{fig:diagrams}
\end{figure}

At leading twist (LT) the quark from the proton scatters once off a gluon from the nucleus, Fig.~\ref{fig:diagrams}(a). The cross section takes the standard collinear factorized form~\cite{Owens:1986mp},
\begin{equation}
E_\gamma\frac{d\sigma^{\rm LT}}{d^3p_\gamma}
=\frac{\alpha_{\rm em}\alpha_s}{N_c}\sum_q e_q^2\int dx_q\, f_{q/p}(x_q)\,
x f_{g/A}(x)\,\frac{\xi^2\left[1+(1-\xi)^2\right]}{p_{\gamma\perp}^4}\,,
\label{eq:cgc-LT}
\end{equation}
where $f_{q/p}$ is the quark distribution of the proton and
\begin{equation}
f_{g/A}(x)=\frac{1}{xp^+}\int\frac{dy^-}{2\pi}e^{-ixp^+y^-}
\langle A|F^{+\alpha}(0)F_\alpha^{\ +}(y^-)|A\rangle \,,
\label{eq:cgc-fgA}
\end{equation}
is the gluon distribution of the nucleus, normalized to the total number of gluons in the nucleus.\footnote{With this normalization $f_{g/A}\simeq A f_{g/N}$ in the absence of nuclear modifications. The quark distribution $f_q^A$ of Eq.~(\ref{eq:nuclear-pdf}) can be normalized in the same way.} The nuclear dependence of this term is entirely contained in the $A$ dependence of $f_{g/A}$.

At twist 4 the quark line exchanges two additional soft gluons with the nucleus. The four gluon insertions can be distributed as two in the amplitude and two in its conjugate for the central cut, or as three on one side of the cut and one on the other for the left and right cuts. Because the photon is colorless, the extra scatterings happen either before the hard $q+g\to q+\gamma$ vertex (initial state), after it on the outgoing quark (final state), or one on each side (initial--final interference). Representative diagrams are shown in Fig.~\ref{fig:diagrams}. The generalized factorization formalism~\cite{Qiu:1990xxa,Luo:1993ui} expresses all of these contributions in terms of four-gluon correlation functions.  

The nuclear-enhanced result at twist 4, including initial-state, final-state and interference contributions for all cuts, reads~\cite{Fu:2024sba}
\begin{equation}
E_\gamma\frac{d\sigma^{\rm T4}}{d^3p_\gamma}
=\frac{4\pi^2\alpha_{\rm em}\alpha_s^2}{N_c^2}\sum_q e_q^2\int dx_q\, f_{q/p}(x_q)\,
\frac{1+(1-\xi)^2}{p_{\gamma\perp}^6}
\sum_X\Big[\mathcal{D}_X\,T_X(x_1,x_2,x_3)\Big]_{x_1=x,\,x_2=x_3=0}\,,
\label{eq:cgc-T4}
\end{equation}
with the twist-4 correlation functions
\begin{equation}
T_X(x_1,x_2,x_3)=\int\frac{dy^-dy_1^-dy_2^-}{(2\pi)^2}\,
e^{ix_1p^+y^-}e^{ix_2p^+(y_1^--y_2^-)}e^{ix_3p^+y_2^-}\,
\frac{1}{p^+}\,\mathcal{F}^4_X\,\Theta_X\,.
\label{eq:cgc-TX}
\end{equation}
Here $X$ labels both the cut and the type of scattering. The object $\mathcal{F}^4_X$ is a nuclear matrix element of four field-strength operators $F^{+\perp}$ at positions $0$, $y^-$, $y_1^-$ and $y_2^-$, with an operator ordering set by the cut. The factor $\Theta_X$ is a product of step functions that orders the scatterings along the light cone, and $\mathcal{D}_X$ is a linear combination of the identity and of the derivatives $\partial/\partial x_i$ and $\partial^2/\partial x_i\partial x_j$. For instance, for initial-state double scattering (central cut)
\begin{align}
\mathcal{F}^4_{C,I}&=\langle A|F^{+\alpha}(y_2^-)F^{+\beta}(0)F_\beta^{\ +}(y^-)F_\alpha^{\ +}(y_1^-)|A\rangle,
\qquad
\Theta_{C,I}=\theta(y^--y_1^-)\theta(-y_2^-)\,,
\nonumber\\
\mathcal{D}_{C,I}&=\xi^4x^2\frac{\partial^2}{\partial x_1^2}-3\xi^4x\frac{\partial}{\partial x_1}+(1-\xi)\xi^3x\frac{\partial}{\partial x_2}+4\xi^4\,,
\label{eq:cgc-DCI}
\end{align}
and the full list can be found in~\cite{Fu:2024sba}. The derivative terms are a generic feature of the collinear expansion, since the transverse momentum $\bm{k}_\perp$ of the soft gluons enters the hard part together with the longitudinal phases, and expanding in $k_\perp$ differentiates those phases, exactly as in the sine term of Eq.~(\ref{eq:collinear-expansion}). The non-derivative term, with $\mathcal{D}_X\to 4\xi^4$, is the one that would be obtained by neglecting the longitudinal momentum exchanged with the medium.

The phase $e^{ixp^+y^-}$ in Eq.~(\ref{eq:cgc-TX}) restricts the separation between the two hard vertices to $y^-\lesssim 1/(xp^+)$, which is the coherence length of the probe. For $x\sim\mathcal{O}(1)$ this length is shorter than the nucleon size, and the hard scattering is localized on a single nucleon.\footnote{In the nuclear rest frame the coherence length is $\lambda_c\simeq 1/(2m_Nx)\approx 0.1\,{\rm fm}/x$~\cite{Qiu:2003vd}. It is shorter than the nucleon radius, $r_0\simeq 1.2$~fm, for $x\gtrsim 0.1$, which is the incoherent regime of this subsection, and longer than the nuclear radius, $r_0A^{1/3}$, for $x\lesssim 0.1A^{-1/3}$, which is the coherent regime of Sec.~\ref{sec:HT-smallx}. In between, the probe interacts coherently with a few nucleons along its path.} The step functions then decide which twist-4 contributions are enhanced by the nuclear size. For final-state scattering the three cuts combine into a single ordered integral, $y^->y_2^->y_1^->0$. The positions $y_1^-$ and $y_2^-$ are then confined to the same nucleon as the hard vertex, and the contribution carries no $A^{1/3}$ enhancement. Such contributions are usually called contact terms, and the same happens for the initial--final interference~\cite{Fu:2024sba}. Only the initial-state double scattering survives, since the soft gluons at $y_1^-$ and $y_2^-$ may then come from a nucleon far from the one struck at the hard vertex. In this limit Eq.~(\ref{eq:cgc-T4}) reduces to
\begin{equation}
E_\gamma\frac{d\sigma^{\rm T4}}{d^3p_\gamma}\bigg|_{x\sim\mathcal{O}(1)}
=\frac{4\pi^2\alpha_{\rm em}\alpha_s^2}{N_c^2}\sum_q e_q^2\int dx_q\, f_{q/p}(x_q)\,
\frac{1+(1-\xi)^2}{p_{\gamma\perp}^6}\,\xi^4
\left[x^2\frac{\partial^2T^{I}(x)}{\partial x^2}-x\frac{\partial T^{I}(x)}{\partial x}+T^{I}(x)\right]\,,
\label{eq:cgc-T4-largex}
\end{equation}
with
\begin{equation}
T^{I}(x)=\int\frac{dy^-dy_1^-dy_2^-}{(2\pi)^2}\,e^{ixp^+y^-}\,
\frac{1}{xp^+}\langle A|F^{+\alpha}(0)F^{+\beta}(y_2^-)F_\beta^{\ +}(y_1^-)F_\alpha^{\ +}(y^-)|A\rangle\,
\theta(y^--y_1^-)\theta(-y_2^-) \,.
\label{eq:cgc-TI}
\end{equation}
This coincides with earlier calculations that kept only initial-state double scattering from the start~\cite{Guo:1995zk,Kang:2013ufa}, which is therefore justified at large $x$, where the final-state and interference contributions reduce to contact terms. The same structure, with second-derivative, first-derivative and non-derivative terms sharing one hard function, was found for single inclusive hadron and prompt photon production at large $x$~\cite{Kang:2013ufa}.

At large $x$, the hard process takes place on one nucleon, and the incoming quark rescatters once on a different nucleon on its way in. The two soft gluon fields at $y_1^-$ and $y_2^-$ carry vanishing momentum fraction, $x_2=x_3=0$. In the maximal two-gluon correlation approximation of Sec.~\ref{sec:TMB}, the matrix element in Eq.~(\ref{eq:cgc-TI}) then factorizes into the gluon distribution of the struck nucleon times the path integral of the same soft gluon density that defines the transport coefficient $\hat q$ in Eq.~(\ref{eq:qhat-TMD}). This is the origin of the parametrization $T^{I}(x)\simeq\lambda^2A^{1/3}f_{g/A}(x)$, with $\lambda^2$ a scale of order $10^{-2}$ to $10^{-1}$~GeV$^2$ extracted from transverse momentum broadening data~\cite{Luo:1993ui,Kang:2008us}. For processes with colored final states, such as semi-inclusive deep inelastic scattering (SIDIS) or hadron production, the final-state double scattering is nuclear-enhanced as well. It is responsible for the energy loss and broadening discussed in Secs.~\ref{sec:eloss} and~\ref{sec:TMB}. While incoherent in the sense that the probe resolves individual nucleons, this regime still contains the interference phases $1-\cos(x_Lp^+y_1^-)$ between emissions from successive scattering centers. Those phases control the formation time of the radiated gluon rather than that of the hard process.

\subsection{High-twist expansion at small $x$: coherent multiple scattering}

\label{sec:HT-smallx}

As $x$ decreases the coherence length $1/(xp^+)$ grows, and once it exceeds the longitudinal extent of the nucleus, $xp^+L_A^-\ll 1$, the phases $e^{ixp^+\Delta y^-}$ in Eq.~(\ref{eq:cgc-TX}) can be set to one across the entire nucleus. The probe no longer resolves individual nucleons, and it interacts coherently with all the nucleons along its path. First, the contact-term argument of the previous subsection no longer applies, because without the localizing phase the ordered integrals over $y_1^-$ and $y_2^-$ extend over the whole nucleus, and the final-state and interference contributions become nuclear-enhanced too. Second, if the derivative terms are subleading at small $x$, in the sense that $x\,\partial T_X/\partial x_i\ll T_X$, the sum over cuts and scattering types in Eq.~(\ref{eq:cgc-T4}) collapses to the non-derivative term,
\begin{align}
E_\gamma\frac{d\sigma^{\rm LT+T4}}{d^3p_\gamma}\bigg|_{x\to 0}
&\simeq\frac{\alpha_{\rm em}\alpha_s}{N_c}\sum_q e_q^2\int dx_q\, f_{q/p}(x_q)\,
\frac{\xi^2\left[1+(1-\xi)^2\right]}{p_{\gamma\perp}^4}
\nonumber\\
&\quad\times\left[xf_{g/A}(x)+\frac{(2\pi)^2\alpha_s}{N_c}\frac{4\xi^2}{p_{\gamma\perp}^2}T_{gg}(x,0,0)+\cdots\right]\,,
\label{eq:cgc-T4-smallx}
\end{align}
with
\begin{equation}
T_{gg}(x_1,x_2,x_3)\equiv\frac{1}{4}\left[T_{C,I}+T_{C,IF}+T_{C,FI}+T_{C,F}\right](x_1,x_2,x_3)\,,
\label{eq:cgc-Tgg}
\end{equation}
provided all four central-cut correlators share the same small-$x$ limit~\cite{Fu:2023jqv}. 

Eq.~\eqref{eq:cgc-T4-smallx} gives the leading-twist and twist-4 terms of the series. The twist-4 term carries a relative factor of $\alpha_sT_{gg}/(xf_{g/A}\,p_{\gamma\perp}^2)$. At fixed $x$, the ratio $T_{gg}/f_{g/A}$ grows as $A^{1/3}$, so the suppression is governed by the transverse momentum scale accumulated along the nuclear path. In the coherent regime each additional pair of soft gluons can attach anywhere along that path and brings one more factor of $A^{1/3}$, so the expansion is in powers of this nuclear scale over the hard scale, and it must be resummed when the two become comparable. For DIS and for forward hadron production in proton--nucleus collisions, the maximally nuclear-enhanced terms have been resummed at leading power in each scattering, which yields the dynamical shadowing used in the global analysis of Sec.~\ref{sec:TMB-global}~\cite{Qiu:2003vd,Qiu:2004da}. The nuclear scale accumulated along the path is the counterpart of $Q_s^2$ in the CGC, and the precise relation between the two expansions at twist 4 is the subject of Secs.~\ref{sec:CGC-LT} and~\ref{sec:CGC-HT}.

A second point of contact is the evolution. At small $x$ the growth of the gluon density is tamed in the collinear framework by gluon recombination, which adds a nonlinear term to the DGLAP equation~\cite{Gribov:1984tu,Mueller:1985wy},
\begin{equation}
\frac{\partial xG(x,Q^2)}{\partial\ln Q^2}=\frac{\alpha_s}{2\pi}\int_x^1\frac{dz}{z}\,P_{gg}(z)\,\frac{x}{z}G\!\left(\frac{x}{z},Q^2\right)
-\frac{81\,\alpha_s^2(Q^2)}{16\,R^2Q^2}\int_x^1\frac{dx'}{x'}\left[x'G(x',Q^2)\right]^2\,,
\label{eq:GLR-MQ}
\end{equation}
written here for $N_c=3$, with quark contributions omitted. The second term is a twist-4 contribution: it is suppressed by $1/Q^2$, and its four-gluon correlator has been approximated by the square of the gluon distribution spread over a transverse area $\pi R^2$. This is the same factorization that underlies the maximal two-gluon correlation approximation of Sec.~\ref{sec:TMB}. For a nucleus, $xG_A\propto A$ and $R_A^2\propto A^{2/3}$, so the recombination term relative to the linear one scales as $A^{1/3}/Q^2$, the same nuclear enhancement discussed above. In the CGC the corresponding equation is the Balitsky--Kovchegov (BK) equation for the dipole amplitude $N(\bm{x}_{01},Y)$ at rapidity $Y=\ln(1/x)$~\cite{Balitsky:1995ub,Kovchegov:1999yj},
\begin{equation}
\frac{\partial N(\bm{x}_{01},Y)}{\partial Y}=\frac{\bar\alpha_s}{2\pi}\int d^2\bm{x}_2\,\frac{x_{01}^2}{x_{02}^2x_{12}^2}
\Big[N(\bm{x}_{02},Y)+N(\bm{x}_{12},Y)-N(\bm{x}_{01},Y)-N(\bm{x}_{02},Y)N(\bm{x}_{12},Y)\Big]\,,
\label{eq:BK}
\end{equation}
with $\bar\alpha_s=\alpha_sN_c/\pi$. Its linear part is the BFKL equation. At small dipole size the amplitude is proportional to $\alpha_s\,x_{01}^2\,xG(x,1/x_{01}^2)/R^2$, and in the double logarithmic approximation, where the daughter dipoles are much larger than the parent, the linear terms reduce to the small-$x$ limit of DGLAP while the quadratic term reproduces the recombination term of Eq.~(\ref{eq:GLR-MQ})~\cite{Kovchegov:1999yj}. The two equations therefore agree on the leading nonlinear correction and differ in everything else. BK keeps all powers of $N$, which in the language of this section are all the maximally nuclear-enhanced twists, and the full BFKL kernel in $\ln(1/x)$, but it contains DGLAP evolution only in the double logarithmic limit. GLR--MQ keeps the full DGLAP kernel but truncates the series at twist 4 and relies on the factorized form of the four-gluon correlator. Neither describes the evolution of a genuine, unfactorized twist-4 correlator such as $T_{gg}$, which is what enters Eq.~(\ref{eq:cgc-T4-smallx}).

\subsection{Matching the CGC to leading-twist factorization}

\label{sec:CGC-LT}

In the CGC the nucleus is described by a classical background field, which in covariant gauge has a single large component $A^+(y^-,\bm{y}_\perp)$ localized in $y^-$. The quark from the proton, moving along the minus light-cone direction, crosses this field at fixed transverse position and acquires an eikonal phase. All its scatterings resum into the light-like Wilson line in the fundamental representation, and the effective quark--nucleus vertex reads~\cite{McLerran:1994vd}
\begin{equation}
\Gamma_q(l)=(2\pi)\delta(l^-)\gamma^-\int d^2y_\perp\,e^{-i\bm{l}_\perp\cdot\bm{y}_\perp}\,V(\bm{y}_\perp),
\qquad
V(\bm{y}_\perp)=\mathcal{P}\exp\left[ig\int_{-\infty}^{\infty}dy^-A^+(y^-,\bm{y}_\perp)\right]\,,
\label{eq:cgc-vertex}
\end{equation}
where $l$ is the momentum transferred from the nucleus to the quark. Two approximations are built into Eq.~(\ref{eq:cgc-vertex}): the nucleus transfers transverse momentum but no longitudinal momentum, hence the $\delta(l^-)$ and the absence of any dependence on $l^+$, and the quark does not radiate while inside the nucleus, since the field is treated as a shockwave of vanishing extent. We work within the hybrid factorization, in which the proton is described by collinear parton distributions and the nucleus by Wilson line correlators~\cite{Dumitru:2005gt}. At leading order the direct photon cross section then reads~\cite{Gelis:2002ki}
\begin{equation}
E_\gamma\frac{d\sigma^{\rm CGC}}{d^3p_\gamma}
=\frac{\alpha_{\rm em}}{2\pi^2}\sum_q e_q^2\int dx_q\, f_{q/p}(x_q)\,
\xi^2\left[1+(1-\xi)^2\right]
\int\frac{d^2l_\perp}{(2\pi)^2}\,
\frac{l_\perp^2\,F(x,\bm{l}_\perp)}{(\xi\bm{l}_\perp-\bm{p}_{\gamma\perp})^2\,p_{\gamma\perp}^2}\,,
\label{eq:cgc-xsec}
\end{equation}
where the dipole distribution is the Fourier transform of the two-point correlator of Wilson lines,
\begin{equation}
F(x,\bm{l}_\perp)=\int d^2y_\perp d^2y'_\perp\,e^{-i\bm{l}_\perp\cdot(\bm{y}_\perp-\bm{y}'_\perp)}\,
\frac{1}{N_c}\left\langle{\rm Tr}\left[V^\dagger(\bm{y}'_\perp)V(\bm{y}_\perp)\right]\right\rangle_x\,.
\label{eq:cgc-dipole}
\end{equation}
The average $\langle\cdots\rangle_x$ is taken over the color sources of the nucleus, and it depends on $x$ through the BK/JIMWLK evolution of the sources. The saturation scale is implicit in $F$, as the value of $l_\perp$ at which $l_\perp^2F(x,\bm{l}_\perp)$ peaks, and it may be interpreted as the typical transverse momentum transferred from the nucleus to the quark--photon pair. Eq.~(\ref{eq:cgc-xsec}) contains the collinear singularities of photon emission from the incoming and outgoing quark. For direct photons they are removed by an isolation criterion or, equivalently for our purpose, by expanding in inverse powers of $p_{\gamma\perp}^2$.

Expanding the denominator of Eq.~(\ref{eq:cgc-xsec}) in $l_\perp^2/p_{\gamma\perp}^2$ organizes the cross section as a series of moments $\int d^2l_\perp\,l_\perp^{2n}F(x,\bm{l}_\perp)$. Each factor of $l_\perp^2$ becomes a pair of transverse derivatives acting on the Wilson lines, and each derivative turns a gauge field $A^+$ into a field strength $F^{+\perp}$, up to terms that rebuild the gauge link. Expanding the Wilson lines in powers of the background field and identifying the CGC average with the nuclear expectation value, $\langle\mathcal{O}\rangle_x=\langle A|\mathcal{O}|A\rangle/\langle A|A\rangle$, then relates the $n$-th moment of the dipole distribution to a collinear matrix element of $2n$ field strengths at vanishing longitudinal momentum~\cite{Dominguez:2011wm,Fu:2024sba}. The second transverse-momentum moment (the first moment in $l_\perp^2$) gives the collinear gluon distribution of the nucleus at small $x$~\cite{Kang:2012vm,Dominguez:2011wm},
\begin{equation}
\lim_{x\to 0}xf_{g/A}(x)\simeq\frac{N_c}{2\pi^2\alpha_s}\int\frac{d^2l_\perp}{(2\pi)^2}\,l_\perp^2\,F(x,\bm{l}_\perp)\,,
\label{eq:cgc-moment2}
\end{equation}
and inserting it into the leading term of the expansion gives
\begin{equation}
E_\gamma\frac{d\sigma^{\rm CGC}}{d^3p_\gamma}\bigg|_{\rm LT}
=\frac{\alpha_{\rm em}\alpha_s}{N_c}\sum_q e_q^2\int dx_q\, f_{q/p}(x_q)\,
\frac{\xi^2\left[1+(1-\xi)^2\right]}{p_{\gamma\perp}^4}\,
xf_{g/A}(x)\Big|_{x\to 0}\,,
\label{eq:cgc-LT-match}
\end{equation}
which coincides with Eq.~(\ref{eq:cgc-LT}) in the strict limit $x\to 0$. More precisely, the two agree when the phase $e^{ixp^+y^-}$ in the definition of $f_{g/A}$ can be neglected over the nuclear size, $xA^{1/3}\ll 1$. This is the dilute limit of the CGC, and it has been verified for a number of observables, among them photon and dilepton production~\cite{Gelis:2002ki,Baier:2004tj}, heavy quark production~\cite{Gelis:2003vh}, the Sivers asymmetry in Drell--Yan~\cite{Kang:2012vm}, and single inclusive hadron production in proton--nucleus collisions. In the latter case the matching at one loop takes the form of a rapidity factorization between the collinear and small-$x$ evolutions~\cite{Chirilli:2011km,Kang:2014lha}. More recently the matching has been formulated at the level of the distributions rather than of a cross section. The collinear quark and gluon distributions of a nucleus can be computed in the shockwave formalism and matched onto collinear factorization for inclusive and exclusive DIS~\cite{Bhattacharya:2025fnz}. The next-to-eikonal corrections to the dipole picture of DIS and SIDIS reproduce the parton-model contribution of the collinear quark distribution, which is absent at eikonal accuracy~\cite{Altinoluk:2025ang}. These results extend Eq.~(\ref{eq:cgc-moment2}) from the gluon to the quark sector, where the leading-twist term is itself a subeikonal effect.

The same logic applies to observables with a semi-hard scale. In the back-to-back limit of dijet or dihadron production, the leading-order CGC cross section can be written as hard factors multiplying a small set of small-$x$ gluon distributions, the Weizs\"acker--Williams distribution and the dipole distribution of Eq.~(\ref{eq:cgc-dipole}) among them~\cite{Dominguez:2011wm,Marquet:2016cgx}. These distributions differ from one another at $l_\perp\lesssim Q_s$, but at $l_\perp\gg Q_s$ all of them reduce to the same perturbative tail, proportional to the collinear gluon distribution and falling as $1/l_\perp^2$~\cite{Dominguez:2011wm,Xiao:2017yya}. This is the semi-hard counterpart of Eq.~(\ref{eq:cgc-moment2}). The improved TMD factorization keeps the full $l_\perp$ dependence of the hard factors together with the small-$x$ distributions and interpolates between the two limits~\cite{Kotko:2015ura,Altinoluk:2019fui}. The extension of this matching beyond leading order and leading power belongs to the small-$x$ TMD program~\cite{Boussarie:2023izj}.

\subsection{Matching the CGC to the high-twist expansion}

\label{sec:CGC-HT}

The next term in the expansion of Eq.~(\ref{eq:cgc-xsec}) in $1/p_{\gamma\perp}^2$ is where the CGC and the HT expansion are expected to meet at twist 4. In this subsection we examine how the fourth moment of the dipole distribution maps onto the four-gluon correlators of Eq.~(\ref{eq:cgc-TX}), why the eikonal CGC reproduces only part of the HT result, and which two ingredients restore the agreement at finite $x$.

In covariant gauge the field strength is $F^{+\alpha}=\partial_\perp^\alpha A^+$, so that a transverse derivative of the Wilson line inserts a field strength between two segments of the line,\footnote{Strictly speaking the transverse components of the gauge field should be kept as well. Gauge invariance then replaces $\partial^\alpha_\perp A^+$ by the full $F^{+\alpha}$, and the identities below hold with that replacement.}
\begin{equation}
\frac{\partial V(\bm{y}_\perp)}{\partial y_\perp^\alpha}
=ig\int_{-\infty}^{\infty}dy_1^-\,\left[\infty,y_1^-\right]_{\bm{y}_\perp}F^{+}_{\ \alpha}(y_1^-,\bm{y}_\perp)\left[y_1^-,-\infty\right]_{\bm{y}_\perp}\,,
\label{eq:cgc-dV}
\end{equation}
where $[a,b]_{\bm{y}_\perp}$ denotes the gauge link along $y^-$ from $b$ to $a$ at fixed transverse position. The second transverse-momentum moment in Eq.~\eqref{eq:cgc-moment2} is the correlator of one derivative of $V$ with one derivative of $V^\dagger$. By Eq.~(\ref{eq:cgc-dV}) it is a two-point function of field strengths joined by gauge links, and its term of order $g^2$ is the gluon distribution at $x=0$.

The fourth moment involves two derivatives on each line, and the second derivative of $V$ contains two kinds of terms: a single insertion of $\partial_\perp F^{+\alpha}$, and a double insertion of two field strengths ordered along $y^-$. The first kind is a higher-derivative operator at a single point. It is not enhanced by the nuclear size and is dropped, consistently with the neglect of such terms in the HT formalism. The second kind, with the ordering supplied by the path ordering of the Wilson line, is what the step functions $\Theta_X$ encode on the HT side. With this selection the identification reads~\cite{Fu:2023jqv,Fu:2024sba},\footnote{Two kinds of power corrections appear in the CGC: kinematic power corrections, of the form $l_\perp/p_{\gamma\perp}$, obtained by expanding the hard factor while keeping the Wilson lines intact, and saturation, or genuine, power corrections, of the form $Q_s/p_{\gamma\perp}$, associated with multi-gluon correlators~\cite{Altinoluk:2019wyu,Boussarie:2021ybe}. Eq.~(\ref{eq:cgc-moment4}) is the genuine twist-4 piece of the fourth moment, which is the part that carries the $A^{1/3}$ enhancement of a second scattering.}
\begin{equation}
\lim_{x\to 0}T_{gg}(x,0,0)\simeq\frac{N_c^2}{2(2\pi)^4\alpha_s^2}
\int\frac{d^2l_\perp}{(2\pi)^2}\,l_\perp^4\,F(x,\bm{l}_\perp)\bigg|_{\rm T4}\,,
\label{eq:cgc-moment4}
\end{equation}
with $T_{gg}$ the average of the four central-cut correlators defined in Eq.~(\ref{eq:cgc-Tgg}). The CGC side produces a single trace of four field strengths in the fundamental representation, whereas the HT correlators carry adjoint color indices contracted pairwise. The two coincide if the color correlations in the nucleus follow the pairing of the Lorentz indices, which is the case in the Gaussian models used in practice but is otherwise an assumption on the nuclear matrix element.

The twist-4 term of the eikonal cross section is proportional to $T_{gg}(x,0,0)$ with the coefficient $4\xi^4$, and it reproduces only the non-derivative term of Eq.~(\ref{eq:cgc-T4-smallx}), under the further assumption that all four central-cut correlators have the same small-$x$ limit. The derivative terms of $\mathcal{D}_X$ are absent since they arise from derivatives of the longitudinal phase factors with respect to the momentum fractions, and those phases were discarded when the field was treated as a shockwave. One could argue that the derivative terms come with powers of $x$ and should be dropped at small $x$. As noted in Sec.~\ref{sec:HT-smallx}, however, this presumes a hierarchy $x\,\partial T_X/\partial x_i\ll T_X$ that has not been tested, and the correlators could well vary as fast as the gluon distribution itself. We conclude that the collinear expansion and the eikonal limit do not commute: the twist-4 expansion of the eikonal CGC result is the small-$x$ limit of the HT result, not the HT result itself~\cite{Fu:2023jqv,Fu:2024sba}.

Two ingredients restore the agreement at finite $x$, and both go beyond the shockwave approximation: i) the longitudinal phases that accompany each insertion of the background field once the nucleus is allowed to transfer longitudinal momentum, and ii) the diagram with photon emission between two scatterings, Fig.~\ref{fig:diagrams}(d), which the shockwave forbids. 

Once a longitudinal momentum $l^+$ is transferred, each insertion of $A^+$ in the expansion of Eq.~(\ref{eq:cgc-vertex}) acquires a factor $e^{il^+y^-}$, and the insertions no longer exponentiate into a Wilson line. The single, double and single--triple interference contributions are therefore computed one by one with the full QCD vertices. For single scattering the replacement is $V(\bm{y}_\perp)-1\to ig\int dy^-\,e^{ixp^+y^-}A^+(y^-,\bm{y}_\perp)$. The squared amplitude, expanded to leading power in $1/p_{\gamma\perp}^2$, contains a factor $l_\perp^2$ that becomes a pair of transverse derivatives acting on $\langle A^+(y)A^+(y')\rangle_x$, exactly as in Sec.~\ref{sec:CGC-LT}, while the phase $e^{ixp^+(y^--y'^-)}$ survives. The result is Eq.~(\ref{eq:cgc-LT}) with the gluon distribution of Eq.~(\ref{eq:cgc-fgA}) at finite $x$ rather than in its $x\to 0$ limit~\cite{Fu:2024sba}. This is the only change at leading twist, and it already shows what the phases do: they keep track of how far apart in $y^-$ the two hard vertices are allowed to be.

The double scattering amplitude has three pieces, according to whether the photon is emitted after both scatterings, before both, or between them, Fig.~\ref{fig:diagrams}(b), (c) and (d). In each piece the integral over the longitudinal momentum of one gluon is done by residues, which puts the quark propagator between the two scatterings on shell. The first two pieces have one pole each. The third has two, one for the propagator before the photon emission and one for the propagator after it, and the two residues differ by a phase,
\begin{equation}
\mathcal{M}_{(d)}\propto 1-e^{-i\Delta y^-/\tau_\gamma}\,,
\qquad
\tau_\gamma=\frac{2x_qP^-\xi(1-\xi)}{(\bm{p}_{\gamma\perp}-\xi\bm{\ell}_\perp)^2}\,,
\label{eq:cgc-tau}
\end{equation}
where $\Delta y^-$ is the separation of the two scattering centers, $\bm{\ell}_\perp$ is the transverse momentum exchanged in the first scattering, and $\tau_\gamma$ is the formation time of the photon, the analogue of Eq.~(\ref{eq:formation-time}). In the high-energy limit $P^-\to\infty$ the formation time exceeds any separation, the bracket vanishes, and the third piece drops out, consistently with a photon that cannot resolve two scattering centers inside the shockwave. At finite $x$ the cancellation is incomplete, and the remainder is the Landau--Pomeranchuk--Migdal (LPM) interference~\cite{Landau:1953gr,Migdal:1956tc} familiar from parton energy loss~\cite{Baier:1996sk,Zakharov:1996fv,Gyulassy:2000fs}. Each of the two residues shares its phase with one of the first two pieces, which suggests combining the emission-after piece with the first residue and the emission-before piece with the second. Each combination vanishes when the momentum of the soft gluon goes to zero, as gauge invariance requires, and after the expansion in $1/p_{\gamma\perp}^2$ the two combinations become the initial-state and final-state contributions of the HT classification, with the soft gluon attached before or after the hard vertex~\cite{Fu:2024sba}.

The expansion in $1/p_{\gamma\perp}^2$ proceeds in two steps. The hard factor of each double scattering contribution depends on the transverse momenta of the two soft gluons and on the momentum $\bm{L}_\perp$ flowing through the hard vertex. Expanding first in the soft momenta turns them into derivatives of transverse delta functions, which promote the two soft gauge fields into field strengths, as for single scattering. Expanding then in $\bm{L}_\perp$ to quadratic order yields two terms: a term without derivatives on the remaining pair of fields, which is the expansion of the gauge link of the leading-twist gluon distribution, and a term with two more derivatives, which is the genuine twist-4 contribution. The coefficient of the latter is a polynomial in the separations along the light cone. For initial-state double scattering with the central cut,
\begin{align}
\mathcal{D}_{C,I}&\to\xi^2\Big[4\xi^2+\xi(1-\xi)\big(ixp^+\Delta y_{12}^-\big)-3\xi^2\big(ixp^+\Delta y^-\big)+\xi^2\big(ixp^+\Delta y^-\big)^2\Big]\,,
\nonumber\\
ixp^+\Delta y^-&\leftrightarrow x\frac{\partial}{\partial x_1}\,,\qquad
ixp^+\Delta y_{12}^-\leftrightarrow x\frac{\partial}{\partial x_2}\,,
\label{eq:cgc-DCI-phases}
\end{align}
where $\Delta y^-$ is the separation of the two hard vertices and $\Delta y_{12}^-$ that of the two soft insertions. Acting on the phases of Eq.~(\ref{eq:cgc-TX}), the powers of $ixp^+\Delta y^-$ are the derivatives with respect to the momentum fractions, and Eq.~(\ref{eq:cgc-DCI-phases}) reproduces the operator $\mathcal{D}_{C,I}$ of Eq.~(\ref{eq:cgc-DCI}) term by term. The same holds for the other cuts and for the final-state and interference contributions, each with its own polynomial and ordering $\Theta_X$~\cite{Fu:2024sba}. The physical content of Eq.~(\ref{eq:cgc-DCI-phases}) is that the derivative terms are an expansion in $\Delta y^-/\lambda_c$, the separation of the scattering centers in units of the coherence length $\lambda_c=1/(xp^+)$ of Sec.~\ref{sec:HT-largex}. The eikonal limit sets this ratio to zero and keeps the constant term, which is the coherent regime of Sec.~\ref{sec:HT-smallx}, whereas the HT expansion keeps the ratio exactly, which is what allows it to interpolate toward the incoherent regime. With the phases and the LPM pieces included, the CGC and HT results agree through twist 4 at leading order,
\begin{equation}
E_\gamma\frac{d\sigma^{\rm CGC}_{\rm sub}}{d^3p_\gamma}\bigg|_{p_{\gamma\perp}>Q_s}
=E_\gamma\frac{d\sigma^{\rm LT}}{d^3p_\gamma}+E_\gamma\frac{d\sigma^{\rm T4}}{d^3p_\gamma}+\cdots\,,
\label{eq:cgc-final}
\end{equation}
for all cuts and for initial-state, final-state and interference contributions~\cite{Fu:2024sba}. Two further terms were left out on both sides, the twist-4 expansion of the single scattering contribution and the interference between single and double scattering, because they produce higher derivatives of the field strength at a single point and carry no nuclear enhancement. The longitudinal phases kept here are the simplest member of the family of next-to-eikonal corrections, and the same corrections have since been computed from the CGC side for forward parton--nucleus scattering, for dijets in DIS and for DIS off a large nucleus in a Gaussian model~\cite{Altinoluk:2024dba,Altinoluk:2024zom}.

\section{Conclusions}
\label{sec:conclusions}
Multiple parton scattering provides a common origin for energy loss and transverse momentum broadening in cold nuclear matter. The HT approach expresses their leading nuclear-enhanced contributions through twist-4 correlation functions, connecting observable modifications of hard probes to the quark and gluon structure of the nucleus. A quantitative description of these effects makes parton propagation a probe of nuclear correlations and constrains the CNM contributions needed to interpret QGP observables.

The energy-loss calculations reviewed here connect medium-induced radiation to the evolution of fragmentation functions, with LPM interference retaining the formation-time dependence. With initial conditions built from multiple induced emissions, they describe much of the pion and kaon attenuation measured by HERMES and yield $\hat q_{F0}\simeq0.020\pm0.005~\mathrm{GeV^2/fm}$ at the center of a large nucleus~\cite{Chang:2014fba}. Applying the same mechanism to the incoming parton in DY production illustrates how constraints from $e+A$ can inform $p+A$ calculations. The additional suppression from energy loss is modest over much of the E866 kinematic range, where target shadowing dominates, but becomes more important at large projectile momentum fraction and lower beam energies~\cite{Xing:2011fb}. 

Broadening provides a complementary test of this description. The NLO calculations of weighted SIDIS and DY cross sections absorb collinear singularities into the appropriate parton distributions, fragmentation functions, and twist-4 matrix elements, leaving finite hard coefficients~\cite{Kang:2013raa,Kang:2014ela,Kang:2016ron}. Their common quark--gluon evolution structure establishes a universality test for initial- and final-state rescattering at this order. Comparisons with quarkonium and particle-pair production extend the sensitivity to other color configurations and the interplay of initial- and final-state rescattering~\cite{Kang:2008us,Xing:2012ii}. Global analyses favor an effective $\hat q(x,\mu^2)$ with kinematic dependence over a single constant across the fitted datasets~\cite{Ru:2019qvz,Ru:2023ars}. Together, these results support comparisons across processes while making clear that the extracted medium properties depend on the resolution scale.

Extending this description toward small $x$ requires control over coherence. At hard scales well above $Q_s$, the leading term of the CGC expansion reproduces the collinear gluon contribution in the small-$x$ limit. At twist 4, the eikonal CGC calculation for direct-photon production reproduces only the non-derivative terms of the HT result. Retaining longitudinal momentum transfer and emission between successive scatterings restores agreement with the full HT result through twist 4 at leading order~\cite{Fu:2023jqv,Fu:2024sba}. The associated LPM interference connects this matching to the formation-time physics of energy loss. It provides a starting point for extending the correspondence to other processes and to radiative corrections.

A common quantitative interpretation of energy loss and broadening still requires several ingredients. The central, local $\hat q_{F0}$ used in the attenuation analyses and the geometry-averaged coefficient used in broadening fits must be compared using consistent density profiles and path-length averaging. The off-diagonal correlations entering radiation and twist-4 evolution remain poorly constrained, while multiple emissions require modeling. Discrepancies for baryons and at large hadron momentum fraction also call for a more complete treatment of hadronization and hadronic interactions. These limitations must be addressed when transferring constraints between reactions or using them to interpret hot-medium effects.

The EIC can sharpen these tests through measurements of attenuation, broadening, and particle-pair imbalance over a broad range of energies, scales, and nuclear species~\cite{Accardi:2012qut,AbdulKhalek:2021gbh,Ru:2023ars}. Combined with $p+A$ measurements at RHIC and the LHC~\cite{Arleo:2025oos}, they can test how far a common set of nuclear correlations accounts for different observables and where coherent multiple scattering requires a resummed description. Such progress would establish a quantitative account of color propagation and coherence in nuclei while strengthening the CNM baseline for determining QGP properties.

\begin{ack}[Acknowledgments]%
 Z.K. is supported by the National Science Foundation under grant No. PHY-2515057. F.~S. is supported by the Laboratory Directed Research and Development of Brookhaven National Laboratory and RIKEN-BNL Research Center, as well as the National Science Foundation (NSF) within the framework of the JETSCAPE collaboration, under grant number OAC-2514008 (CSSI:C-SCAPE). H.X. is supported by the National Natural Science Foundation of China under Project No. 12525508 and No. 12475139. X.-N.~W. is supported by the National Natural Science Foundation of China under Project No. 12535010.

 \end{ack}

\bibliographystyle{Numbered-Style} 
\bibliography{reference}

\end{document}